\documentclass[fleqn,usenatbib]{mnras}
\usepackage[T1]{fontenc}
\usepackage{ae,aecompl}
\usepackage{graphicx}
\usepackage{amsmath}
\usepackage{amssymb}
\usepackage{lineno}

\usepackage{mathrsfs}
\usepackage{hyperref}
\usepackage{pdflscape}
\usepackage{float}
\usepackage{here}

\newcommand\xmm{{\it XMM-Newton}}

\newcommand\nustar{{\it NuSTAR}}

\newcommand\keV{{\rm~keV}}
\newcommand\ev{{\rm~eV}}

\title[High-density disc reflection spectroscopy]{High-density disc reflection spectroscopy of low-mass active galactic nuclei}

\author[L. Mallick et al.]{L. Mallick$^{1}$ \thanks{Email: lmallick@caltech.edu}, 
A. C. Fabian$^{2}$, J. A. Garc\'ia$^{1,3}$, J. A. Tomsick$^{4}$, M. L. Parker$^{2}$, T. Dauser$^{3}$, D. R. Wilkins$^{5}$,   
\newauthor B. De Marco$^{6,7}$, J. F. Steiner$^{8}$, R. M. T. Connors$^{1}$, G. Mastroserio$^{1}$, A. G. Markowitz$^{6,9}$, C. Pinto$^{10}$, 
\newauthor W. N. Alston$^{11}$, A. M. Lohfink$^{12}$, P. Gandhi$^{13}$   \\
$^{1}$ Cahill Center for Astronomy and Astrophysics, California Institute of Technology, Pasadena, CA 91125, USA \\
$^{2}$ Institute of Astronomy, University of Cambridge, Madingley Road, Cambridge CB3 0HA, UK\\
$^{3}$ Dr. Karl Remeis-Observatory and Erlangen Centre for Astroparticle Physics, Sternwartstr.~7, 96049 Bamberg, Germany \\
$^{4}$ Space Sciences Laboratory, 7 Gauss Way, University of California, Berkeley, CA 94720-7450, USA \\
$^{5}$ Kavli Institute for Particle Astrophysics and Cosmology, Stanford University, 452 Lomita Mall, Stanford, CA 94305, USA \\
$^{6}$ Nicolaus Copernicus Astronomical Center, Polish Academy of Sciences, Bartycka 18, 00-716, Warsaw, Poland \\
$^{7}$ Departament de F\'isica, EEBE, Universitat Polit\'ecnica de Catalunya, Av. Eduard Maristany 16, E-08019, Barcelona, Spain \\
$^{8}$ Center for Astrophysics, Harvard \& Smithsonian, 60 Garden St. Cambridge, MA 02138, USA \\
$^{9}$ University of California, San Diego, Center for Astrophysics and Space Sciences, 9500 Gilman Dr, La Jolla, CA 92093, USA \\
$^{10}$ INAF - IASF Palermo, Via U. La Malfa 153, I-90146 Palermo, Italy \\
$^{11}$ European Space Agency (ESA), European Space Astronomy Centre (ESAC), E-28691 Villanueva de la Ca\~{n}ada, Madrid, Spain \\
$^{12}$ eXtreme Gravity Institute, Department of Physics, Montana State University, Bozeman, MT 59717, USA \\
$^{13}$ School of Physics and Astronomy, University of Southampton, Highfield, Southampton SO17 1BJ, UK }

\begin{document}

\date{Accepted 2022 April 5. Received 2022 April 5; in original form 2021 December 9}


\maketitle

\label{firstpage}

\begin{abstract}
The standard alpha-disc model predicts an anti-correlation between the density of the inner accretion disc and the black hole mass times square of the accretion rate, as seen in higher mass ($M_{\rm BH}>10^{6} M_{\odot}$) active galactic nuclei (AGNs). In this work, we test the predictions of the alpha-disc model and study the properties of the inner accretion flow for the low-mass end ($M_{\rm BH}\approx 10^{5-6}M_{\odot}$) of AGNs. We utilize a new high-density disc reflection model where the density parameter varies from $n_{\rm e}=10^{15}$ to $10^{20}$~cm$^{-3}$ and apply it to the broadband X-ray (0.3$-$10\keV{}) spectra of the low-mass AGN sample. The sources span a wide range of Eddington fractions and are consistent with being sub-Eddington or near-Eddington. The X-ray spectra reveal a soft X-ray excess below $\sim 1.5\keV{}$ which is well modeled by high-density reflection from an ionized accretion disc of density $n_{\rm e}\sim 10^{18}$~cm$^{-3}$ on average. The results suggest a radiation pressure-dominated disc with an average of 70~per~cent fraction of the disc power transferred to the corona, consistent with that observed in higher mass AGNs. We show that the disc density higher than $10^{15}$~cm$^{-3}$ can result from the radiation pressure compression when the disc surface does not hold a strong magnetic pressure gradient. We find tentative evidence for a drop in black hole spin at low-mass regimes.

\end{abstract}


\section{Introduction}
Active galactic nuclei (AGNs) are luminous nuclear sources powered by the accretion of matter from the host galaxy (see e.g. \citealt{ss19}) onto a supermassive black hole (SMBH; \citealt{ly69,re94}) of mass $\sim 10^{5}-10^{10} M_{\odot}$ (see \citealt{re13,ba15,eht19}). They are characterized by complex spatial and time-variable emission observed over the entire electromagnetic spectrum, from very high energy gamma-rays to radio wavebands. The X-ray energy band provides a unique window as the X-ray photons come from the immediate vicinity of the central SMBH and help us probe the extreme gravity just outside the event horizon of the black hole (e.g. \citealt{fa89,fa02,wi21}). We can characterize black holes fundamentally by two parameters: mass and spin, which play a crucial role in understanding the physical nature of the AGN central engine and the cosmological evolution of galaxies which is directly associated with the energy output from the SMBHs (e.g. \citealt{fa12,kh13}). However, most of our knowledge on the physical processes occurring in the AGN central engine is confined to active galaxies hosting SMBHs of mass greater than $10^{6} M_{\odot}$. The low-mass end ($M_{\rm BH}\lesssim 10^{6} M_{\odot}$) of AGNs is a less explored, yet intriguing domain that serves as the missing link to probe whether the accretion process is the same at all mass scales. The least massive AGNs can play a significant role in probing the nature of primordial SMBHs, which are crucial to constrain the cosmological black hole growth models.

X-ray spectroscopy is a powerful tool to probe strong-field accretion in the innermost regions of the AGN central engine. It allows us to determine the physical conditions of the plasma surrounding the central X-ray source. The properties of the inner accretion discs of AGNs are measured through X-ray reflection spectra, which show a soft X-ray excess below $\sim 2$\keV{}, a broad emission profile due to Fe~K lines at $\sim 6.4-7.1$\keV{}, and Compton hump above 10\keV{} (e.g. \citealt{mar14,mat14,zo15,ga19,xu21}). The physical modeling of these features provides direct probes for the inner disc atmosphere and black hole spin (e.g. \citealt{br06,wa13,ga14,re19,re21,ba21}), the measurement of which is essential to study the cosmic evolution of spin and the long-term growth of black holes (e.g. \citealt{vo13}). 

Traditional reflection models assume that the density of the disc atmosphere is constant and fixed at $n_{\rm e}=10^{15}$~cm$^{-3}$ (e.g. \citealt{rf93,rf05,ga13,ga14}). These constant density disc reflection models work well for high-mass ($M_{\rm BH}>10^{7} M_{\odot}$) black holes and higher accretion rates. However, the standard \citet{ss73} (hereafter SS73) disc model predicts that the density of the inner accretion disc is likely to be higher than $10^{15}$~cm$^{-3}$ for lower-mass ($\log M_{\rm BH}\sim 5-7$) SMBHs and stellar-mass BH X-ray binaries (BHXRBs). The theoretical calculations of \citet{sz94} (hereafter SZ94) reveal that the inner disc density is even higher than that predicted from the SS73 disc model if we consider the fraction of the transferred power from the accretion disc into the corona. \citet{ga16} investigated the effects of the gas density on the X-ray reflection spectra and provided a model ({\tt{relxillD}}) where the density is a free parameter varying from $n_{\rm e}=10^{15}$ to $10^{19}$~cm$^{-3}$. High-density disc reflection models successfully explain the soft X-ray excess, broad Fe~K emission line, and Compton hump in low-mass ($M_{\rm BH}\sim 10^{6-7} M_{\rm \odot}$) AGNs (e.g. \citealt{ma18,ji18}). Some studies challenge the constant density ($n_{\rm e}=10^{15}$~cm$^{-3}$) disc reflection origin of the soft excess and prefer an additional warm Comptonization component (e.g. \citealt{ma17,po18,pe18}). Application of the variable density reflection model to the X-ray spectra of several AGNs (e.g. \citealt{ji19,ga19,xu21}) and XRBs (e.g. \citealt{tom18,ji20}) has lately become more prevalent.

In this paper, we apply our recently developed variable density ($\log[n_{\rm e}$/cm$^{\rm -3}$]$=15-20$) reflection model to the broad-band (0.3--10\keV{}) X-ray spectra of a sample of low-mass ($M_{\rm BH}\approx 10^{5-6} M_{\rm \odot}$) AGNs, study their spectral properties and compare them with higher mass AGNs from the literature. \citet{mi09} and \citet{lu15} studied some of these sources and detected a soft X-ray excess which was modeled by a blackbody component. The detection of soft X-ray reverberation time lags in some of these low-mass AGNs \citep{ma21} is the strongest supporting evidence for the reflection origin of the soft X-ray excess. 

We organize the paper as follows. In Section~2, we describe the details of the sample and data reduction techniques. In Section~3, we perform the spectral analysis of the low-mass AGN sample using the variable density reflection model. We interpret the results of the spectral modeling and compare the measured accretion properties of the sample with their higher mass counterparts in Section~4. We conclude the results in Section~5. Finally, we outline the future work in Section~6.

\section{Sample selection and data reduction}
\label{sec2}
To study the accretion properties of the low-mass AGN end, we obtained a sample of the least massive AGNs from the \citet{gh07} (hereafter GH07) catalog and one AGN, POX~52, from \citet{ba04}, with central BH masses of $\sim 10^{5-6} M_{\rm \odot}$. We searched the \xmm{} \citep{ja01} archive and found that a total of 29 AGNs from the sample are present in the field of view of the European Photon Imaging Camera (EPIC). 

We reduced all the available observations in the Scientific Analysis System ({\tt{SAS}}~v.18.0.0) with the most recent (as of 2021 February) calibration files. We generated raw event files by processing the EPIC pn and MOS data with the {\tt{SAS}} tasks {\tt{EPPROC}} and {\tt{EMPROC}}, respectively. To filter the raw EPIC pn and MOS events, we employed {\tt{PATTERN}}$\leq$4 and {\tt{PATTERN}}$\leq$12, respectively, and removed bad pixel events with {\tt{FLAG}}$==$0. We produced flare-filtered clean event files by excluding the proton flare intervals using the method described in \citet{ma21}. We confirmed the absence of any pile-up effects using the task {\tt{EPATPLOT}}. We extracted the source and background spectra from a circular region centered on the source and nearby source-free zone of radius 20~arcsec and 40~arcsec, respectively. The RMF (Redistribution Matrix File) and ARF (Ancillary Response File) for each observation were generated with the {\tt{SAS}} tasks {\tt{RMFGEN}} and {\tt{ARFGEN}}, respectively. We then grouped the spectral data to have a minimum of 10 counts per bin and to oversample by at most a factor of 3 using the {\tt{SPECGROUP}} task. We excluded sources that do not have EPIC-pn counts above 4\keV{} after background subtraction. To obtain a clear view of the central engine, we selected only Seyfert~1 class AGNs. These criteria resulted in a final sample of 13 AGNs. The observation log and source properties are presented in Table~\ref{tab1} and Table~\ref{tab2a}, respectively.

\begin{table*}
\caption{Summary of \xmm{} observations of the sample used in this work. The source name, right ascension and declination are given in columns~(1), (2) and (3), respectively. The observation ID and start date are shown in columns~(4) and (5), respectively. Total net exposure time for EPIC pn+MOS is listed in column~(6). The net exposure time is the background filtered, live time of the instrument. Column~(7) shows the 0.3$-$10\keV{} EPIC pn+MOS background-subtracted net source count rate.}
\begin{center}
\scalebox{0.9}{%
\begin{tabular}{ccccccccccccccc}
\hline 
Source Name [Short] & RA    & DEC    & Obs.~ID &   Start date  & Net total      &  Net source count  \\
                    & (Deg) & (Deg)  &         &  (yyyy-mm-dd) & exp. (ks)   &   rate (ct/s)  \\  
(1)                 & (2)   & (3)    & (4)     &      (5)      &  (6)      &   (7)    \\                                                    
\hline 
SDSS~J010712.03$+$140844.9 [J0107] & 16.800 & 14.146 & 0305920101 & 2005-07-22  & 68.8 & 0.33    \\ [0.1cm]

SDSS~J022849.51$-$090153.7 [J0228] & 37.206 & $-$9.032 & 0674810101 & 2012-01-17 & 28.5 & 0.24  \\ [0.1cm]

SDSS~J094057.19$+$032401.2 [J0940] & 145.238 & 3.400  & 0306050201 & 2005-10-30  & 72.8 & 0.27   \\ [0.1cm]
       
SDSS~J102348.44$+$040553.7 [J1023] & 155.952 & 4.098 & 0108670101 & 2000-12-05  & 148.3  & 0.08    \\ [0.1cm]
                                  &          &       & 0605540201 & 2009-12-13  & 306.6  & 0.11      \\ [0.1cm]
                                  &          &       & 0605540301 & 2009-05-08  & 129.1  & 0.08     \\ [0.1cm]

SDSS~J114008.71$+$030711.4 [J1140] & 175.036 & 3.120 & 0305920201 & 2005-12-03  & 111.9 & 0.72    \\ [0.1cm]
                                 &          &        & 0724840101 & 2013-12-18  & 110.6 & 0.33  \\ [0.1cm]
                                 &          &        & 0724840301 & 2014-01-01  & 146.3 & 0.64    \\ [0.1cm]

SDSS~J134738.23$+$474301.9 [J1347] & 206.909 & 47.717 & 0744220701 & 2014-11-22 & 72.1 & 0.59 \\ [0.1cm]
   
SDSS~J135724.52$+$652505.8 [J1357] & 209.352 & 65.418 & 0305920601 & 2005-06-23 & 40.3 & 0.48  \\ [0.1cm]   
       
SDSS~J143450.62$+$033842.5 [J1434] & 218.711 & 3.645  & 0305920401 & 2005-08-18 & 61.5  & 0.13  \\ [0.1cm]
                                   &         &        & 0674810501 & 2011-08-16 & 35.4  & 0.13    \\ [0.1cm]
       
SDSS~J154150.85$+$310037.2 [J1541] & 235.462 & 31.010 & 0744220401 & 2015-01-31 & 30.3  & 0.33  \\ [0.1cm]

SDSS~J155909.62$+$350147.4 [J1559] & 239.790 & 35.030  & 0744290101 & 2015-03-02  & 204.5 & 2.7    \\ [0.1cm]
                                   &         &         & 0744290201 & 2015-02-24  & 236.3  & 2.5    \\ [0.1cm]

SDSS~J162636.40$+$350242.0 [J1626] & 246.652 & 35.045 & 0674811001 & 2012-01-17  & 30.1  & 0.11  \\ [0.1cm]
                                 
SDSS~J163159.59$+$243740.2 [J1631] & 247.998 & 24.628  & 0674810601 & 2011-08-28 & 39.1 & 0.07  \\ [0.1cm]                                     

POX52 & 180.737 & $-$20.934  & 0302420101 & 2005-07-08  & 245.4 & 0.07 \\ [0.1cm]
\hline 
\end{tabular}}
\end{center} 
\label{tab1}           
\end{table*}

\begin{table*}
\caption{The source name and redshift are given in columns~(1) and (2), respectively. Columns~(3) and (4) represent the black hole mass and Eddington ratio, respectively, and are obtained from \citet{th08} for POX52; for all other sources from the GH07 catalog. The BH mass uncertainty is $\sim 0.3$~dex for the GH07 sample. Columns~(5) and (6) show the optical class of each source obtained from NED/SIMBAD and the Galactic hydrogen column density \citep{wil13}, respectively. The maximum disc temperature for Schwarzschild and Kerr black holes are presented in columns (7) and (8), respectively. Column~(9) shows the temperature of the soft X-ray excess component when fitted with a disc blackbody model.}
\begin{center}
\scalebox{0.9}{%
\begin{tabular}{cccccccccc}
\hline 
Source & $z$ & $M_{\rm BH}$ & $L_{\rm bol}/L_{\rm E}$ & Optical & $N_{\rm H, Gal}$  & $T_{\rm max}$ ($a=0$) & $T_{\rm max}$ ($a=0.998$) & $T_{\rm SE}$ \\[0.1cm]
    &   & ($10^{5}M_{\rm \odot}$)  & &   Type    &  (10$^{20}$~cm$^{-2}$) &      (eV)   &   (eV) &   (eV) \\  [0.2cm]
(1)    &   (2)   &   (3)   &   (4)  &  (5)   & (6) &   (7)  &    (8) &    (9)   \\                                                      
\hline 
J0107 & 0.0767  & 16.0$^{+16.0}_{-8.0}$ &  0.40 &  NLSy1    & 3.94 & 29.5 &  58.6  & 186$^{+16}_{-17}$  \\ [0.25cm]

J0228 & 0.0722  & 3.2$^{+3.1}_{-1.6}$ & 0.25  & BLSy1 & 3.88  & 39.3 & 78.1 &   151$^{+15}_{-17}$  \\ [0.25cm]

J0940 & 0.0606  & 16.0$^{+16.0}_{-8.0}$ &  0.40  & NLSy1  & 3.73  & 29.5 &  58.6 &  150$^{+12}_{-13}$ \\ [0.25cm]

J1023 & 0.0989  & 5.0$^{+5.0}_{-2.5}$ &  0.50 & NLSy1  & 2.9  & 41.7   &  82.7 & 156$^{+20}_{-25}$ \\ [0.25cm]
                   
J1140 & 0.081  & 12.6$^{+12.5}_{-6.3}$ & 0.63 & NLSy1  & 2.06  & 35.0 & 69.6 & 170$^{+4}_{-5}$ \\ [0.25cm]
                             
J1347 & 0.0643  & 10.0$^{+10.0}_{-5.0}$ &  0.50 & NLSy1 & 1.9  & 35.0 & 69.6 & 166$^{+6}_{-6}$ \\ [0.25cm]
   
J1357 & 0.106 & 16.0$^{+16.0}_{-8.0}$ & 0.50 & NLSy1  & 1.45  & 31.2 & 62.0 & 150$^{+11}_{-10}$  \\ [0.25cm] 
       
J1434 & 0.0283 & 6.3$^{+6.3}_{-3.1}$ &  0.10 & Sy1  & 2.76   & 26.3 & 52.2 & 173$^{+28}_{-30}$ \\ [0.25cm]

J1541 & 0.0684  & 16.0$^{+16.0}_{-8.0}$  & 0.25 &  BLSy1 & 2.67 & 26.3 & 52.2 & 114$^{+20}_{-19}$ \\ [0.25cm]  
       
J1559 & 0.031 & 16.0$^{+16.0}_{-8.0}$ & 0.63 &  NLSy1  & 2.28  & 33.1 & 65.7& 144$^{+2}_{-3}$ \\ [0.25cm]

J1626 & 0.0341  & 5.0$^{+5.0}_{-2.5}$ & 0.32 &  Sy~1.5  &  1.52 & 37.1 & 73.7 & 103$^{+25}_{-22}$   \\ [0.25cm]
                                 
J1631 & 0.0433  & 6.3$^{+6.3}_{-3.1}$ & 0.32 &  BLSy1 & 4.08  & 35.0 & 69.6 & 128$^{+31}_{-29}$ \\ [0.25cm]                                   

POX52  & 0.021  & 3.2$^{+1.0}_{-1.0}$ & 0.35 &  Sy~1.8  & 4.41 & 42.6 & 84.6 & 243$^{+36}_{-34}$ \\ [0.25cm]
\hline 
\end{tabular}}
\end{center} 
\label{tab2a}           
\end{table*}

\section{Variable density disc reflection Modeling}
\label{sec3}
We explore our physically motivated variable density disc reflection model {\tt{relxillDCp}} to study the broad-band (0.3$-$10\keV{}) X-ray spectra of the sample. The model allows a higher density for the accretion disc varying from $n_{\rm e}=10^{15}$ to $10^{20}$~cm$^{-3}$ and includes a Comptonization continuum ({\tt{nthComp}}) with a variable coronal temperature as the incident spectrum instead of a simple power-law. The high-energy cutoff of the {\tt{relxillDCp}} model is parameterized by the electron temperature ($kT_{\rm e}$) of the hot corona and spans within the range of 1--400\keV{}. This new version of the model will soon be included in the next official release of {\tt{RELXILL}}~v.1.5.0. 

The spectral fitting was performed in the {\tt{XSPEC}} (v.12.11.1) software \citep{ar96}. We employed C-stat \citep{ca79} to find the best-fit values of the model parameters and $\chi^{2}$-statistics to test the goodness-of-fit. All the reported errors are estimated through Markov Chain Monte Carlo (MCMC) simulations and correspond to the 90~per~cent confidence intervals unless otherwise specified.

First, we fit the hard band (2$-$10\keV{}) continuum spectra with the phenomenological power-law ({\tt{zpowerlw}}) model modified by the absorption due to the Galactic interstellar medium (ISM). We used the Tuebingen-Boulder ISM absorption model ({\tt{Tbabs}}) to account for the Galactic absorption. The model considers the solar ISM abundances of \citet{wi00} and the photoionization cross-section table of \citet{ve96}. We employed the Galactic absorption column density ($N_{\rm H, Gal}$) calculated by \citet{wil13},  which accounts for the column density of both the atomic ($N_{\rm {HI}}$) and molecular ($N_{\rm H2}$) hydrogen. The Galactic hydrogen column density ($N_{\rm H, Gal}$) for each source, as listed in Table~\ref{tab2a}, was kept fixed throughout the spectral modeling. The absorbed power-law model describes the hard X-ray emission well for all AGNs except J1559 which showed a narrow Fe~K$_{\alpha}$ emission line at 6.4\keV{}. The presence of a narrow Fe~K$_{\alpha}$ line cannot be tested in most of the sources due to the low statistics. We modeled the Fe~K$_{\alpha}$ emission profile of J1559 by a narrow Gaussian ({\tt{zgauss}}) line with the width fixed at 10\ev{}. The addition of the narrow Gaussian line improved the test statistics by $\Delta \chi^{2}=21$ for 2 degrees of freedom. The significance of the emission line is $\ge 99.99$~per~cent, estimated by the maximum likelihood ratio (MLR) test. We find that the hard X-ray spectrum of J1559 is well explained by an absorbed power-law model along with a narrow Fe~K$_{\alpha}$ emission line. It is likely that any of these spectra do not have sufficient counts in the 2--10\keV{} band to significantly detect the broad Fe~K$_{\alpha}$ emission line. A study by \citet{de10} demonstrated that at least $1.5\times 10^{5}$ hard X-ray (2$-$10\keV{}) counts are needed for a significant detection of the relativistic Fe~K$_{\alpha}$ line in Type~1 AGNs.

The extrapolation of the hard X-ray absorbed power-law model to the soft X-ray (0.3$-$2\keV{}) band revealed an excess emission component below $\sim 1.5$\keV{} for all the sources in the sample. However, we see an absorption curvature at around $1-2$\keV{} for POX~52, which could be a signature of partial covering absorption. The broad-band (0.3$-$10\keV{}) X-ray spectra, hard band (2$-$10\keV{}) best-fit extrapolated models ({\tt{Tbabs$\times$[zgauss$+$zpowerlw]}} for J1559 and {\tt{Tbabs$\times$zpowerlw}} for all other sources), and residual plots for each observation are shown in Fig.~A1. To fit the observed soft X-ray excess, we first use a simple disc blackbody ({\tt{diskbb}}) model. We find that the temperature of the soft excess emission for the sample is in the range of $\sim 0.1-0.3$\keV{}, which is consistent with Type~1 class AGNs having a wide range of black hole mass (see \citealt{lu15} for low-mass AGNs and \citealt{gd04} for high-mass quasars). To verify whether the soft X-ray excess can be the direct thermal emission from the accretion disc in low-mass AGNs, we calculate the maximum temperature of the SS73 disc for both Schwarzschild ($a=0$) and maximally spinning ($a=0.998$) Kerr black holes. The effective temperature of thermal radiation from the disc atmosphere at radius $r$ is given by (see e.g. \citealt{re21}):

\begin{equation}
T(r)=3.3\times 10^{7}\eta^{\frac{-1}{4}}\Big(\frac{M_{\rm BH}}{10M_{\odot}}\Big)^{\frac{-1}{4}}\Big(\frac{L_{\rm bol}}{L_{\rm E}}\Big)^{\frac{1}{4}}\Big(\frac{r}{r_{\rm g}}\Big)^{\frac{-3}{4}}\Big(1-\sqrt{\frac{r_{\rm isco}}{r}}\Big)^{\frac{1}{4}}{\rm K}
\label{Tr}
\end{equation}
where $M_{\rm BH}$ is the central BH mass in units of $M_{\odot}$, and $\frac{L_{\rm bol}}{L_{\rm E}}$ is the Eddington ratio of the system, which are presented in Table~\ref{tab2a}, $r_{\rm isco}$ is the innermost stable circular orbit in units of $r_{\rm g}$, radiative efficiency $\eta$ is solely a function of spin and ranges from $\eta \approx 0.057$ for $a=0$ to $\eta \approx 0.42$ for $a=0.998$. The disc temperature attains the maximum value at $r=\frac{49}{36} r_{\rm isco}$.

\begin{equation}
T_{\rm max}=1.61\times 10^{7}\eta^{\frac{-1}{4}}\Big(\frac{M_{\rm BH}}{10M_{\odot}}\Big)^{\frac{-1}{4}}\Big(\frac{L_{\rm bol}}{L_{\rm E}}\Big)^{\frac{1}{4}}\Big(\frac{r_{\rm isco}}{r_{\rm g}}\Big)^{\frac{-3}{4}}{\rm K}
\label{Tr_max}
\end{equation}
The temperature of the soft X-ray excess emission and the maximum temperature expected from the accretion disc for Schwarzschild ($a=0$) and maximally spinning ($a=0.998$) Kerr black holes are reported in Table~2. The temperature ($T_{\rm max}$) of the thermal disc emission varies from 25\ev{} to 85\ev{}, with a median value of around 50\ev{} for the low-mass sample. On the other hand, the soft excess temperature ($T_{\rm SE}$) is found to be constant within the range of $\sim 0.1-0.3$\keV{}, which agrees with higher mass AGNs. Moreover, there is no correlation observed between the soft excess and disc peak temperature. We also verified that the high-energy tail of the standard disc emission does not contribute to the soft X-ray (0.3--2\keV{}) band for any of these AGNs. Therefore, the soft X-ray excess cannot be assigned to the direct thermal emission from the accretion disc for the sample.

\begin{figure*}
\centering
\begin{center}
\includegraphics[scale=0.26,angle=-0]{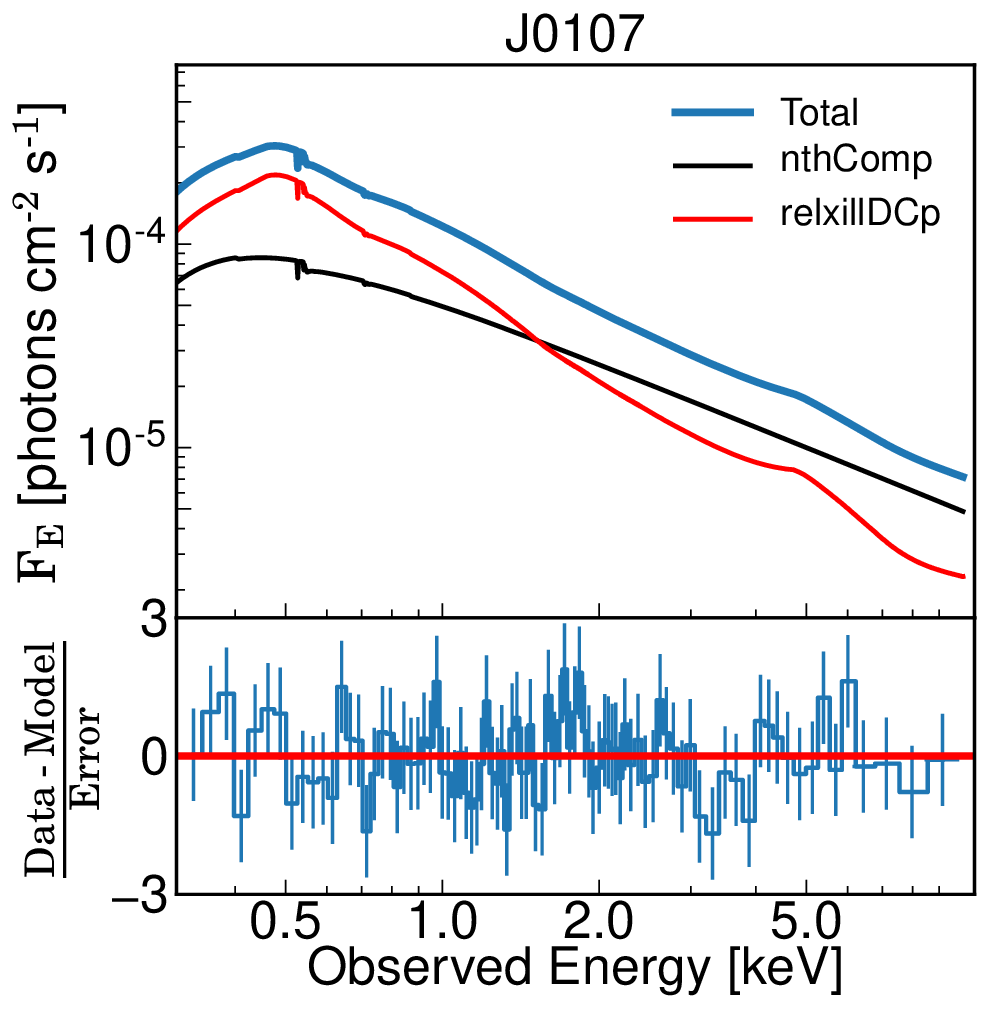}
\includegraphics[scale=0.26,angle=-0]{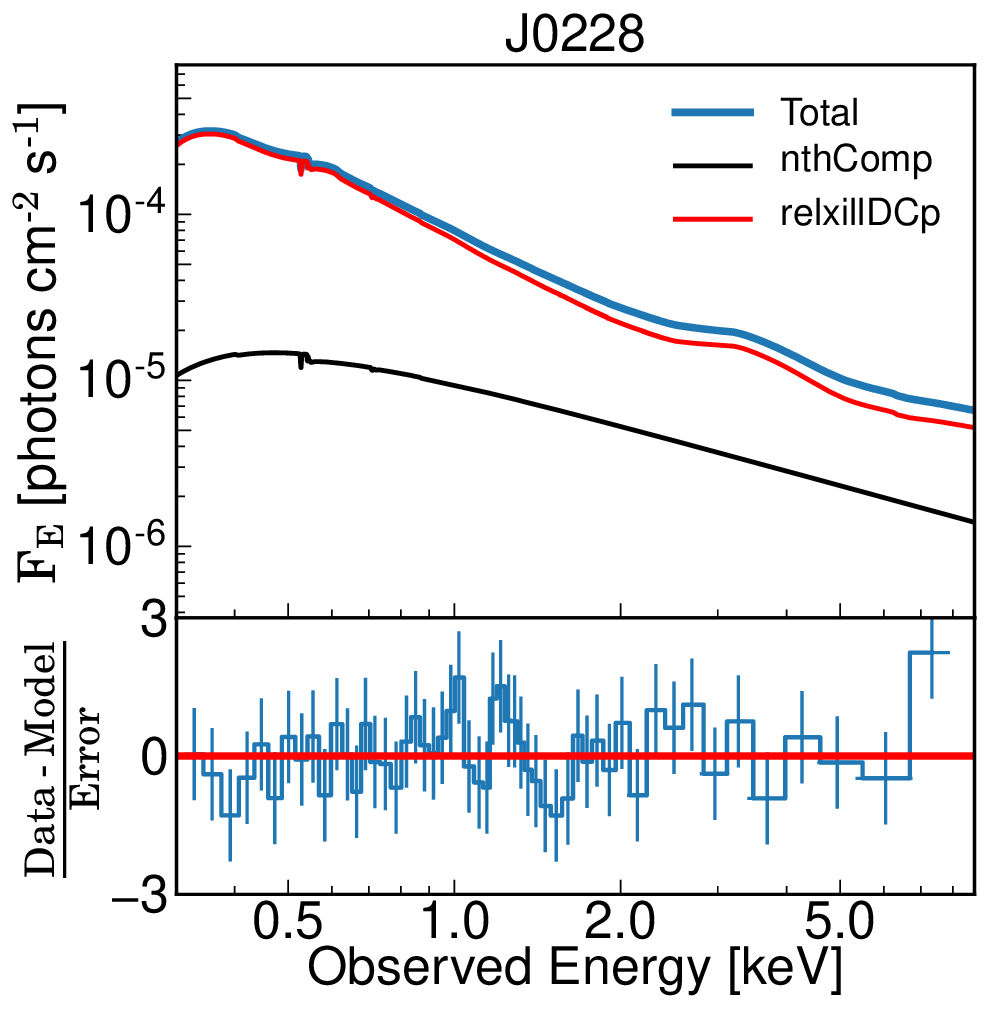}
\includegraphics[scale=0.26,angle=-0]{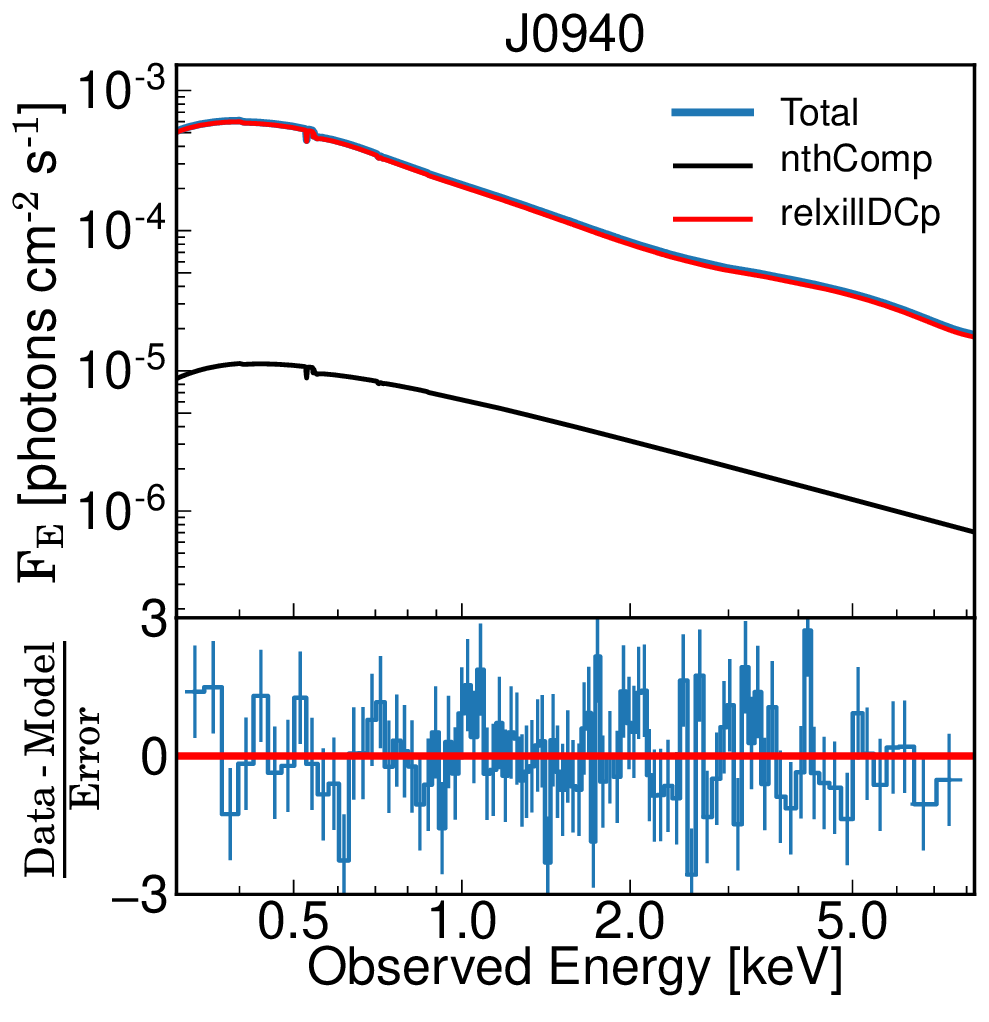}
\includegraphics[scale=0.26,angle=-0]{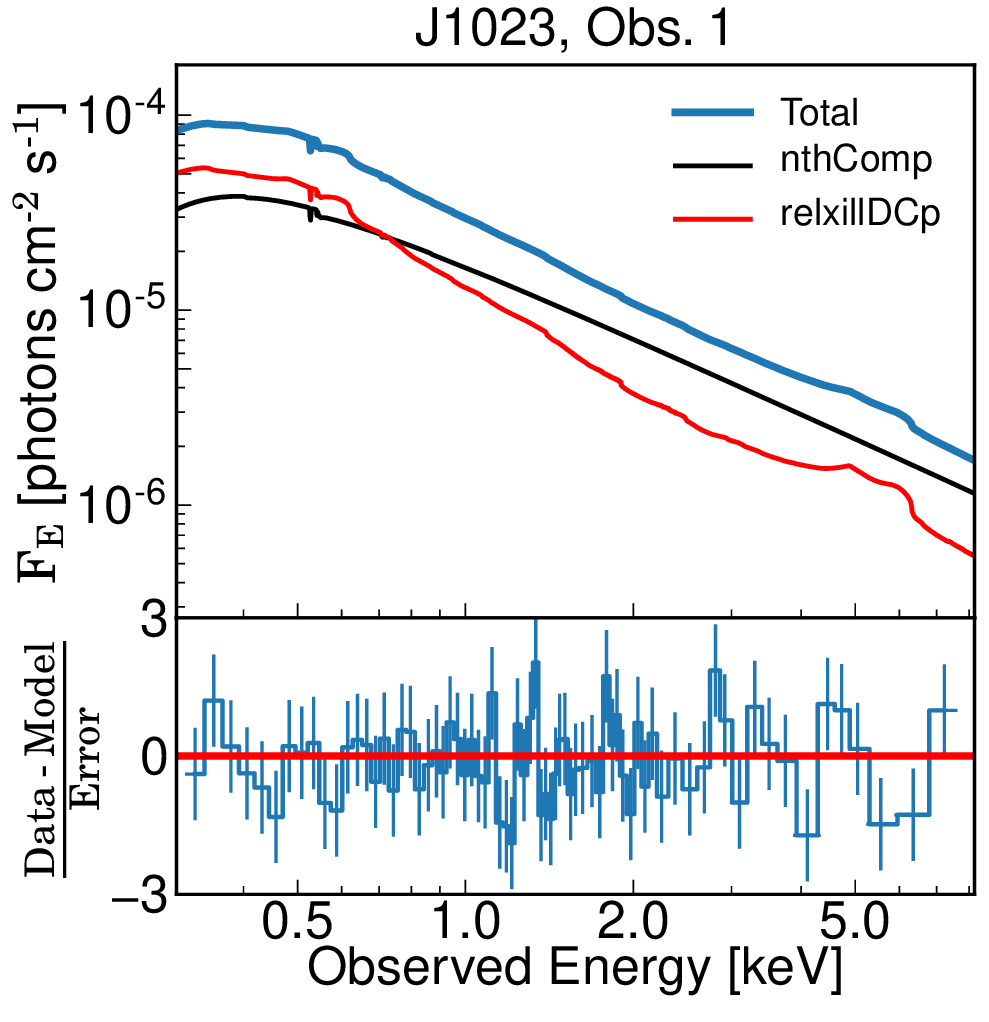}
\includegraphics[scale=0.26,angle=-0]{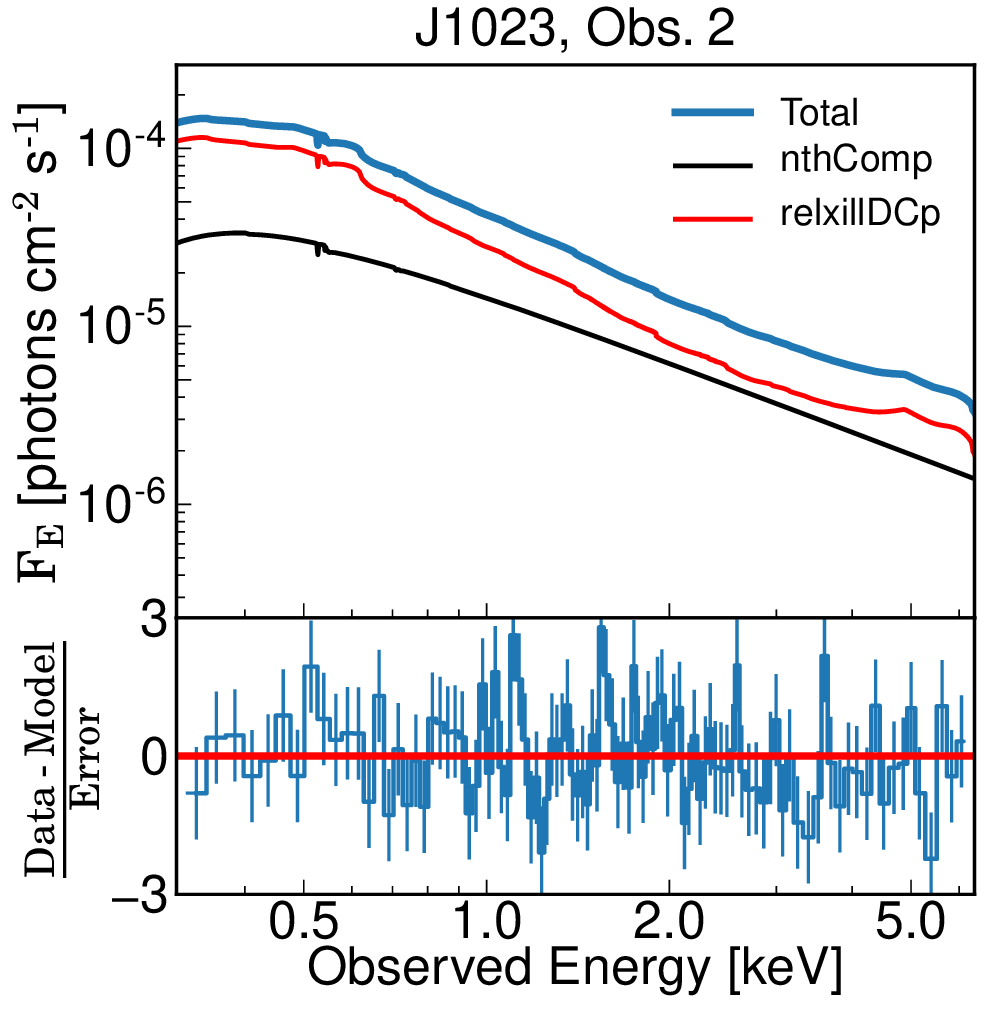}
\includegraphics[scale=0.26,angle=-0]{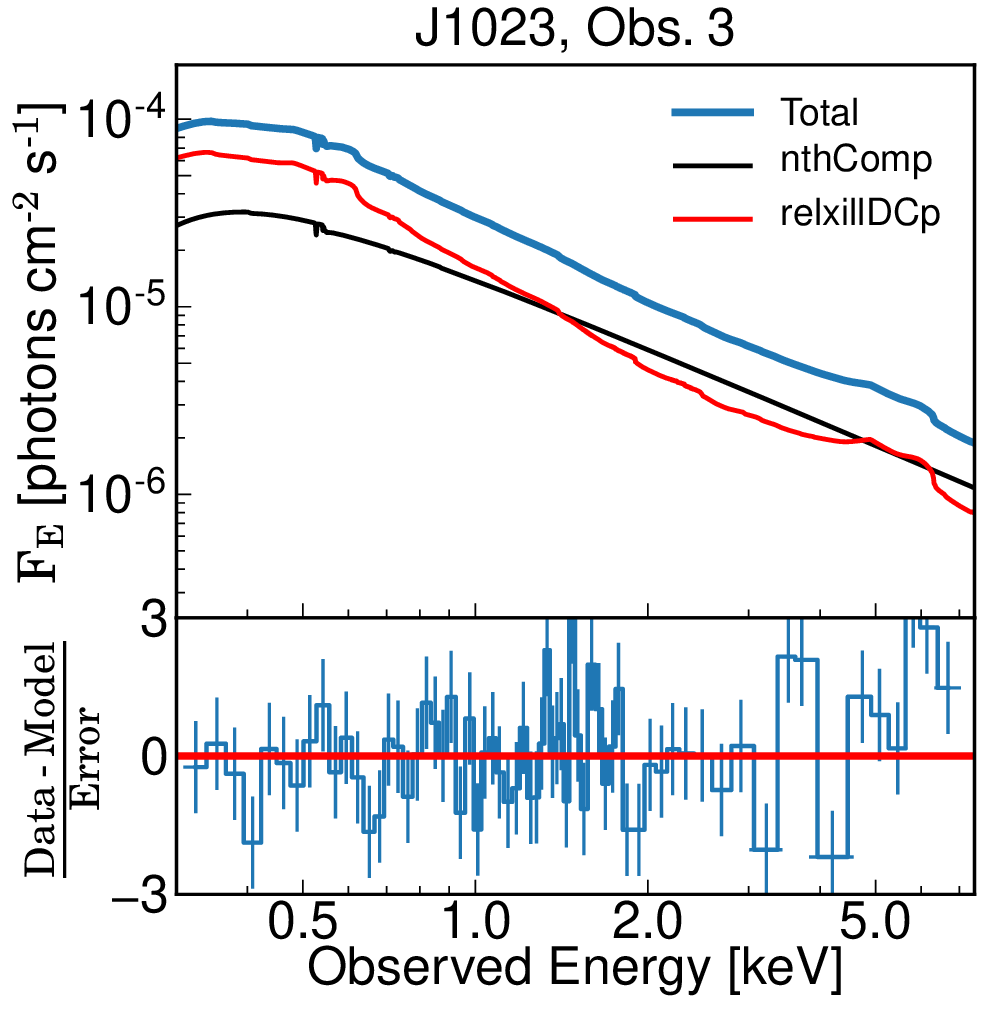}
\includegraphics[scale=0.26,angle=-0]{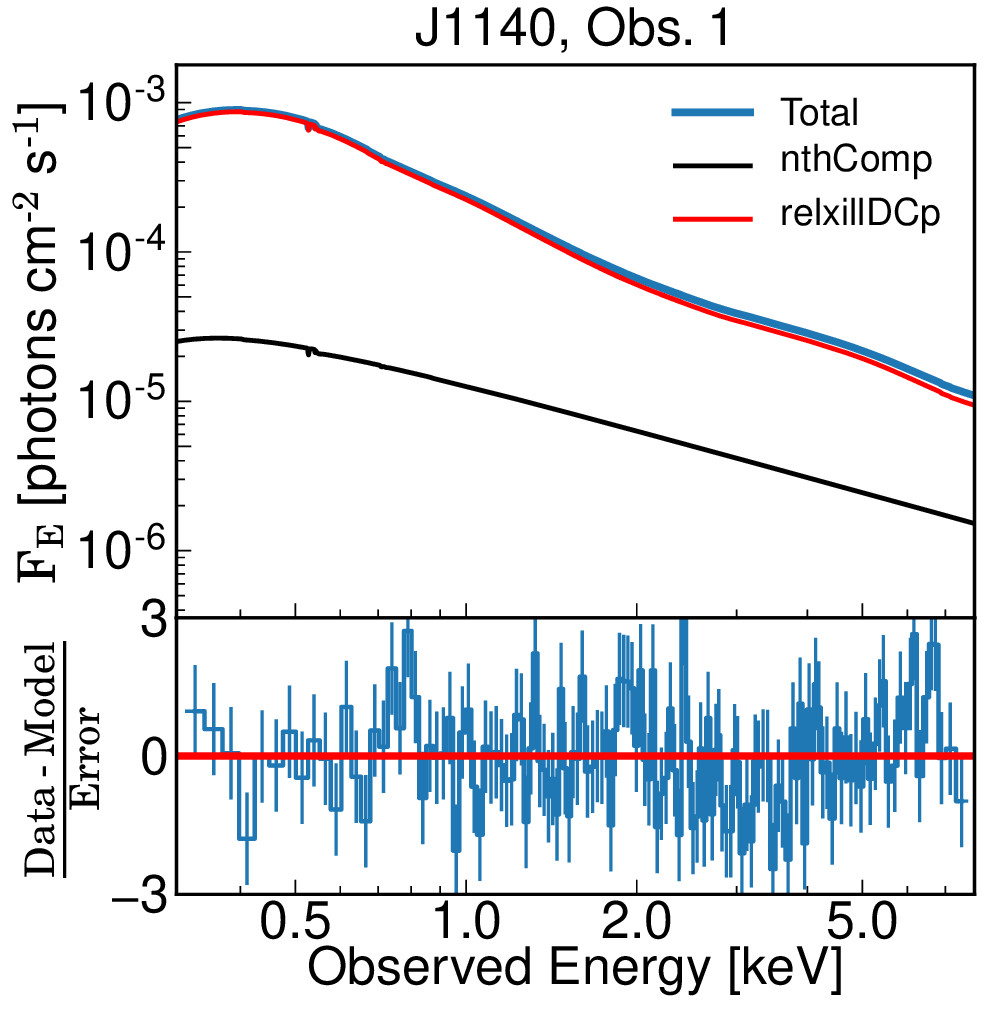}
\includegraphics[scale=0.26,angle=-0]{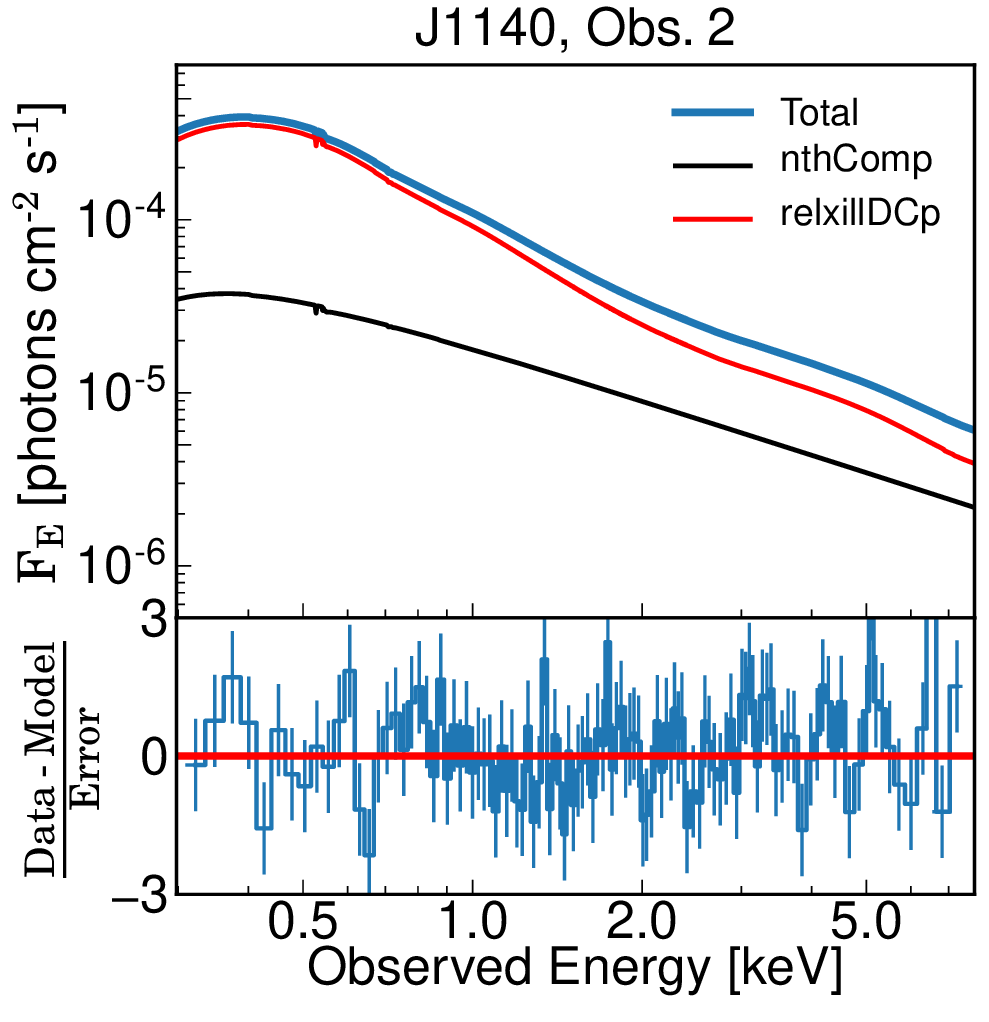}
\includegraphics[scale=0.26,angle=-0]{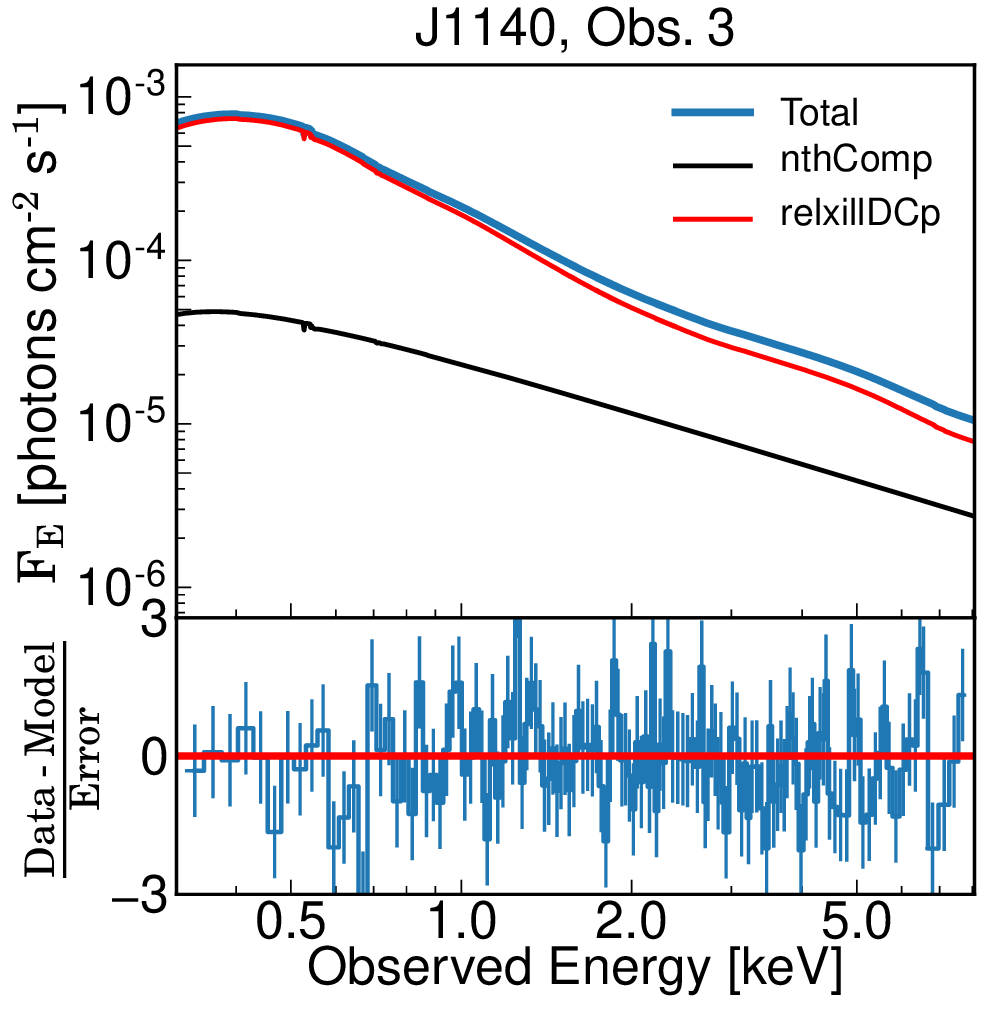}
\includegraphics[scale=0.26,angle=-0]{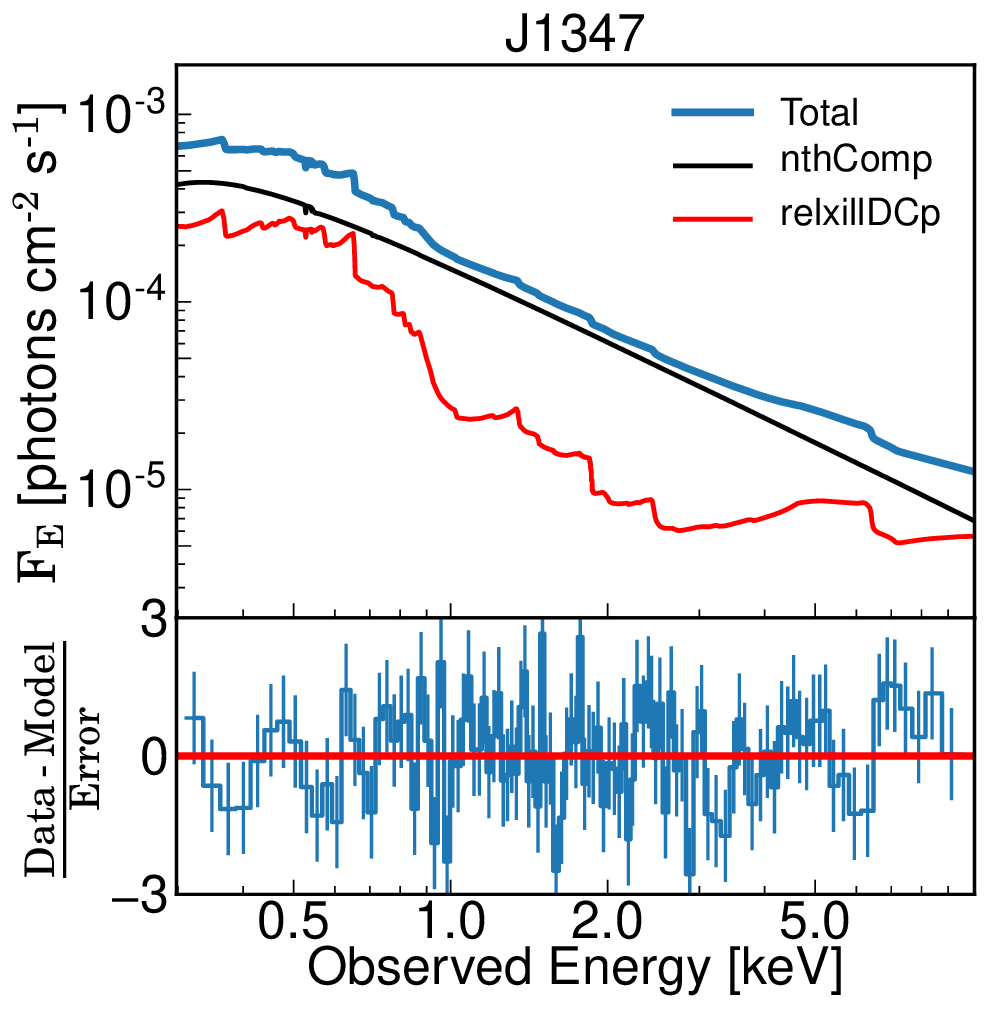}
\includegraphics[scale=0.26,angle=-0]{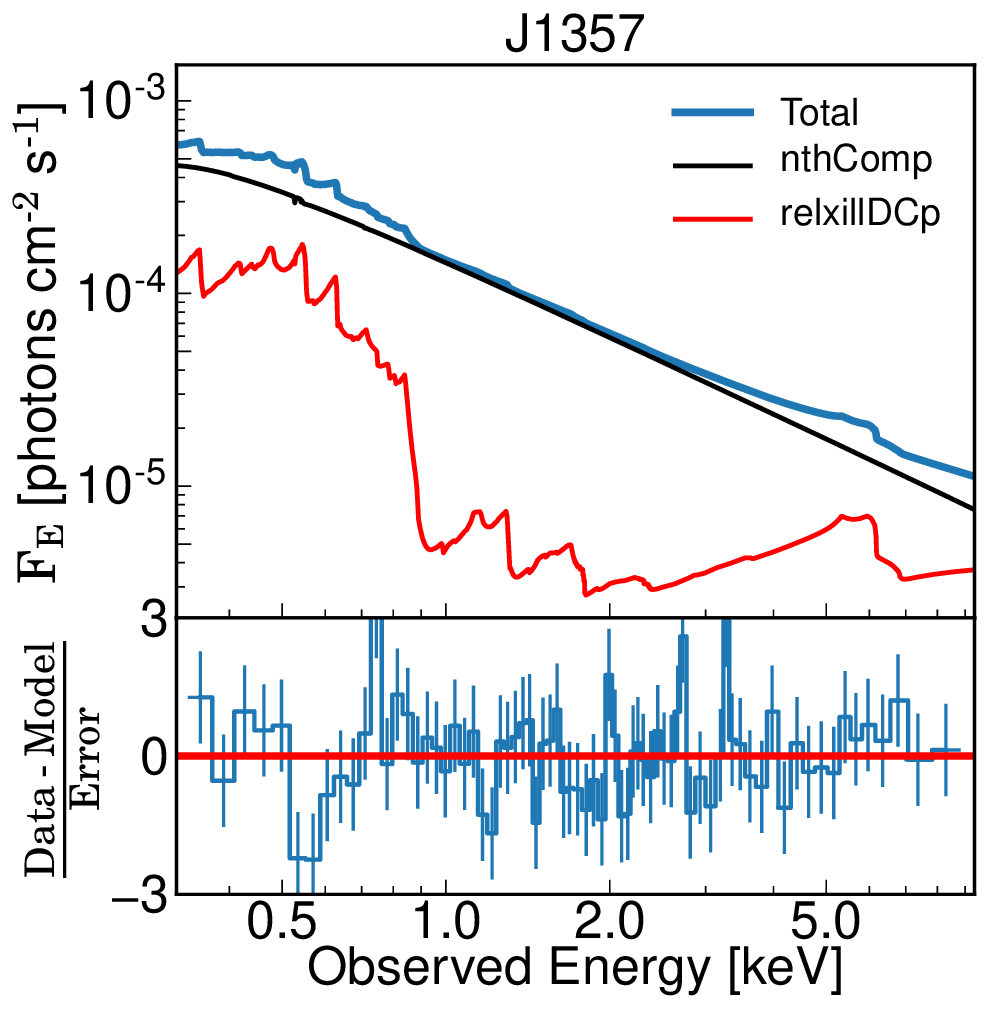}
\includegraphics[scale=0.26,angle=-0]{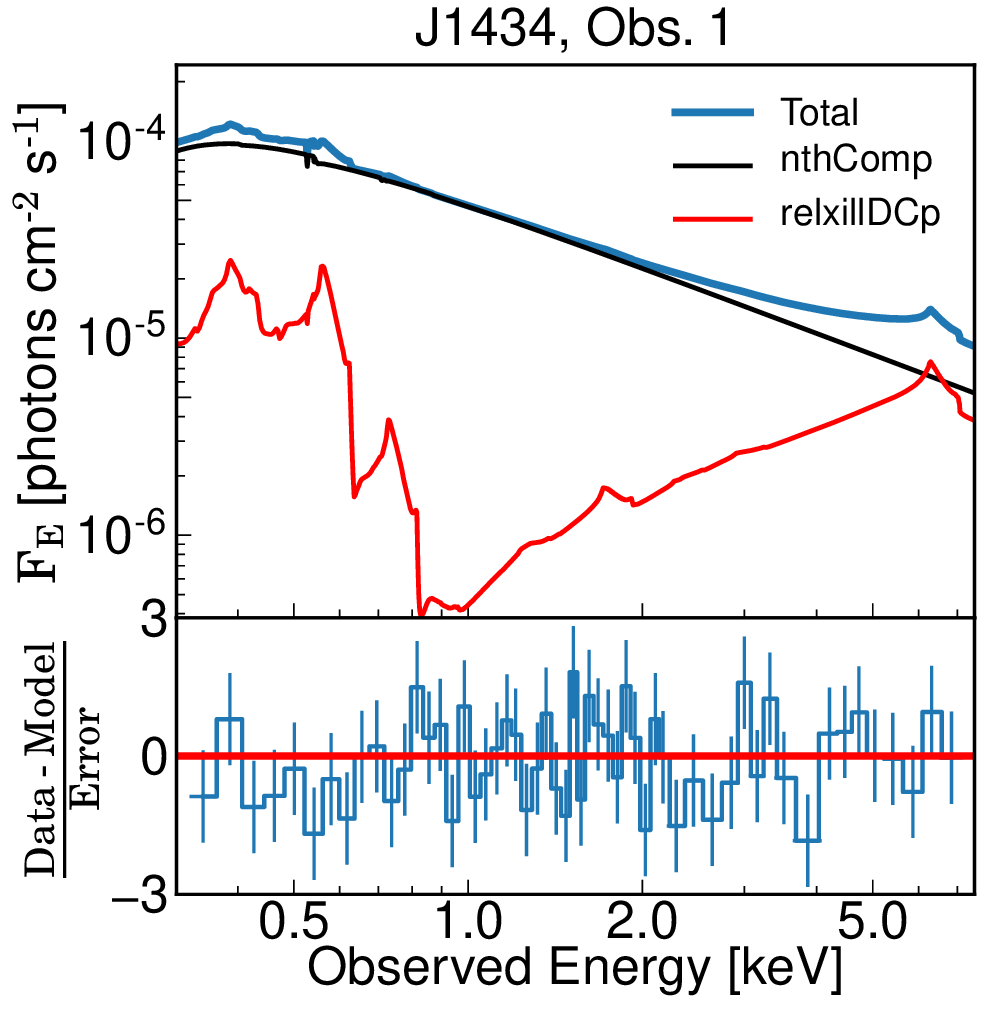}
\includegraphics[scale=0.26,angle=-0]{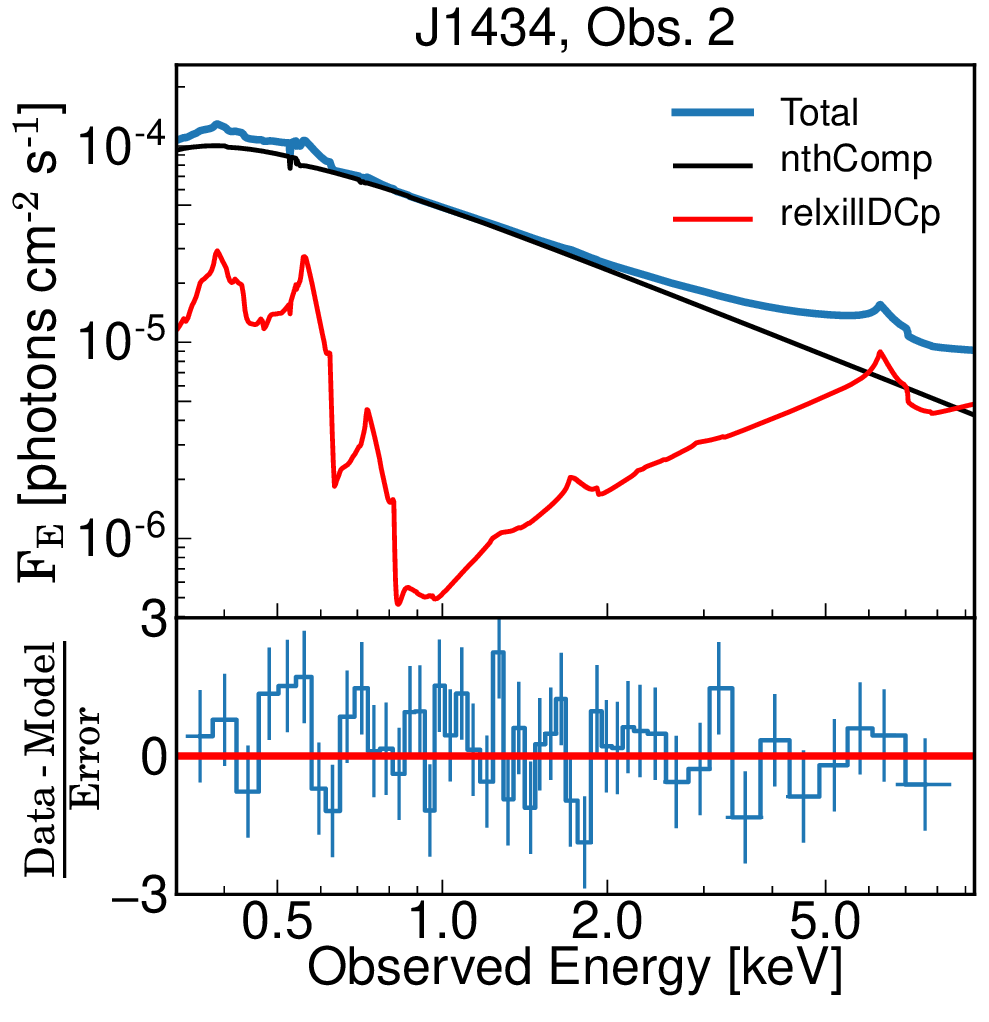}
\includegraphics[scale=0.26,angle=-0]{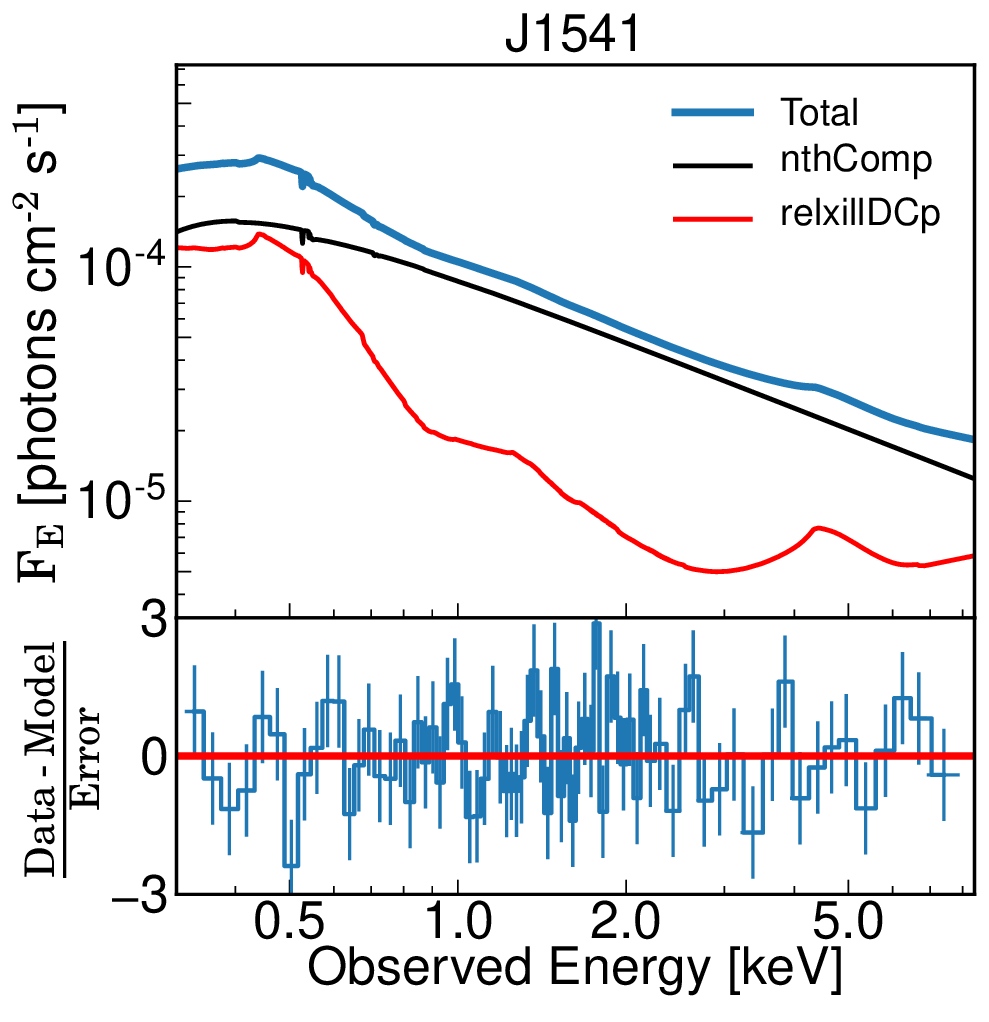}
\includegraphics[scale=0.26,angle=-0]{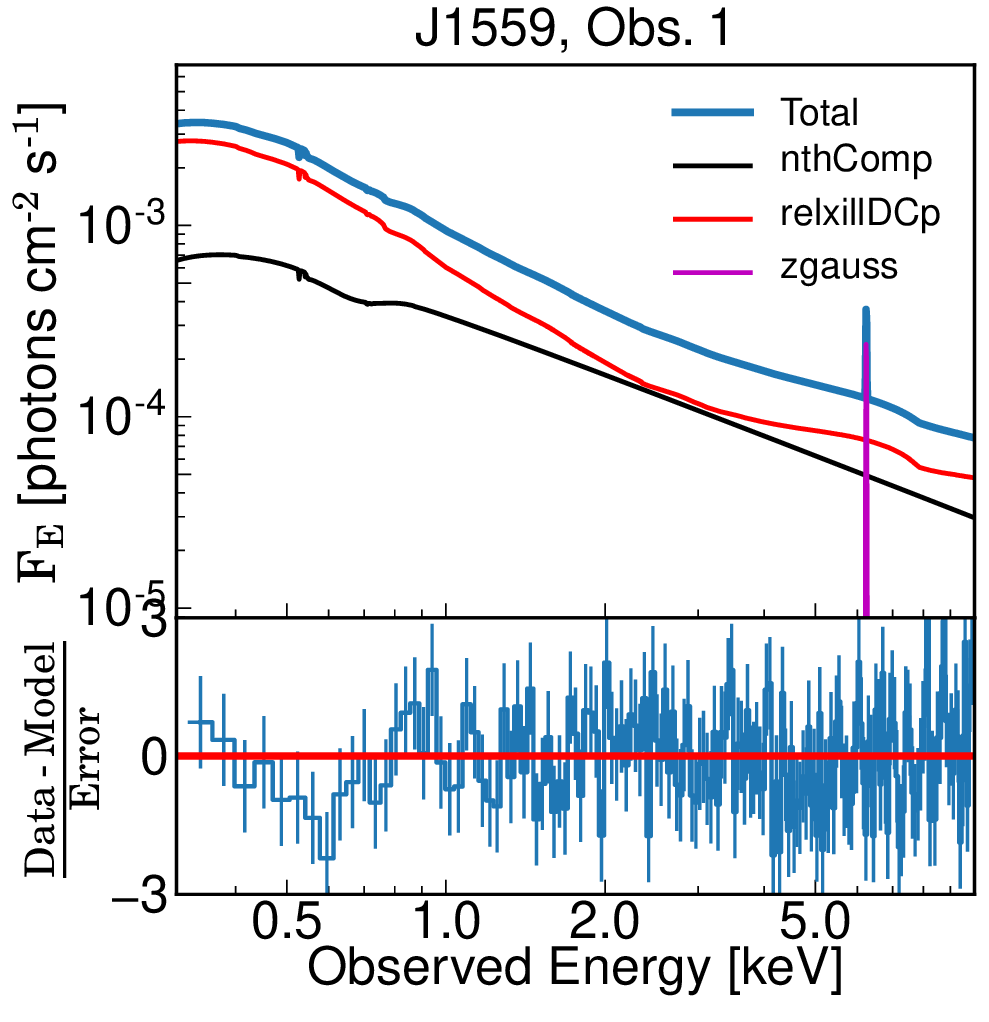}
\includegraphics[scale=0.26,angle=-0]{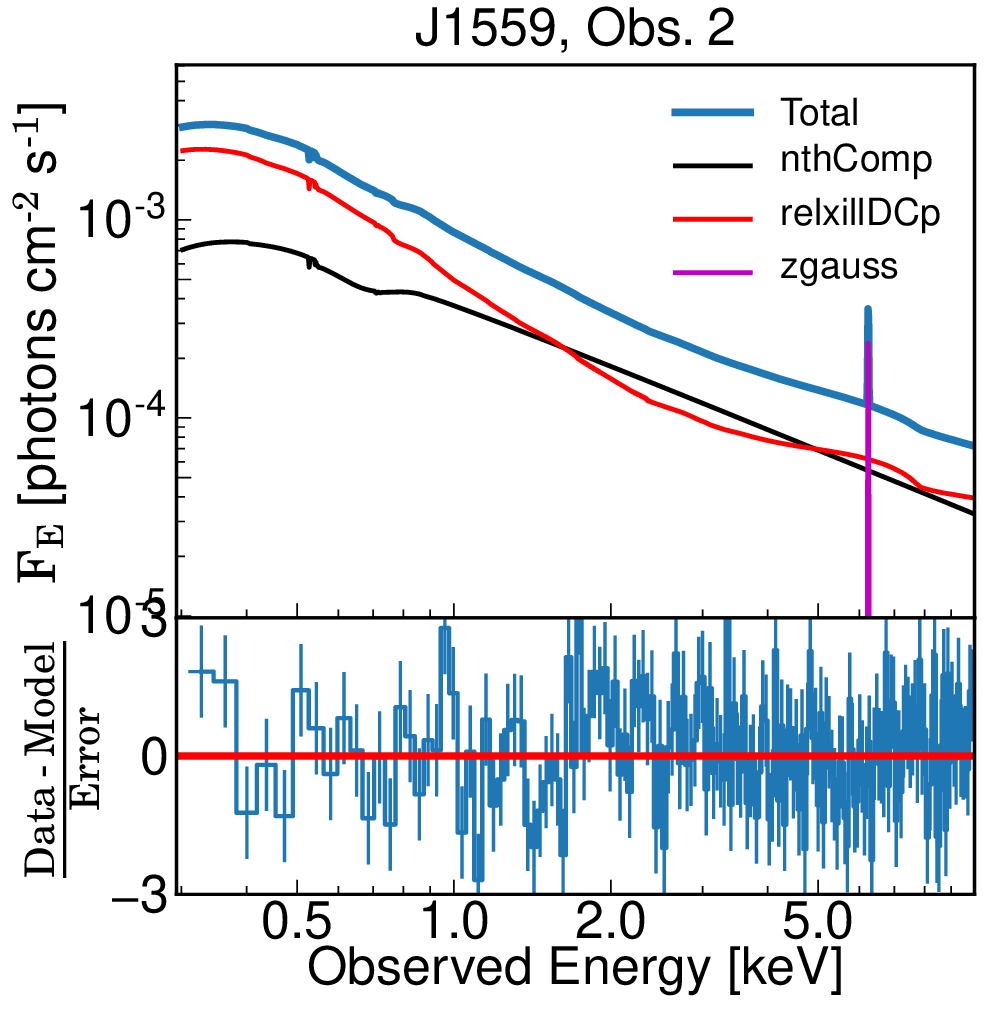}
\includegraphics[scale=0.26,angle=-0]{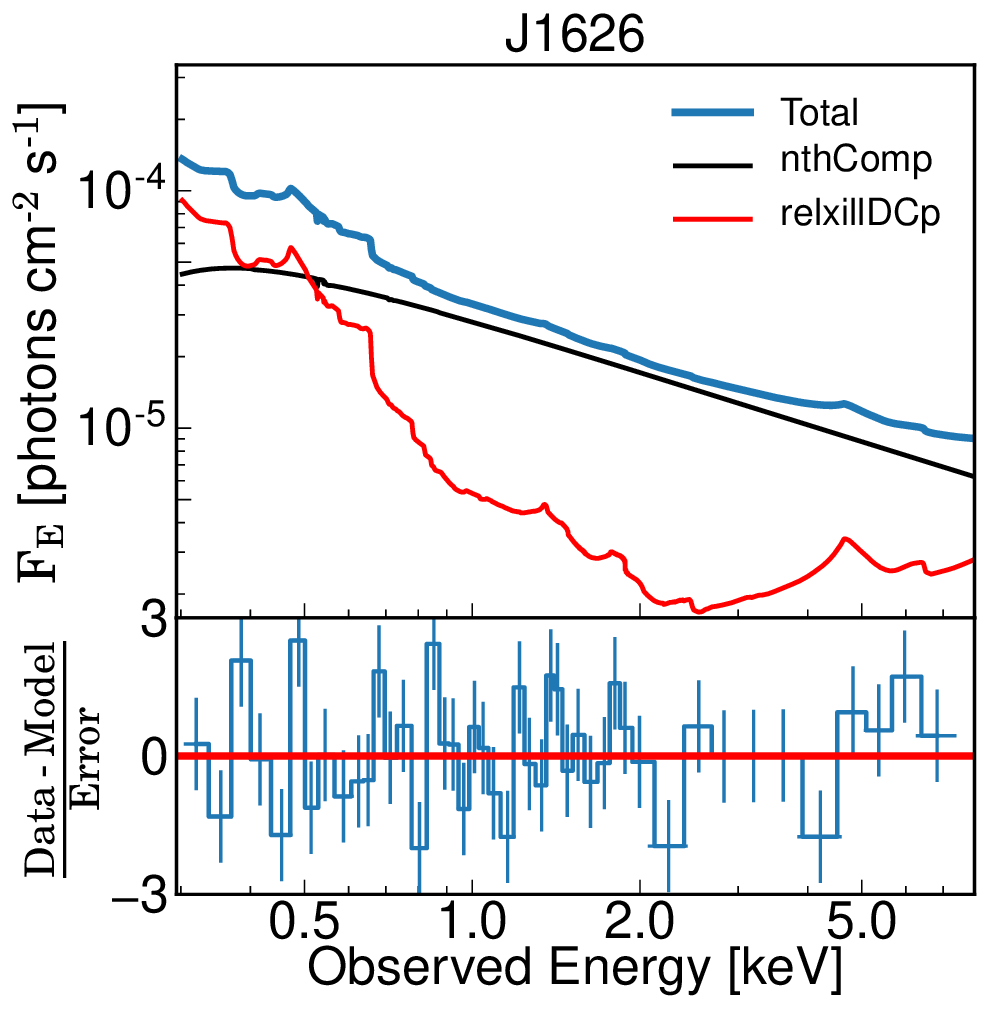}
\includegraphics[scale=0.26,angle=-0]{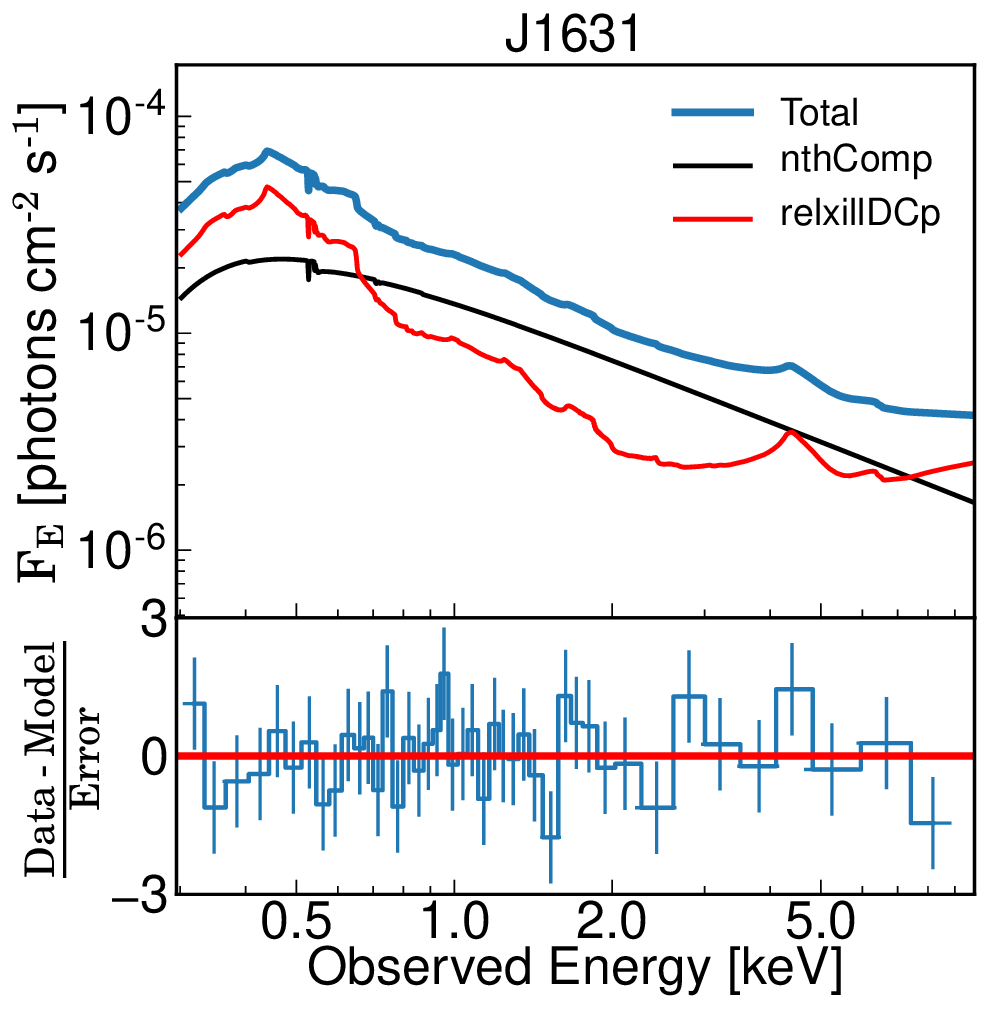}
\includegraphics[scale=0.26,angle=-0]{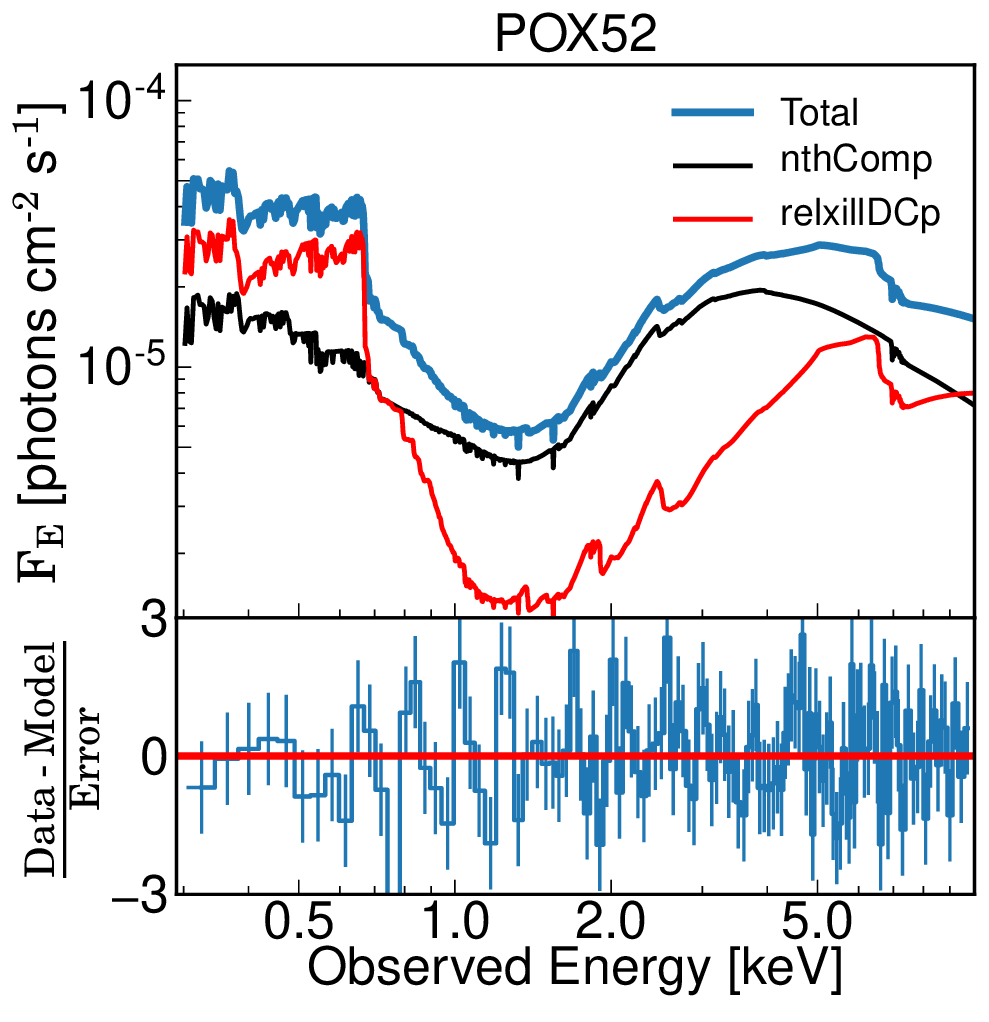}
\caption{The top panels represent the best-fit flux-energy spectral model (in blue) along with model components (black: {\tt{nthComp}}, red: {\tt{relxillDCp}}, purple: {\tt{zgauss}}). The total model expressions are {\tt{Tbabs$\times$gabs$\times$(zgauss$+$relxillDCp$+$nthComp)}}, {\tt{Tbabs$\times$zxipcf$\times$(relxillDCp$+$nthComp)}}, and {\tt{Tbabs$\times$(relxillDCp$+$nthComp)}}, respectively, for J1559, POX~52, and all other AGNs. The bottom panels show the corresponding residual plots.}
\end{center}
\label{spec3}
\end{figure*}

\begin{table*}
\caption{The 2--10\keV{} intrinsic luminosity at the source frame, average bolometric correction factor and the Eddington ratio as measured from the bolometric corrected 2--10\keV{} source luminosity are presented in columns~(2), (3) and (4), respectively. The innermost stable circular orbit $r_{\rm isco}$ in units of $r_{\rm g}$ and radiative efficiency $\eta=r_{\rm g}/2r_{\rm isco}$ are listed in columns~(5) and (6), respectively. Columns~(7) and (8) represent the dimensionless mass accretion rate $\dot{m}=\frac{\dot{M}}{\dot{M}_{\rm E}}=\frac{L_{\rm bol}}{\eta L_{\rm E}}$ and Eddington accretion limit $\dot{m}_{\rm E}=1/\eta$, respectively.}
\begin{center}
\scalebox{0.95}{%
\begin{tabular}{ccccccccccc}
\hline 
Source & $L_{\rm 2-10 keV}$  & $\kappa$ & $L_{\rm bol}/L_{\rm E}$ & $r_{\rm isco}$ & $\eta$ & $\dot{m}$ & $\dot{m}_{\rm E}$  \\[0.1cm]
    &  ($10^{41}$~erg~s$^{-1}$) &   &    & ($r_{\rm g}$) &     &    &    \\  [0.2cm]
(1)    &   (2)   &   (3)   &   (4)  &  (5)   & (6) &   (7)  &    (8)     \\                                                      
\hline 
J0107 & $31.3^{+0.7}_{-0.7}$ & 20$^{+7}_{-7}$ & 0.28$^{+0.48}_{-0.19}$ &  $2.5_{-0.6}^{+1.2}$  & $0.20_{-0.06}^{+0.06}$ &  $1.4_{-1.1}^{+4.2}$  &  $5.0_{-1.1}^{+2.4}$     \\ [0.25cm]

J0228 & $18.4^{+0.7}_{-0.7}$ & 18$^{+6}_{-6}$ & 0.75$^{+1.24}_{-0.50}$ & $2.8_{-1.2}^{+0.4}$  & $0.18_{-0.02}^{+0.14}$ & $4.2_{-3.4}^{+8.7}$  &  $5.6_{-2.5}^{+0.9}$     \\ [0.25cm]

J0940 & $37.7^{+0.8}_{-0.8}$ & 21$^{+7}_{-7}$ & 0.36$^{+0.59}_{-0.24}$ &  $1.3_{-0.05}^{+0.3}$  & $0.38_{-0.07}^{+0.02}$ & $0.9_{-0.6}^{+2.1}$  &  $2.6_{-0.1}^{+0.6}$     \\ [0.25cm]

J1023 & $12.8^{+0.2}_{-0.2}$ & 16$^{+6}_{-5}$ & 0.29$^{+0.51}_{-0.19}$   &  $4.1_{-1.9}^{+0.6}$ & $0.12_{-0.02}^{+0.11}$ &  $2.4_{-2.0}^{+5.1}$  &  $8.3_{-3.9}^{+1.2}$     \\ [0.25cm]
                   
J1140 & $38.2^{+0.3}_{-0.3}$ & 21$^{+7}_{-8}$ & 0.45$^{+0.76}_{-0.31}$ & $1.7_{-0.2}^{+0.2}$ & $0.3_{-0.03}^{+0.03}$ &  $1.5_{-1.1}^{+3.0}$  &  $3.4_{-0.3}^{+0.4}$     \\ [0.25cm]
                             
J1347 & $33.0^{+0.5}_{-0.5}$ & 20$^{+7}_{-7}$ & 0.47$^{+0.80}_{-0.32}$ &  $3.1_{-1.2}^{+1.8}$ & $0.16_{-0.06}^{+0.10}$ &  $2.9_{-2.3}^{+9.4}$  &  $6.2_{-2.4}^{+3.6}$     \\ [0.25cm]
   
J1357 & $85.3^{+1.9}_{-1.9}$ & 26$^{+9}_{-9}$ & 1.0$^{+1.7}_{-0.7}$ & $4.8_{-0.6}^{+0.3}$ & $0.10_{-0.01}^{+0.01}$ &  $9.6_{-6.8}^{+17.9}$  &  $9.6_{-1.2}^{+0.6}$     \\ [0.25cm] 
       
J1434 & $3.0^{+0.1}_{-0.1}$ & 13$^{+5}_{-4}$ & 0.04$^{+0.08}_{-0.03}$ & $3.7_{-1.4}^{+1.7}$  & $0.14_{-0.04}^{+0.08}$ &  $0.3_{-0.2}^{+1.0}$  &  $7.4_{-2.8}^{+3.4}$     \\ [0.25cm]

J1541 & $37.5^{+1.2}_{-1.2}$ & 21$^{+7}_{-8}$ & 0.35$^{+0.59}_{-0.24}$ & $2.3_{-0.7}^{+1.2}$ & $0.22_{-0.08}^{+0.10}$ & $1.6_{-1.2}^{+4.9}$  &  $4.5_{-1.4}^{+2.3}$     \\ [0.25cm]  
       
J1559 & $40.1^{+0.1}_{-0.1}$ & 21$^{+7}_{-7}$ & 0.38$^{+0.63}_{-0.25}$ & $1.24_{-0.0}^{+0.4}$ & $0.40_{-0.11}^{+0.0}$ &  $0.9_{-0.6}^{+2.5}$  &  $2.5_{-0.0}^{+0.9}$     \\ [0.25cm]

J1626 & $3.8^{+0.2}_{-0.2}$ & 14$^{+4}_{-5}$ & 0.08$^{+0.12}_{-0.05}$ & $3.5_{-1.7}^{+0.9}$ & $0.14_{-0.03}^{+0.13}$ &  $0.5_{-0.4}^{+1.2}$  &  $7.0_{-3.3}^{+1.8}$     \\ [0.25cm]
                                 
J1631 & $3.1^{+0.2}_{-0.2}$ & 13$^{+5}_{-4}$ & 0.05$^{+0.08}_{-0.03}$ & $3.2_{-0.9}^{+0.8}$ & $0.16_{-0.03}^{+0.06}$ & $0.3_{-0.2}^{+0.7}$  &  $6.4_{-1.8}^{+1.6}$ \\ [0.25cm]                                   

POX52  & $4.8^{+0.1}_{-0.1}$ & 14$^{+5}_{-5}$ & 0.15$^{+0.15}_{-0.08}$ & $4.0_{-1.8}^{+1.7}$ & $0.12_{-0.04}^{+0.10}$ &  $1.2_{-0.9}^{+2.1}$  &  $8.0_{-3.6}^{+3.3}$     \\ [0.25cm]
\hline 
\end{tabular}}
\end{center} 
\label{tab2b}           
\end{table*}


We then model the observed soft X-ray excess using the new variable density relativistic reflection model ({\tt{relxillDCp}}), which assumes a sandwich configuration for corona that encircles the surrounding disc. The heating of the inner disc by coronal X-rays leads to quasi-thermal emission in the soft X-rays, which is fully represented by the high-density effects in the model (see \citealt{ga16}). As the density parameter is increased, the effective temperature of the reprocessed emission at soft energies increases, leading to a quasi-thermal emission component that appears in the soft band. The model adjusts the temperature of the reflector so that flux in equals flux out, and hence it does not require further thermalisation of the disc.

The {\tt{relxillDCp}} model combines the ionized disc reflection code {\tt{xillverDCp}} \citep{ga13,ga16} with the convolution model {\tt{relconv}} \citep{da13}. The {\tt{relconv}} model determines the relativistic effects in the reflection spectrum and assumes a broken power-law emissivity profile for the illumination of the disc by an X-ray corona. The emissivity profile has the following form: $\epsilon(r)\propto r^{-q_{{\rm in}}}$ for $r_{\rm in}\leq r\leq r_{\rm br}$, and $\epsilon(r)\propto r^{-q_{{\rm out}}}$ for $r_{\rm br}\leq r\leq r_{{\rm {\rm out}}}$, where $q_{\rm in}$ and $q_{\rm out}$ represent inner and outer emissivity indices, respectively, $r_{\rm br}$ is the break radius, $r_{\rm in}$ and $r_{\rm out}$ are the inner and outer disc radii, respectively. The other parameters that characterize the disc reflection model are: black hole spin ($a$), the inclination angle ($\theta^{\circ}$) of the disc to the observer, reflection fraction ({\tt{refl$\_$frac}}), iron abundance ($A_{{\rm Fe}}$), ionization state ($\log\xi$), and number density ($n_{\rm e}$) of electrons in the disc atmosphere. The disc irradiation profile is considered Newtonian over the outer regions of the disc, and hence we fixed the outer emissivity index at 3. We fixed the break radius at 6$r_{{\rm g}}$, which is a typical AGN coronal radius (e.g. \citealt{ma21,wf11}). We assume the solar abundance of iron and fixed the inner and outer radii of the accretion disc at the innermost stable circular orbit ($r_{\rm {isco}}$) and 1000$r_{{\rm g}}$ ($r_{\rm g}=GM_{\rm BH}/c^{2}$), respectively. We set {\tt{refl$\_$frac}}$=-1$ in the {\tt{relxillDCp}} model to solely describe the disc reflection. Since {\tt{relxillDCp}} considers the thermal Comptonization model {\tt{nthComp}} \citep{zd96,zy99} as the irradiating (primary) continuum, we replaced the {\tt{zpowerlw}} continuum with {\tt{nthComp}}. The slope of the {\tt{nthComp}} and {\tt{relxillDCp}} components are tied. The seed photon temperature in the {\tt{nthComp}} model was set at the maximum possible disc temperature (see column~(8) of Table~\ref{tab2a}) for each source. We fixed the electron temperature of the hot coronal plasma at a typical value of 100\keV{}. The relative strength of reflection was measured as a ratio of the disc reflected flux to the irradiating primary source flux in the 0.3--10\keV{} band. We find that the broad-band best-fit spectral model is {\tt{Tbabs$\times$(relxillDCp$+$nthComp)}} for all the sources in the sample, except for J1559 and POX~52. The source, J1559, showed an absorption feature at $\sim 0.7$\keV{}, which was modeled by a Gaussian absorption line {\tt{gabs}}. The {\tt{gabs}} component improves the fit statistic by $\Delta \chi^{2}=92.9$ for 3 free parameters. The expression for the broad-band best-fit model of J1559 is {\tt{Tbabs$\times$gabs$\times$(zgauss$+$relxillDCp$+$nthComp)}}. To model the absorption curvature present in the $1-2$\keV{} band of POX~52, we included an ionized partial covering absorption component ({\tt{zxipcf}}; \citealt{re08}). The {\tt{zxipcf}} model has three free parameters: column density ($N_{\rm {H, wa}}$), ionization state ($\log{\xi_{\rm {wa}}}$) of the absorbing medium and covering fraction ($C_{\rm f}$). The inclusion of the ionized absorption component provided an improvement of $\Delta \chi^{2}=237.5$ for 3 additional free parameters. The maximum likelihood ratio (MLR) test suggests that the {\tt{zxipcf}} model component is $\ge 99.99$~per~cent significant. The best-fit model for POX~52 has the following expression: {\tt{Tbabs$\times$zxipcf$\times$(relxillDCp$+$nthComp)}}. We find that the relativistic reflection from an ionized, higher density accretion disc can self-consistently explain the soft X-ray excess emission for the low-mass AGN sample. The best-fit unfolded spectral models, model components, and data-to-model ratio plots are shown in Fig.~1. We show the broad-band photon count spectra, best-fit models, and residuals plots in Fig.~A2.

To check the robustness of the parameter measurements and determine the constraints on each free parameter, we conducted a Markov Chain Monte Carlo (MCMC) analysis utilizing the Goodman-Weare algorithm in {\tt{XSPEC}}. We run the Markov Chain for 150000 iterations with 100 walkers. We burned the initial 50000 events to obtain a steady chain and used 100000 steady-state samples to determine the parameter distributions and their 90~per~cent confidence intervals. We performed Gelman-Rubin's MCMC convergence test and ensured that the chains have converged with the potential scale reduction factor below 1.2 for each parameter. The contour plots of various parameters infer that the parameter space is not degenerate (see Fig.~A3. A4, A5). The best-fit spectral model parameters, their 90~per~cent confidence intervals and fit statistics for the sample are presented in Table~A1 and A2.

\section{Results and Discussion}
\label{sec4}

\subsection{Eddington ratio and limit}
We measured the Eddington ratio ($L_{\rm bol}/L_{\rm E}$) of the sample through the bolometric correction factor, $\kappa=L_{\rm bol}/L_{\rm 2-10keV}$. The hard X-ray (2$-$10\keV{}) absorption corrected luminosity of the sample is found to be in the range of $L_{\rm 2-10keV}\approx 3\times 10^{41}-9\times 10^{42}$~erg~s$^{-1}$. The average bolometric correction factor and associated uncertainties for each source were obtained from the bolometric correction factor vs. 2--10\keV{} luminosity curve \citep{ho07,vf07}. The measured Eddington ratio is in the range of $L_{\rm bol}/L_{\rm E}=\kappa L_{\rm 2-10keV}/L_{\rm E}\approx 0.01-1.0$, considering the uncertainties of the BH mass estimates. We then estimated the dimensionless mass accretion rate using the formula $\dot{m}=\frac{\dot{M}}{\dot{M}_{\rm E}}=\frac{L_{\rm bol}}{\eta L_{\rm E}}$, where the radiative efficiency $\eta=r_{\rm g}/2r_{\rm isco}$. The innermost stable circular orbit $r_{\rm isco}$ was computed using the equation~(2) of \citet{re21}. The mass accretion rate is below the Eddington accretion limit of $\dot{m}_{\rm E}=1/\eta$ for the sample except for J1357 which has $\dot{m}\sim 10.0$ (see Table~\ref{tab2b}). Therefore, we conclude that the discs of these low-mass AGNs are accumulating matter at sub-Eddington or near-Eddington accretion rates. The measurements are in agreement with the Eddington ratio estimated from the luminosity of the $H_{\beta}$ line for POX~52 \citep{th08} and $H_{\alpha}$ line for the GH07 sample. The hard band absorption-corrected source luminosity, bolometric correction factor, Eddington ratio, innermost stable circular orbit, radiative efficiency, accretion rate and Eddington limit for each AGN are presented in Table~\ref{tab2b}.

\begin{figure*}
\centering
\begin{center}
\includegraphics[scale=0.43,angle=-0]{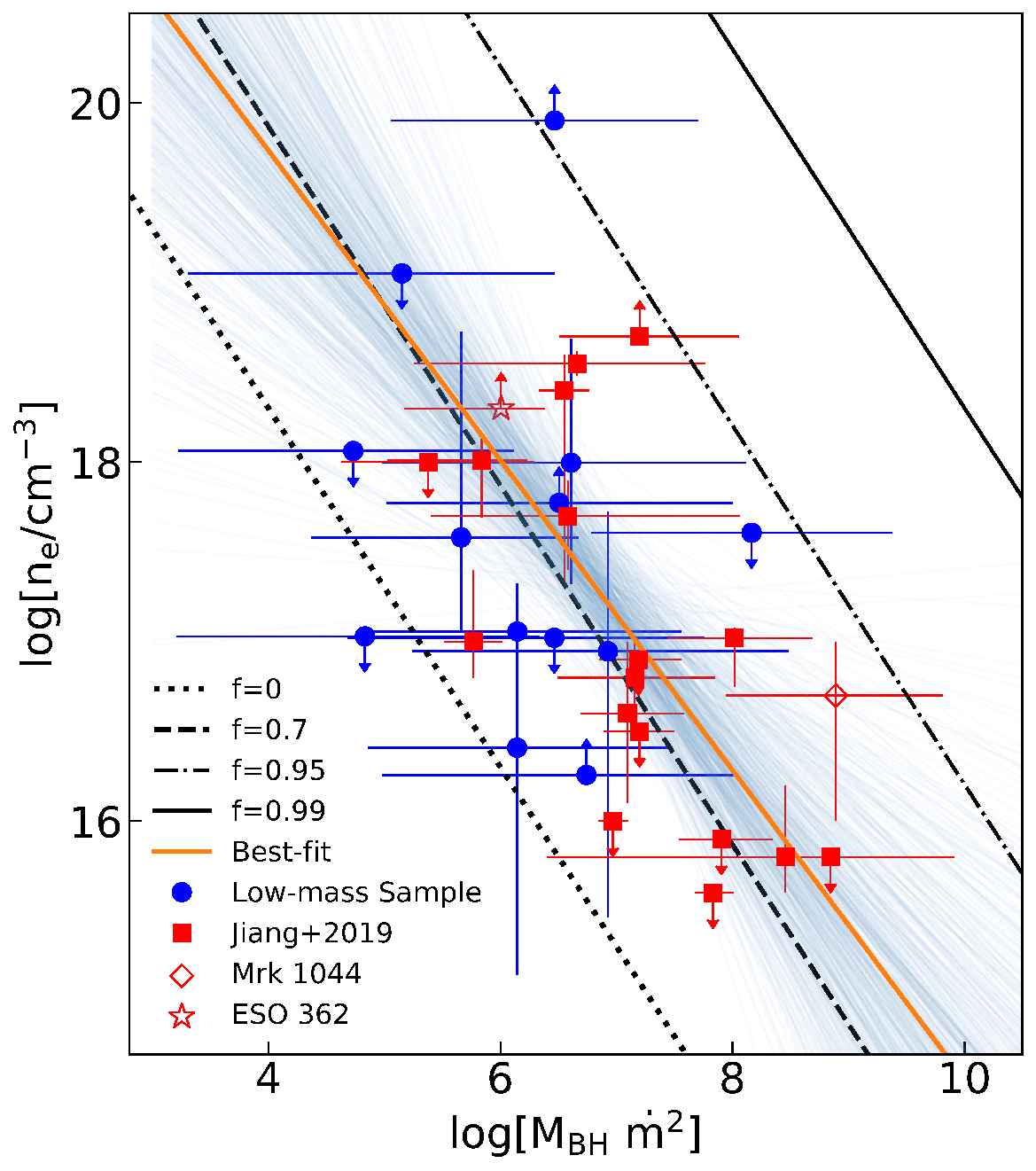}
\includegraphics[scale=0.43,angle=-0]{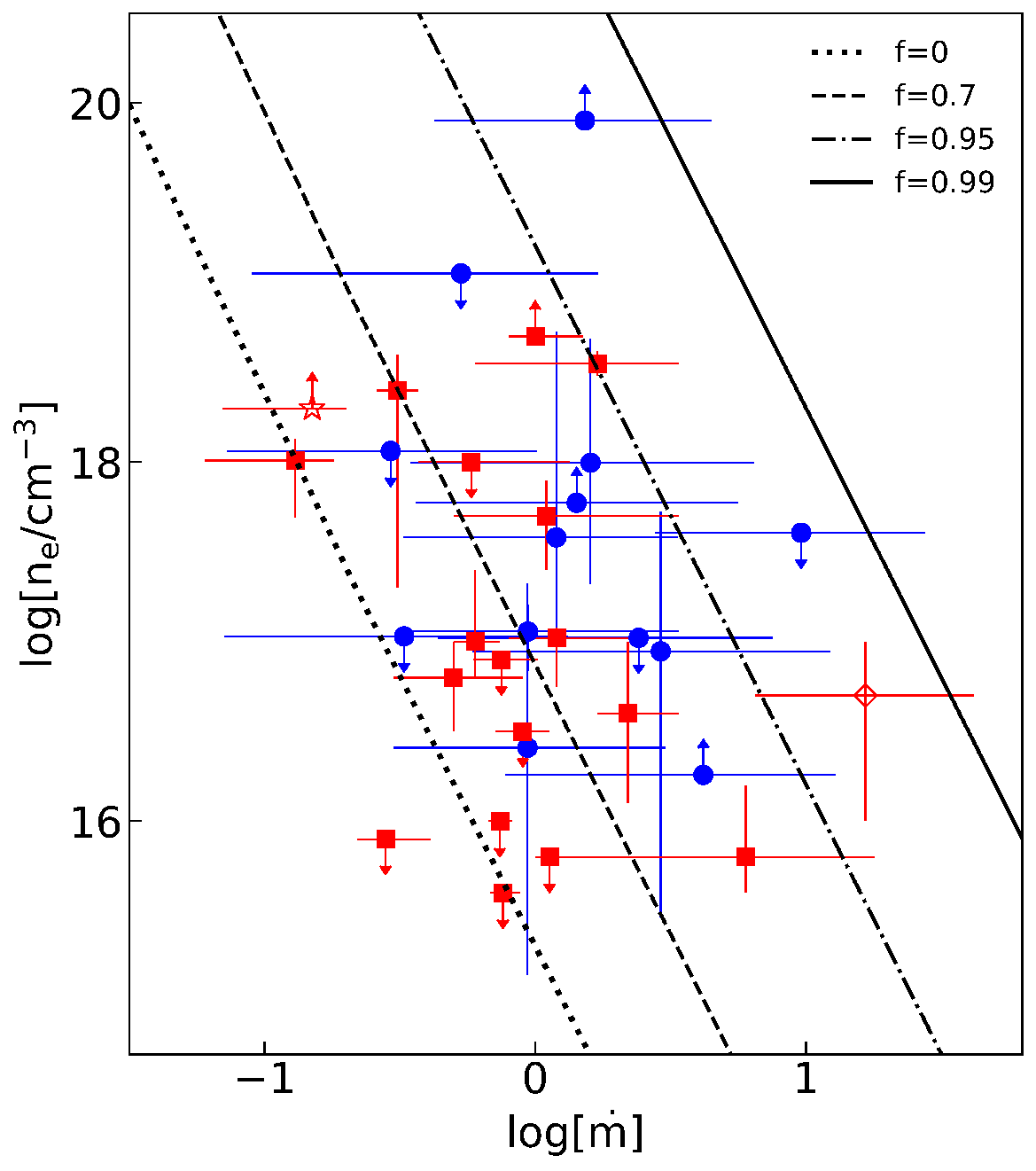}
\includegraphics[scale=0.43,angle=-0]{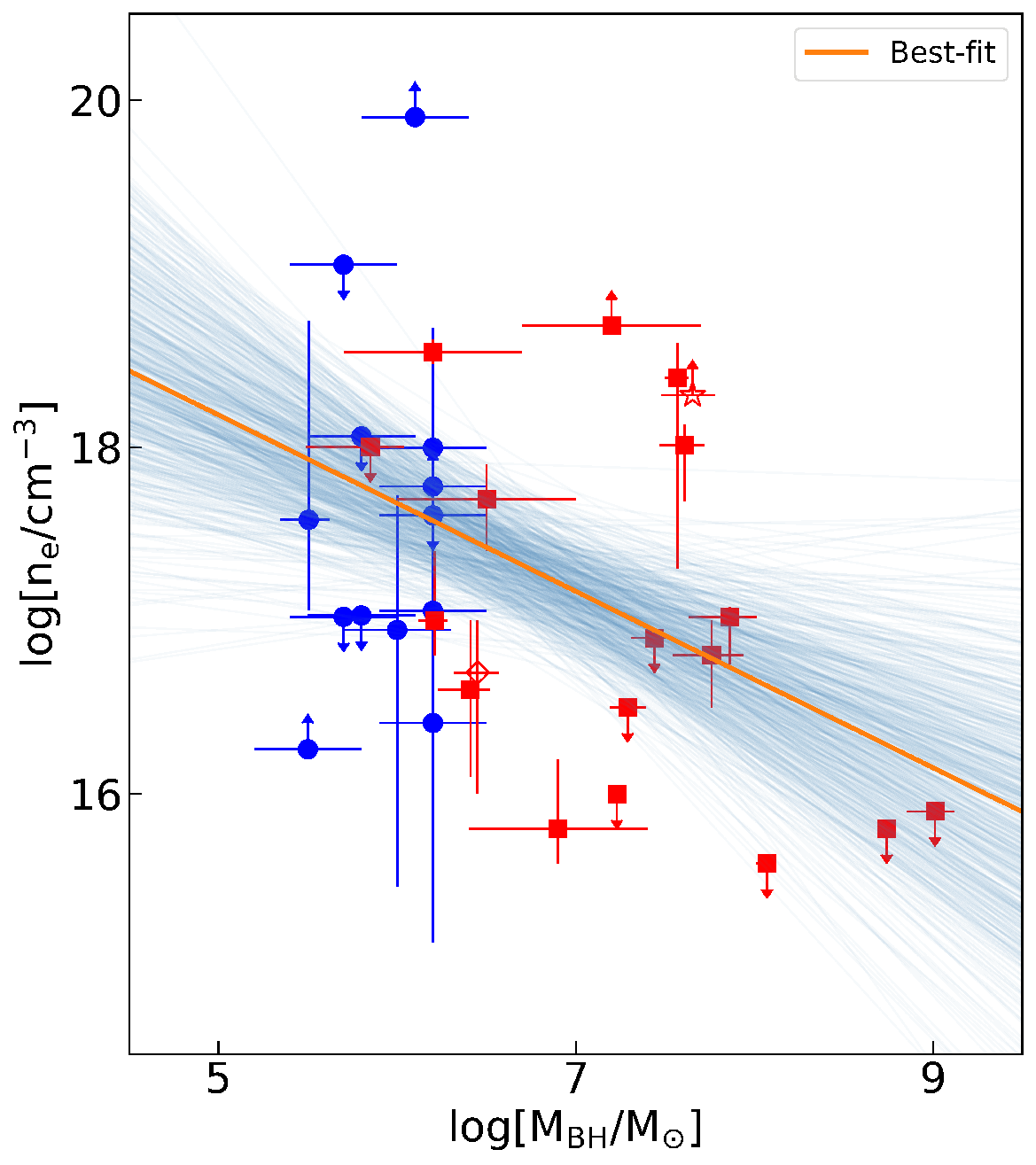}
\includegraphics[scale=0.43,angle=-0]{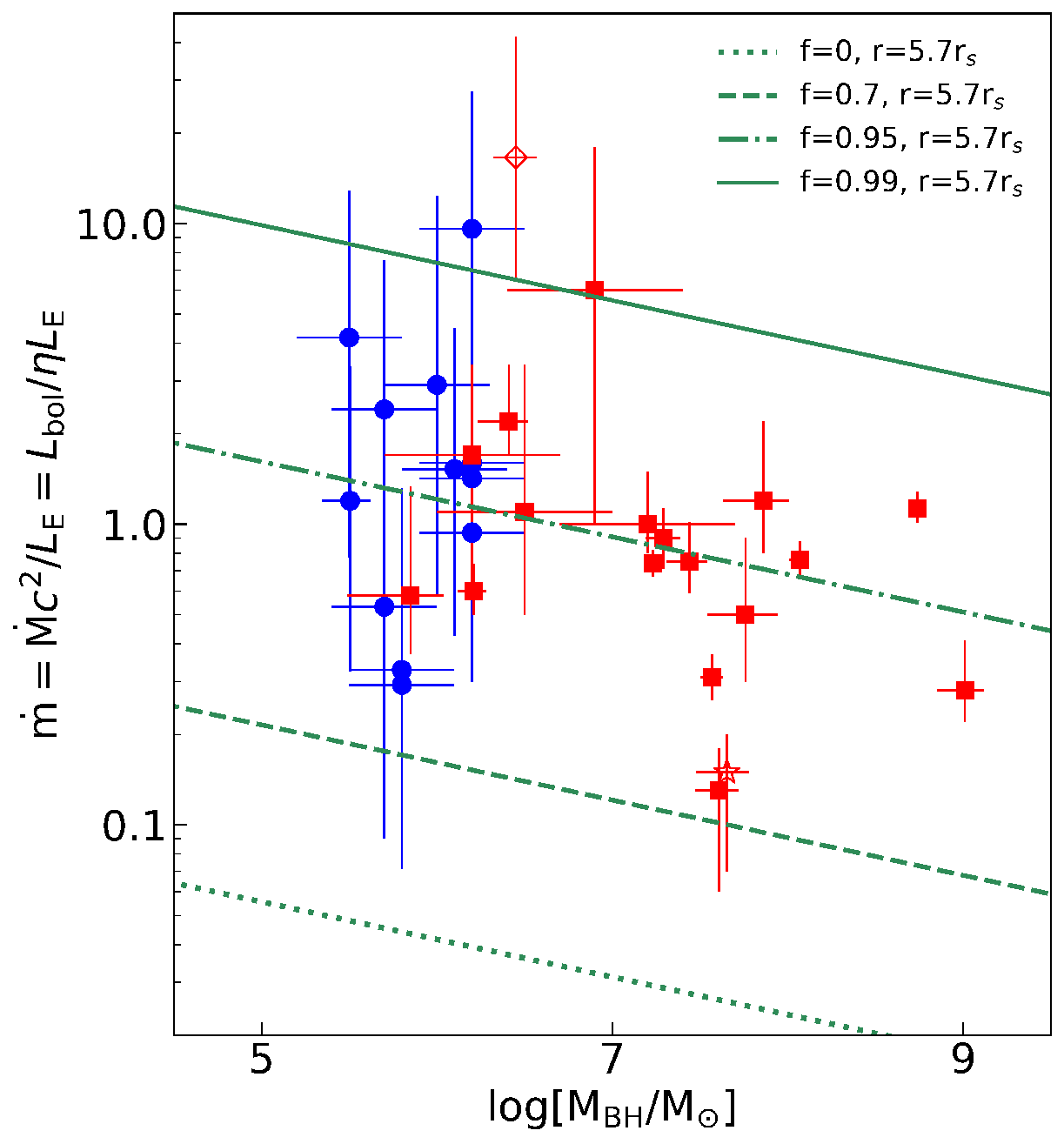}
\caption{The top left and right panels show the variation of disc density $\log n_{\rm e}$ with $\log(M_{\rm BH}\dot{m}^{2})$ and $\log\dot{m}$, respectively, where $\dot{m}=\frac{\dot{M}c^{2}}{L_{\rm E}}=L_{\rm bol}/\eta L_{\rm E}$ is the dimensionless mass accretion rate. The dotted, dashed, dash-dotted, and solid black lines represent solutions of a radiation pressure-dominated disc at the radius of $r=r_{\rm s}$ for $f=0$, 0.7, 0.95, 0.99, respectively, where $f$ is the fraction of the total power released by the disc into the corona. The circle, square, diamond, and star-shaped data points represent density measurements of 13 AGNs from this work, 17 AGNs from \citet{ji19}, the Sy~1 AGN Mrk~1044 from \citet{ma18}, and Sy~1.5 AGN ESO~362$-$G18 from \citet{xu21}, respectively. The bottom left panel depicts the variation of disc density with $\log M_{\rm BH}$. The best-fit linear model and corresponding $1\sigma$ confidence area are presented by the orange solid line and skyblue shaded area, respectively. The bottom right panel shows the variation of dimensionless mass accretion rate with $\log M_{\rm BH}$. The dotted, dashed, dash-dotted, and solid green lines denote the SZ94 solutions at the radius of $r=5.7r_{\rm s}$ for $f=0$, 0.7, 0.95, 0.99, respectively, below which gas thermal pressure dominates the radiation pressure.} 
\end{center}
\label{logne}
\end{figure*}

\begin{figure*}
\centering
\begin{center}
\includegraphics[scale=0.41,angle=-0]{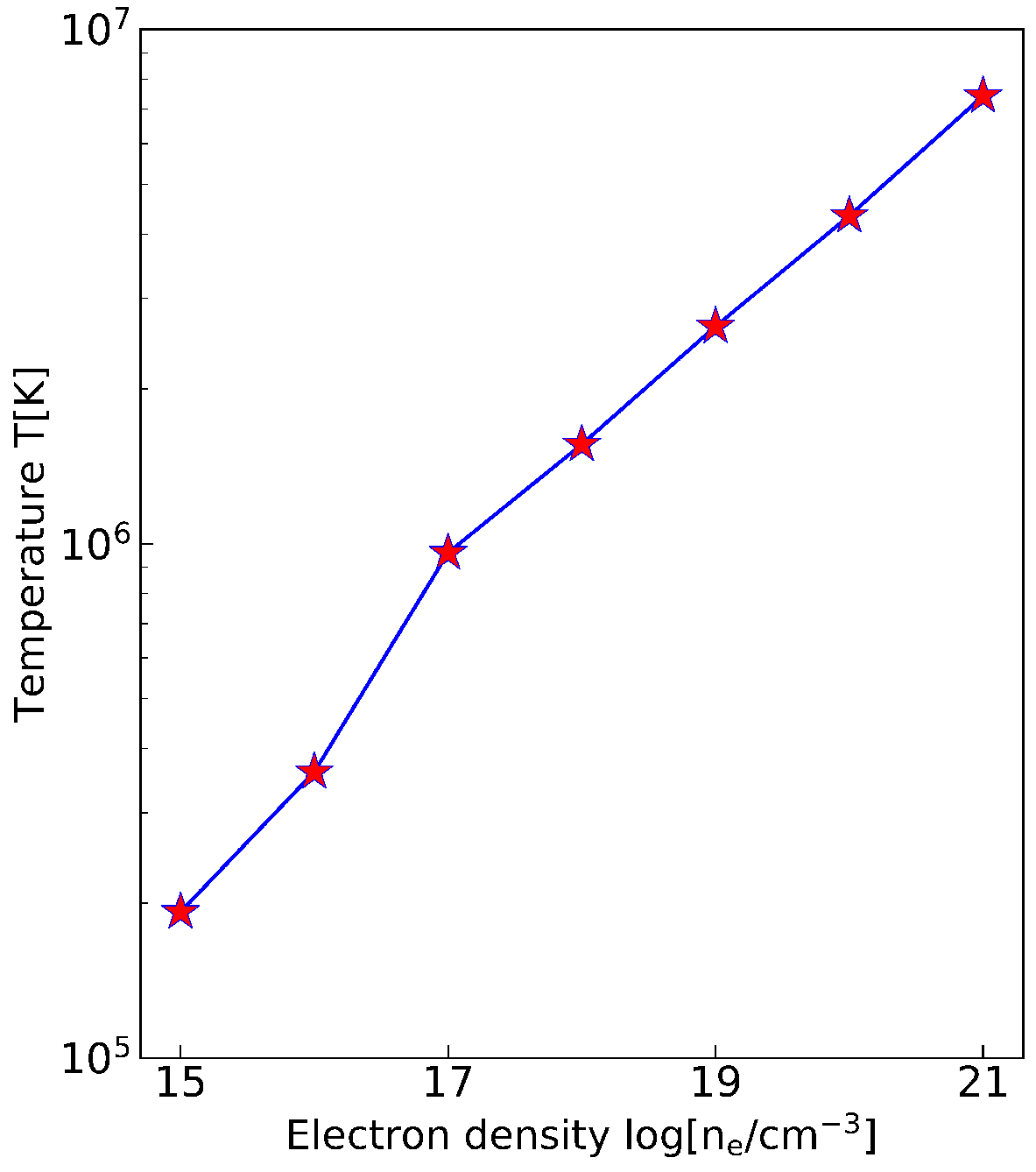}
\includegraphics[scale=0.41,angle=-0]{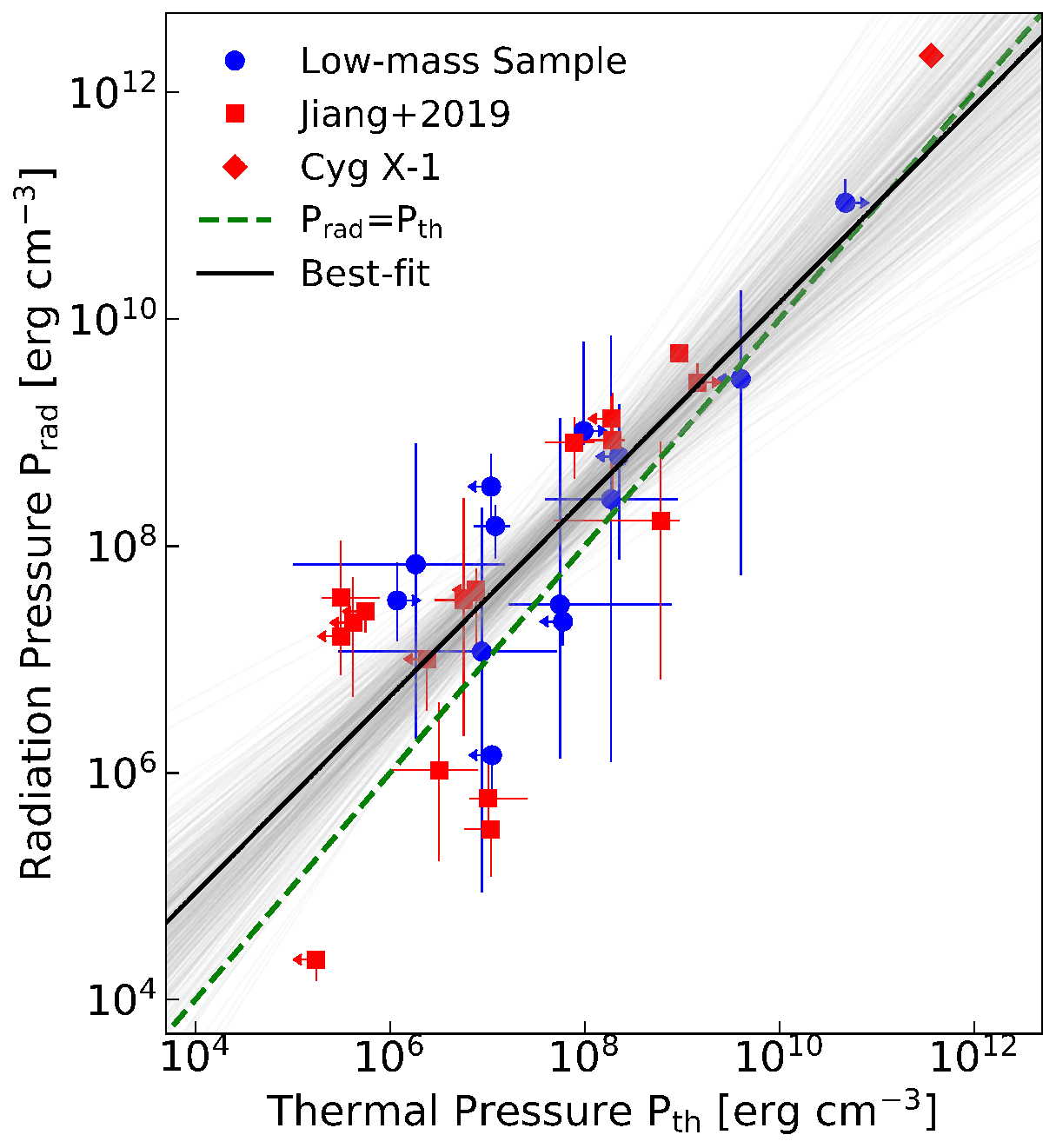}
\caption{Left: temperature at the illuminated atmosphere of the accretion disc as a function density calculated at the Thomson depth of $\tau=1$. Right: variation of ionizing radiation pressure with gas thermal pressure for the low-mass AGN sample from this work, 17 Sy~1 galaxies from \citet{ji19} and Cyg~X-1 from \citet{tom18}. The best-fit linear model and corresponding 1$\sigma$ confidence level for the combined sample are shown by the solid black line and gray shaded area, respectively. The dashed green line shows the $P_{\rm rad}=P_{\rm th}$ solution. The plot is mostly intended to show agreement and consistency with the reflection model, rather than as a correlation (note that the density contributes to both axes).}
\end{center}
\label{logne}
\end{figure*}

\begin{table}
\caption{$\Delta \chi^{2}$ denotes the changes in the chi-squared statistic compared to a constant density ($10^{15}$~cm$^{-3}$) disc. C.I.-MLR and C.I.-FTEST, respectively, represent the significance of the high-density over a constant density ($10^{15}$~cm$^{-3}$) reflection model estimated by the maximum likelihood ratio test and F-test. We did not perform these significance tests for five sources where we obtain only an upper limit on the disc density.}
\begin{center}
\scalebox{0.9}{%
\begin{tabular}{ccccccccc}
\hline 
Source & $\Delta \chi^{2}$  &  C.I.-MLR (~per~cent)  &  C.I.-FTEST (~per~cent) \\
\hline 
J0107 & 8.8  & 99.7  & 99.9  \\ [0.2cm]

J0228 & 3.0   & 91.8 & 96.1   \\ [0.2cm]

J0940 & 4.5 & 96.6 & 95.4  \\ [0.2cm]

J1023 & $-$  & $-$ & $-$  \\ [0.2cm]
                   
J1140 & 14.0 & 99.98 & 99.98    \\ [0.2cm]
                             
J1347 & 16.5  & 99.99 & 99.99   \\ [0.2cm]
   
J1357 & $-$ & $-$ & $-$    \\ [0.2cm]   
       
J1434 & $-$  & $-$ & $-$   \\ [0.2cm]

J1541 & 6.2  & 98.7 & 99.2   \\ [0.2cm]   
       
J1559 & 23.5  & 99.99 & 99.99    \\ [0.2cm]

J1626 & $-$   & $-$ & $-$ \\ [0.2cm]
                                 
J1631 & $-$   & $-$ & $-$   \\ [0.2cm]                                     

POX52 & 8.0   & 99.5 &  99.0    \\ [0.2cm]
\hline 
\end{tabular}}
\end{center} 
\label{tab}           
\end{table}

\subsection{Disc gas density}
We first verified the significance of higher disc densities by fitting all the spectra with a constant disc of density $\log[n_{\rm e}$/cm$^{\rm -3}]=15$, which resulted in a steeper photon index, higher disc ionization state, and worsened the statistics. The changes in the $\chi^2$-statistics are quoted in Table~\ref{tab}. We excised five sources for which we only obtain an upper limit on the disc density. We then estimated the confidence levels by performing the maximum likelihood ratio (MLR) test and F-test. These tests suggest that a higher density disc is preferred for eight out of thirteen AGNs with $\ge 90$~per~cent confidence.

According to SZ94, the density profile of electrons at the illuminated atmosphere of a radiation pressure-dominated disc in the SS73 model follows

\begin{equation}
n_{\rm e}=\frac{1}{\sigma_{{\rm T}}r_{{\rm s}}}\frac{256\sqrt{2}}{27}\alpha^{-1}r^{3/2}\dot{m}^{-2}\Big[1-(r_{{\rm in}}/r)^{1/2}\Big]^{-2}(1-f)^{-3}.
\label{density}
\end{equation}
The disc viscosity parameter $\alpha$ is assumed to be 0.1 in the SS73 model; $\sigma_{\rm T}=6.64\times10^{-25}$~cm$^2$ is the Thomson scattering cross section; $r$ is the disc radius in units of $r_{\rm s}$, where $r_{\rm s}$ is the Schwarzschild radius; the inner disc radius $r_{\rm in}$ is fixed at $r_{\rm isco}$, the average of which is estimated to be $\sim 1.4r_{\rm s}$ for our sample; $\dot{m}$ is the dimensionless mass accretion rate; $f$ is the fraction of the total power transported from the disc to the corona and ranges from 0 to 1.

We assessed the SS73 disc model by exploring the possible correlation between the disc density and BH mass times the accretion rate squared ($M_{{\rm BH}}\dot{m}^{2}$), as observed in higher mass AGNs \citep{ji19}. We show the variation of $\log n_{\rm e}$ with $\log(M_{\rm BH}\dot{m}^{2})$ for the combined sample consisting of 13 AGNs from this work, 17 AGNs from \citet{ji19}, the Sy~1 galaxy Mrk~1044 from \citet{ma18}, and Sy~1.5 galaxy ESO~362$-$G18 from \citet{xu21} in Figure~2 (top left). At the disc radius of $r = r_{\rm s}$, the solutions of the radiation pressure-dominated disc for $f=0$, 0.7, 0.95, 0.99 are represented by the dotted, dashed, dash-dotted, and solid black lines, respectively. We observed a relatively strong anti-correlation between $\log n_{\rm e}$ and $\log(M_{\rm BH}\dot{m}^{2})$ for the combined sample with a Spearman correlation coefficient of $\rho_{\rm s}=-0.59$ and a p-value of $3.5\times 10^{-4}$. We then performed a linear regression analysis using a Bayesian method which accounts for errors both in the independent and dependent variables \citep{ke07}. The best-fit $\log n_{\rm e}$ vs. $\log(M_{{\rm BH}}\dot{m}^{2})$ relation and associated 1$\sigma$ confidence interval for the combined sample is:

\begin{equation}
\log n_{\rm e}=(-0.9\pm 0.3)\log(M_{\rm BH}\dot{m}^{2})+(23.2\pm 1.9).
\label{logn_logmbh.mdot2}
\end{equation}
The solid orange line represents the best-fit regression model and is found to be consistent with the SZ94 density solution for $r\approx r_{\rm s}$ and $f\approx 0.7$:

\begin{equation}
\log n_{\rm e}=-\log(M_{\rm BH}\dot{m}^{2})+23.9.
\end{equation}

We also showed the variation of disc density with the accretion rate for the combined sample in Fig.~2 (top right). We did not find any correlation between the disc density and accretion rate and hence did not fit a linear model to the $\log n_{\rm e}$ vs. $\log(\dot{m})$ plane. The SZ94 solutions for various values of $f$ are shown by black lines in the top right panel of Fig.~2.

The bottom left panel of Fig.~2 plots $\log n_{\rm e}$ vs. $\log M_{\rm BH}$ for the combined sample. The disc density falls off with increasing BH mass which is in agreement with the SS73 model. We obtain a relatively weak anti-correlation ($\rho_{\rm s}=-0.37$, p=0.04) between the disc density and BH mass. The best-fit relation obtained from our Bayesian linear regression analysis is:

\begin{equation}
\log n_{\rm e}=(-0.5\pm 0.2)\log M_{\rm BH}+(20.7\pm 1.5).
\label{logn_mbh}
\end{equation}
The orange solid line and skyblue shaded area, respectively, show the best-fit regression model and associated $1\sigma$ uncertainties for the combined sample. 

We then explored the possibility of a gas pressure-dominated disc. The radius at which gas thermal pressure balances the radiation pressure in the accretion disc \citep{sz94} is:

\begin{equation}
\frac{r}{(1-r^{-1/2})^{16/21}}=188\Big[\alpha\Big(\frac{M_{{\rm BH}}}{10^{8}}\Big)\Big]^{2/21}\dot{m}^{16/21}(1-f)^{6/7},
\label{radius}
\end{equation}
The L.H.S of the equation~(\ref{radius}) has a minimum at $r\approx 5.7 r_{\rm s}$ and the gas thermal pressure dominates the radiation pressure for certain combinations of $M_{\rm BH}$, $\dot{m}$ and $f$. The bottom right panel of Fig.~2 depicts $\dot{m}$ vs. $\log M_{\rm BH}$. The SZ94 disc solutions for $f=0$, 0.7, 0.95, 0.99 are shown by the dotted, dashed, dash-dotted, and solid green lines, respectively, below which the accretion disc is dominated by the gas thermal pressure. We find that the accretion rates of the low-mass sample are well above the thresholds for $f\gtrsim 0.7$, and the thermal pressure begins to dominate the radiation pressure below $f\approx 0.7$.

At the surface of the disc where reflection is taking place the incident downward radiation pressure is balanced by the upward gas thermal pressure, together with any magnetic gradient acting on the gas. This may result in compression of the gas which is known as Radiation Pressure Compression (RPC; see e.g. \citealt{do02,bi19}). Ignoring the magnetic field and any upward intrinsic radiation from the disc since $f$ is high, the radiation pressure $P_{\rm rad}=\frac{L}{4\pi r^{2}c}=\frac{\xi n_{\rm e}}{4\pi c}$ balances the gas thermal pressure $P_{\rm th}=n_{\rm e}k_{\rm B}T$, where $T$ is the temperature at the illuminated atmosphere. This means that $\xi/4\pi ck_{\rm B}T$ should be close to unity, as found for values typical of our sample of a few 100~erg~cm~s$^{-1}$ for $\xi$ and a few million $K$ for $T$. Exploring this further, we followed the prescription of \citet{ga16} and derived the temperature profile in the vertical direction of the illuminated zone for various densities, $\log n_{\rm e}=15-21$ at the Thomson depth of $\tau=1$. The temperature solutions were obtained from reflection calculations using input parameters appropriate for the sample on average, i.e., $\Gamma=2$, $\xi=500$~erg~cm~s$^{-1}$, $A_{\rm Fe}=1$, and $E_{\rm cut}=300$\keV{}. The left panel of Fig.~3 shows the derived $T$ vs. $\log n_{\rm e}$ profile. We then estimated the ionizing radiation pressure and gas thermal pressure for our sample along with 17 Sy~1 AGNs from \citet{ji19} and Cyg~X-1 from \citet{tom18}, and show the results in the right hand panel of Fig.~3. The dashed green line depicts the solution for pressure equality, i.e., $P_{\rm rad}=P_{\rm th}$. The small offset of the best-fit line (solid black) from the $P_{\rm rad}=P_{\rm th}$ relation suggests that a magnetic pressure gradient may exist across the disc surface so that the incident radiation pressure is balanced by a combination of thermal pressure and excess magnetic pressure. Therefore, the energy density associated with this excess is essentially the difference between the incident radiation pressure and thermal pressure:

\begin{equation}
|P_{\rm rad}-P_{\rm th}|=\Delta P_{\rm mag}=\frac{B^{2}}{8\pi},
\end{equation}
where $\Delta P_{\rm mag}$ is the magnetic pressure gradient across the surface and measured in units of erg~cm$^{-3}$, $B$ is the magnetic flux density in units of Gauss. The average excess magnetic flux density across the disc surface for the high-mass AGNs of \citet{ji19} and our low-mass AGN sample is estimated to be $\sim 7\times 10^{4}$~Gauss and $\sim 1.5\times 10^{5}$~Gauss, respectively. The measured magnetic flux density of Cyg~X-1 is even higher, $B\sim 7\times 10^{6}$~Gauss. Presumably the absolute value of magnetic field either side of the surface is much higher than this. RPC may have the effect of increasing the surface density above that prescribed by the equations of SZ94 and contribute to some of the spread seen in the plots.

\begin{figure}
\centering
\begin{center}
\includegraphics[scale=0.43,angle=-0]{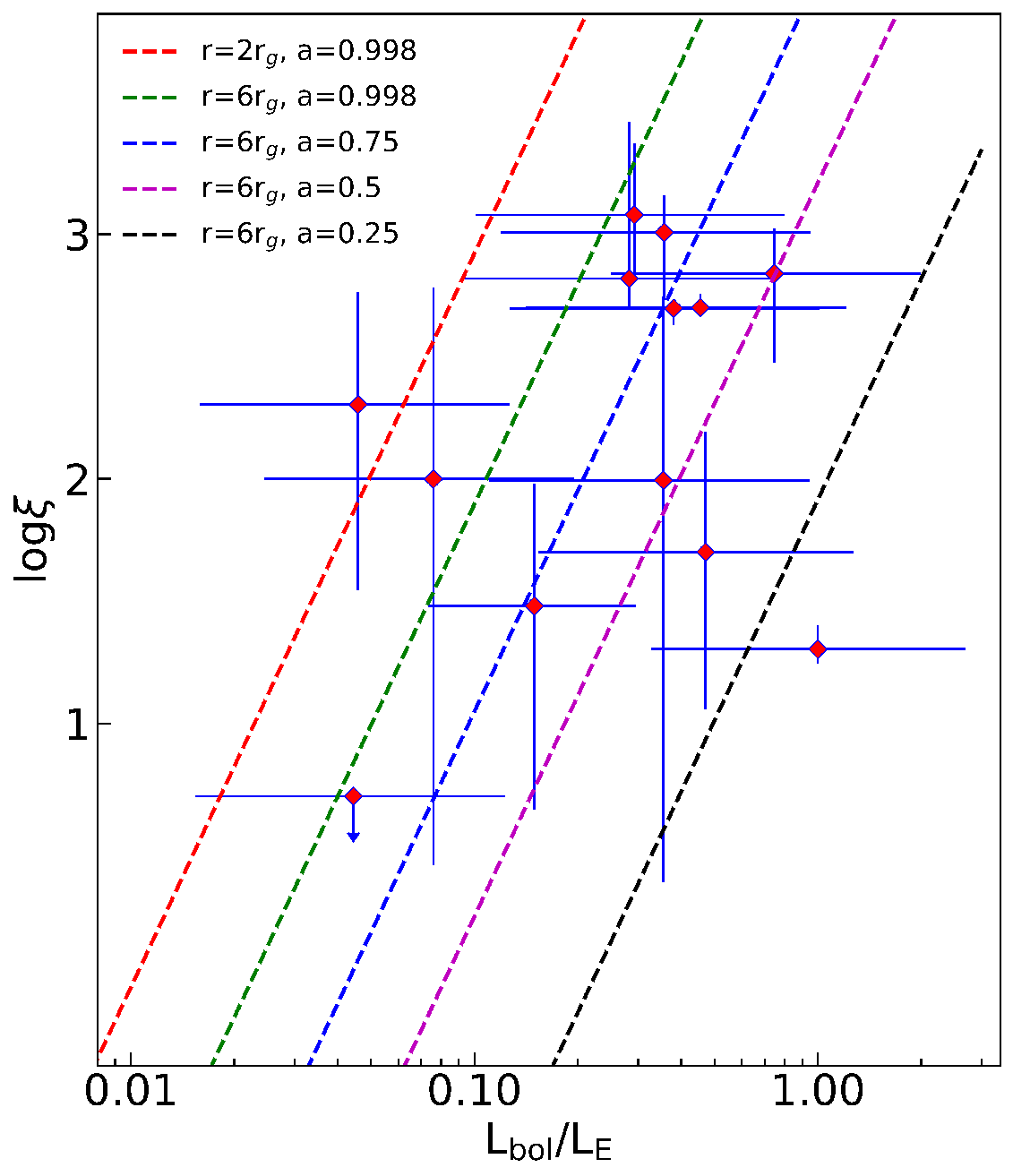}
\caption{Disc ionization parameter $\log\xi$ vs. Eddington ratio $L_{\rm bol}$/$L_{\rm E}$ for the sample. The dashed red, green, blue, magenta, and black lines show model $\log\xi$ as a function of $L_{\rm bol}$/$L_{\rm E}$ for ($a=0.998,r=2r_{\rm g}$), ($a=0.998,r=6r_{\rm g}$), ($a=0.75,r=6r_{\rm g}$), ($a=0.5,r=6r_{\rm g}$), ($a=0.25,r=6r_{\rm g}$), respectively.}
\end{center}
\label{xi}
\end{figure}

\subsection{Disc ionization state}
We measured the ionization state of the irradiated inner accretion disc from the reflection model and found that the inner accretion discs of this low-mass sample are moderate to highly ionized ($\log[\xi_{\rm fit}/{\rm erg~cm~s^{-1}}]\sim 1.0-3.0$). We first checked whether the ionization parameter $\log\xi_{\rm fit}$ inferred from the fitting is consistent with that expected from the inferred disc density, irradiating flux, and inner disc radius. We can calculate the disc ionization parameter $\log\xi_{\rm cal}$ at the inner radius using the following formula:

\begin{equation}
\xi_{\rm cal}=4\pi F_{\rm irr}(r_{\rm in})/n_{\rm e},
\end{equation}
where $F_{\rm irr}(r_{\rm in})$ is the source flux irradiating at the inner disc and calculated using the equation~(4) of \citet{zd20}:

\begin{equation}
F_{\rm irr}(r_{\rm in})\approx4.6\times10^{24}\frac{L_{\rm 0}/L_{\rm E}}{(M/10M_{\odot})(r_{\rm in}/2r_{\rm g})^{2}} \text{erg cm\ensuremath{^{-2}}s\ensuremath{^{-1}}},
\end{equation}
where $M_{\rm BH}$ is the BH mass in units of $M_{\odot}$, $r_{\rm in}$ is the inner disc radius in units of $r_{\rm g}$ and fixed at $r_{\rm isco}$\footnote[1]{See column~(5) of Table~3 for $r_{\rm isco}$.}, $L_{\rm 0}=4\pi D_{\rm L}^{2}F_{\rm obs,0}$ and $F_{\rm obs,0}$ is the observed irradiating source flux in the 0.01$-$100\keV{} band. Our calculations suggest that the average ionization parameter of the disc at the ISCO is $\log[\frac{\xi_{\rm cal}}{\rm erg~cm~s^{-1}}]=2.17_{-0.60}^{+0.58}$ and consistent with the model fitted average ionization parameter of $\log[\frac{\xi_{\rm fit}}{\rm erg~cm~s^{-1}}]=2.21_{-0.41}^{+0.36}$ at the 2$\sigma$ confidence level \footnote[2]{We used the Monte-Carlo bootstrapping method to determine the mean and corresponding 2$\sigma$ confidence limits.}.

The photoionization of the disc surface is characterized by two main parameters: irradiating X-ray continuum flux and disc density. Thus the measurement of the disc ionization parameter can address various aspects of the disc/corona interplay. According to \citet{ba11}, the ionization parameter of a radiation pressure supported disc illuminated by a geometrically thick corona in the SS73 model can be approximated as:

\begin{equation}
\xi\approx 4.33\times 10^{9}\left(\frac{\eta}{0.1}\right)^{-2}\left(\frac{\alpha}{0.1}\right)\left(\frac{r}{r_{\rm g}}\right)^{-\frac{7}{2}}f_{\rm c}(1-f_{\rm c})^{3}\Big(\frac{L_{{\rm bol}}}{L_{{\rm E}}}\Big)^{3}GR_{{\rm corr}},
\end{equation}
where radiative efficiency $\eta\approx 0.1$ \citep{dl11}; viscosity parameter $\alpha=0.1$ \citep{ss73}; coronal dissipation fraction $f_{\rm c}\approx 0.45$ \citep{vf07}; $r$ is the disc radius in units of $r_{\rm g}$; $GR_{{\rm corr}}=R_{{\rm R}}^{3}R_{{\rm z}}^{-2}R_{{\rm T}}^{-1}$ is a general relativistic correction factor and is solely dependent on the dimensionless black hole spin $a$ (e.g. \citealt{nt73,kr99}). Therefore, the dependence of disc ionization $\log\xi$ on the Eddington ratio ($L_{\rm bol}$/$L_{\rm E}$) is predominantly determined by two parameters: $r$ and $a$:

\begin{equation}
\log\xi\simeq\varPsi(r,a)\Big(\frac{L_{\rm bol}}{L_{\rm E}}\Big).
\end{equation}
Fig.~4 shows the dependence of the disc ionization parameter on the Eddington ratio for the low-mass sample. We show the SS73 model predicted $\log\xi-L_{\rm bol}$/$L_{\rm E}$ relationships for five different combinations of BH spin and inner disc radius: ($a=0.998,r=2r_{\rm g}$), ($a=0.998,r=6r_{\rm g}$), ($a=0.75,r=6r_{\rm g}$), ($a=0.5,r=6r_{\rm g}$), and ($a=0.25,r=6r_{\rm g}$). We noticed that the inferred ionization states of low-mass AGN discs are consistent with the SS73 model predicted solutions. The derived $\log\xi-L_{\rm bol}$/$L_{\rm E}$ plane suggests that if the relativistic reflection originated from within $6r_{\rm g}$ of the inner accretion disc, then the measured ionization parameters of the low-mass sample require spins to be in the range of $a\in [0.25,0.998]$ with a median value of $\sim 0.75$.

\begin{figure}
\centering
\begin{center}
\includegraphics[scale=0.43,angle=-0]{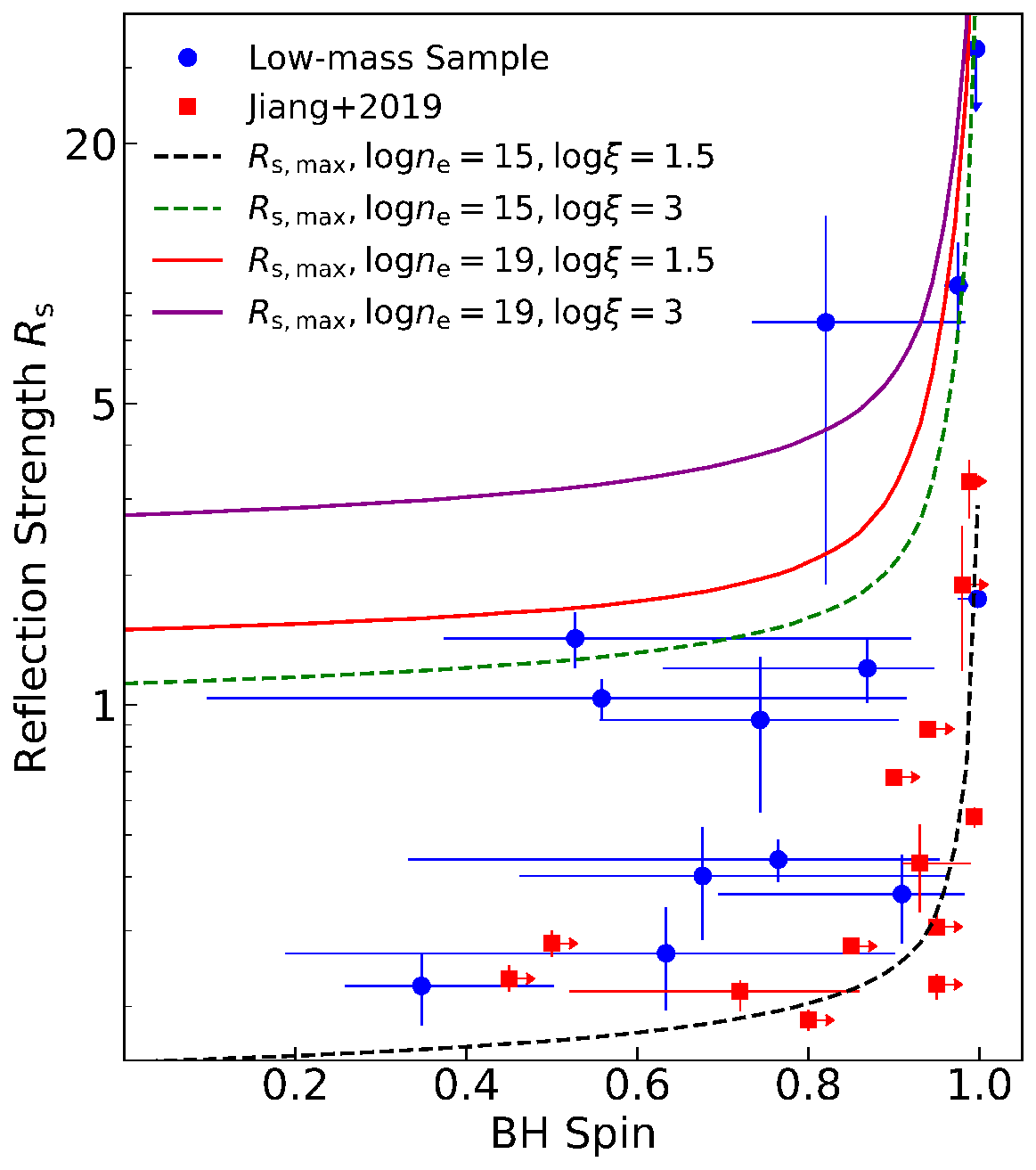}
\caption{Reflection strength $R_{\rm s}$ as a function of the BH spin. The circles and squares denote AGNs from our work and \citet{ji19}, respectively. For sources with multiple Obs.~IDs, we computed the mean value of $R_{\rm s}$. The black, green, red, and purple lines show theoretical models of the maximum possible reflection strength vs. spin for ($\log n_{\rm e}=15,\log\xi=1.5$), ($\log n_{\rm e}=15,\log\xi=3$), ($\log n_{\rm e}=19,\log\xi=1.5$) and ($\log n_{\rm e}=19,\log\xi=3$), respectively. The units of $n_{\rm e}$ and $\xi$ are ${\rm cm^{-3}}$ and ${\rm erg~cm~s^{-1}}$, respectively.}
\end{center}
\label{rf_spin}
\end{figure}

\begin{figure*}
\centering
\begin{center}
\includegraphics[scale=0.46,angle=-0]{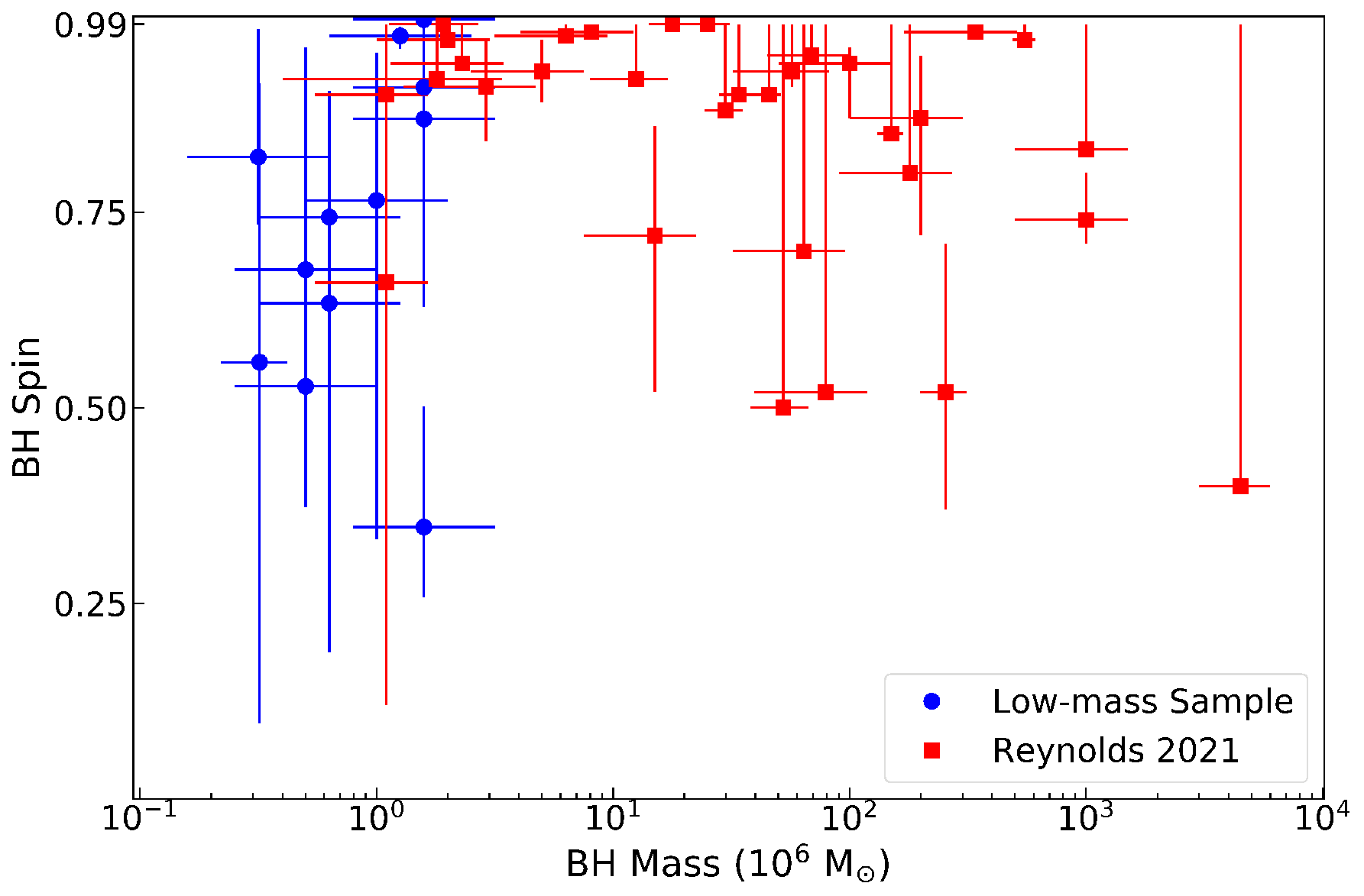}
\caption{Evolution of black hole spin as a function of mass. The circles and squares represent spin measurements for 13 Seyfert~1 galaxies from this work and 32 AGNs from \citet{re21}, respectively. The error bars on the spin parameter correspond to the 90~per~cent confidence intervals for both samples.}
\end{center}
\label{spin}
\end{figure*}

\begin{figure*}
\centering
\begin{center}
\includegraphics[scale=0.39,angle=-0]{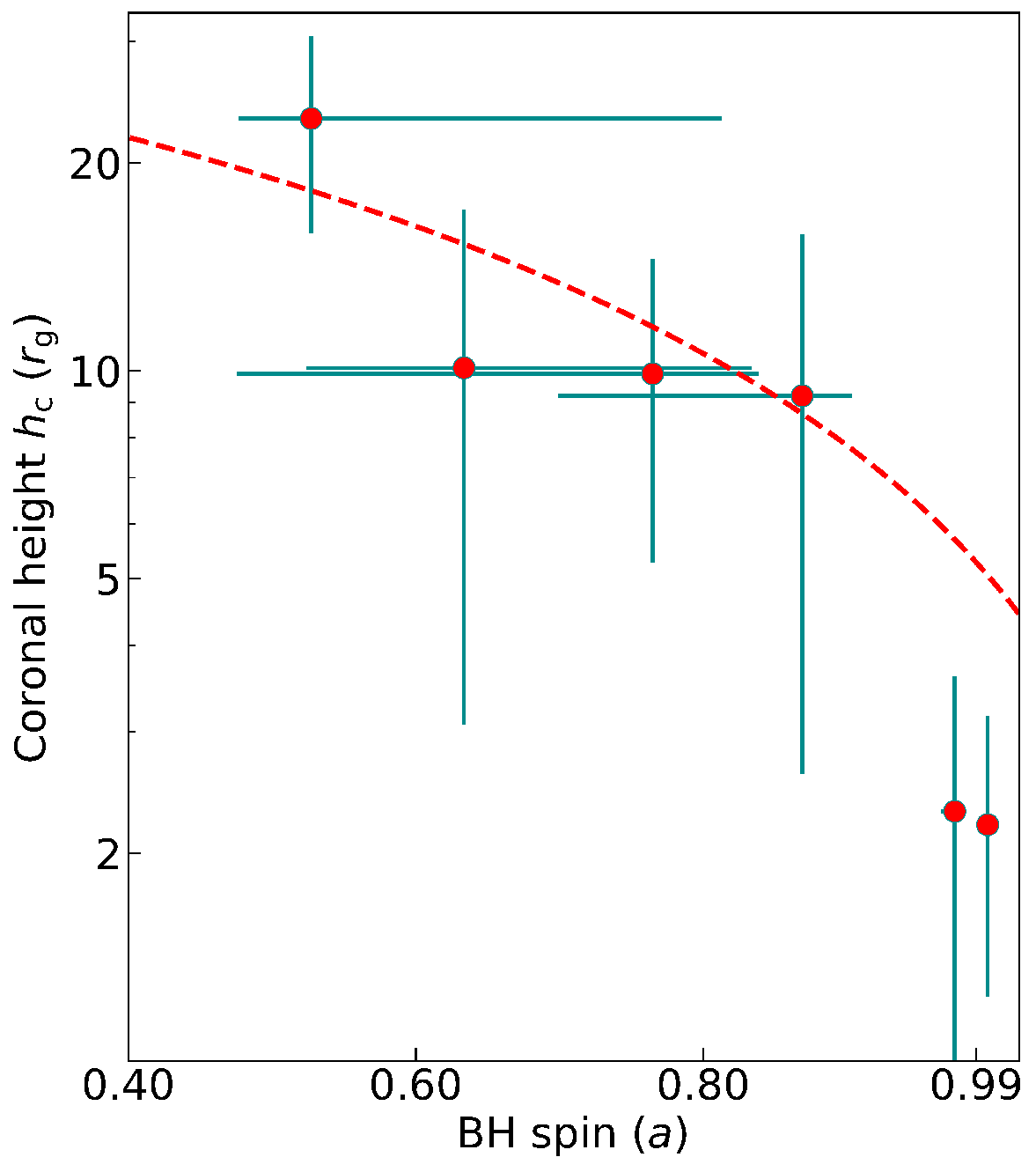}
\includegraphics[scale=0.39,angle=-0]{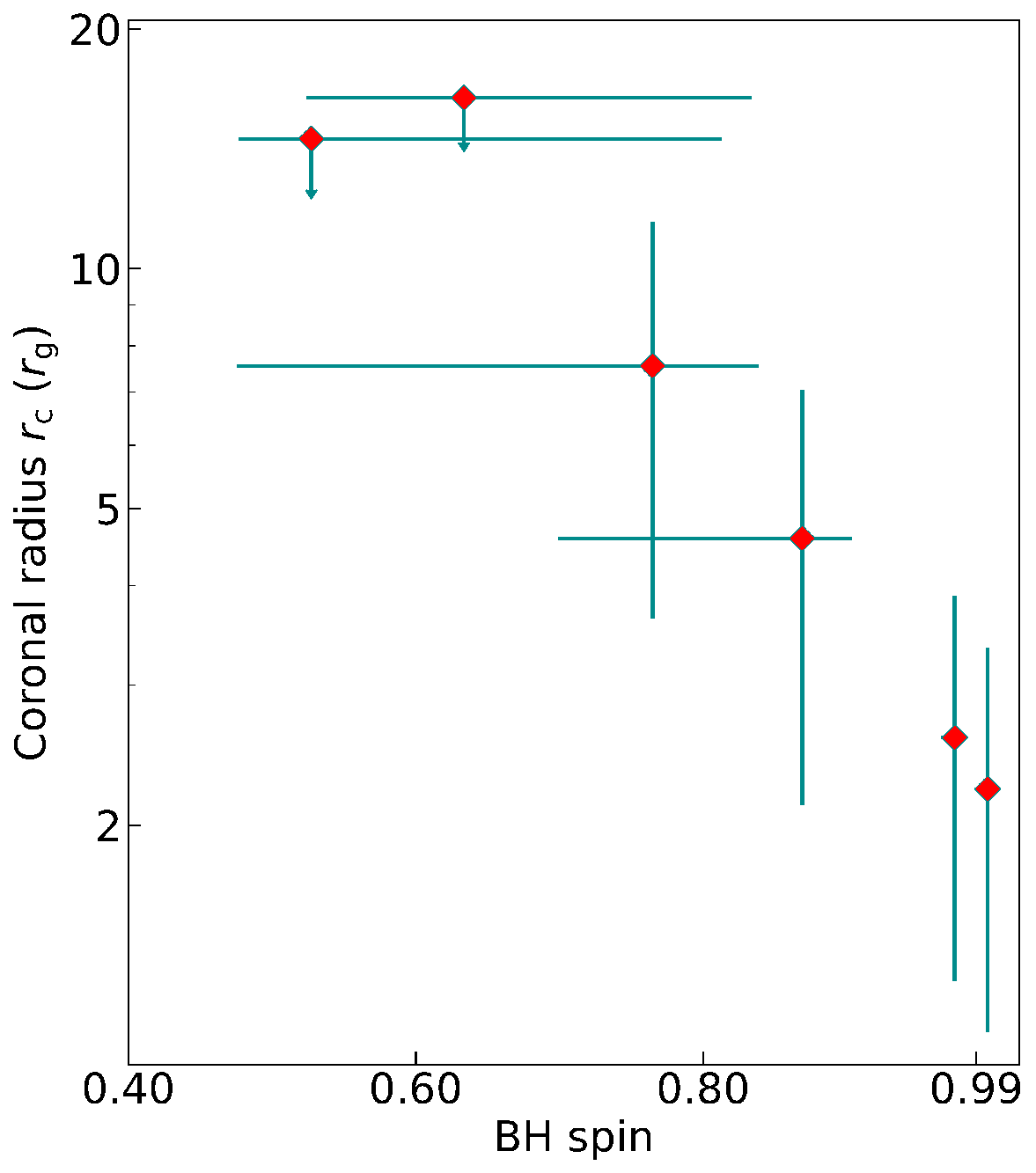}
\caption{Variation of coronal height ($h_{\rm c}$) and radius ($r_{\rm c}$) as a function of black hole spin for six low-mass AGNs where the detection significance of soft reverberation lags exceeded 90 per~cent (see Table~2 of \citealt{ma21}). Error bars on both X and Y variables represent 1$\sigma$ confidence levels. The dashed red line represents the dependence of coronal height on the black hole spin for a Novikov-Thorne disc.}
\end{center}
\label{hc_rc_a}
\end{figure*}

\subsection{Black hole spin}
We measured the black hole spin of the low-mass AGN sample from the modeling of the relativistic reflection continuum. \citet{wa13} employed a similar approach to estimate the BH spin of bare AGNs. We find that the black holes at the center of these low-mass AGNs are spinning with an average value of $\sim 0.75$. The BH spin determines the location of the ISCO and hence the inner edge of the accretion disc. Therefore, the spin has a direct impact on the area of the reflector, which affects the reflection strength. \citet{da14} had shown that it is possible to check for any unphysical solutions of BH spin using the measured strength of reflection in the $20-50$\keV{} band. However, we determine the reflection strength as the ratio between the disc reflected flux and irradiating source flux in the observed $0.3-10$\keV{} band to be consistent with \citet{ji19}. In Fig.~5, we show the dependence of the reflection strength $R_{\rm s}$ on the BH spin for our sample together with the AGN sample from \citet{ji19}, where both the black hole spin and reflection strength were calculated using the high-density disc reflection model. We then overplotted the \citet{da14} model predicted maximum possible reflection strength vs. spin curves for four different combinations of disc density and ionization parameters: ($\log n_{\rm e}=15,\log\xi=1.5$), ($\log n_{\rm e}=15,\log\xi=3$), ($\log n_{\rm e}=19,\log\xi=1.5$) and ($\log n_{\rm e}=19,\log\xi=3$), where the units of $n_{\rm e}$ and $\xi$ are ${\rm cm^{-3}}$ and ${\rm erg~cm~s^{-1}}$, respectively. We assume a lamppost configuration for corona to compute the maximum strength of reflection. We find that the low-mass AGN sample rules out the $\log n_{\rm e}=15, \log\xi=1.5$ combination and prefers higher density or higher ionization for the accretion disc. Additionally, all four theoretical $R_{\rm s}-a$ curves disfavor unrealistic solutions of low spin and large $R_{\rm s}$ for both samples and support the reflection origin of their broad-band spectra.

To study the evolution of BH spin as a function of mass, we constructed the BH spin vs. mass plane for our low-mass sample along with 32 AGNs from \citet{re21}, which is shown in Fig.~6. For the low-mass AGNs, the spin parameter ranges from $a_{\rm min}\sim 0.25$ to $a_{\rm max}\sim 0.998$, suggesting the presence of slow to rapidly spinning black holes at the center of these AGNs. We noticed a population of more slowly spinning black holes below $M_{{\rm BH}}\sim 10^{6}M_{\odot}$ and above $M_{{\rm BH}}\sim 5\times 10^{7}M_{\odot}$. To test the significance of the differences between the low ($M_{{\rm BH}}\le 10^{6}M_{\odot}$), medium ($10^{6}M_{\odot}\le M_{{\rm BH}}\le 10^{8}M_{\odot}$), and high-mass ($10^{8}M_{\odot}\le M_{{\rm BH}}\le 10^{10}M_{\odot}$) black hole spin populations, we performed a Kolmogorov-Smirnov (KS) test between the low-mass and medium-mass bins, and between the medium-mass and high-mass bins. The KS tests suggest that samples were drawn from distinct populations with 99~per~cent significance. Therefore, we conclude that the variation of black hole spin as a function of mass is intrinsic and not driven by the randomness of samples. One possible explanation for the low spin is that the black holes at the center of the low-mass AGNs could result from the collision or merger of intermediate-mass black holes with mass $\sim 10^{5}M_{\odot}$ or lower. The detection of gravitational waves from intermediate-mass black holes with the laser interferometer space antenna ({\it LISA}) can confirm the collision/merger as one of the possible formation mechanisms for the low-mass end of supermassive black holes. The objects with high spin have higher radiative efficiency for the same mass accretion rate and can be approximately five times more luminous than those with zero spins and hence tend to dominate flux-limited samples \citep{va16}.

In \citet{ma21}, we estimated the height and radius of the X-ray emitting corona through time lag analysis for six low-mass AGNs, namely J0107, J1023, J1140, J1347, J1434, and J1559, in which we detected reverberation lags with $>90\%$ confidence. The coronal height was measured as the light-crossing distance between the corona and accretion disc, and it depends on the amplitude of the soft reverberation lag and black hole mass. On the other hand, the coronal radius was determined by the hard viscous propagation time delay, coronal height, and black hole mass. Both the observed soft and hard X-ray lags were dilution-corrected. Additionally, the reverberation lags were corrected for the `Shapiro' delays. Here we investigate the dependence of coronal extents on the black hole spin. Fig.~7 depicts the variation of coronal height and radius with the spin of the central black hole for these six AGNs. The right-hand panel of Fig.~7 has a strikingly similar shape to the half-light radius of a Novikov-Thorne (NT) disc (see Fig.~1 of \citealt{ak00}). This is the radius within which half of the radiation flux is emitted by the disc. Our results suggest $h_{c}\sim r_{c}$, which is relevant for radio-quiet AGNs. Hence, the coronal height can be approximated as $h_{c}\approx 5+28(1-a)$ \citep{fa14}. The dashed red line shows model $h_{c}$ as a function of spin for the NT disc. The variation of coronal size with black hole spin indicates that the most compact corona can be found only around the most rapidly spinning black holes for which the ISCO is at a smaller radius. If the corona is formed from magnetic field lines anchored at the disc, we can expect a higher density of field lines at a smaller radius. The results also point out that spin cannot be well measured unless the innermost parts of the disc are well illuminated, which means the coronal extent has to be $\lesssim 10r_{g}$ for rapidly spinning black holes and $\lesssim 25r_{g}$ for moderately rotating black holes. Presumably, an observational selection effect is at play. However, precise measurements of coronal height or radius require detailed physical modeling of X-ray time lag spectra. Therefore, the observed spin dependence of coronal extents needs to be tested through simultaneous time lag and flux-energy spectral modeling.

\section{Conclusions}
\label{sec5}
In this paper, we study the broad-band (0.3--10\keV{}) \xmm{} spectra of a sample of 13 low-mass ($M_{\rm BH}\approx 10^{5-6} M_{\rm \odot}$) AGNs using the new high-density disc reflection model and determine the detailed accretion properties of the low-mass end of SMBHs in the X-ray energy band. We conclude the main results of this work below:

\begin{itemize}
\item The X-ray spectra of the sample reveal excess emission below $\sim 1.5$\keV{}. Additionally, we detect a narrow Fe~K$_\alpha$ emission line in the NLS1 galaxy J1559 and an ionized, partial covering absorption in the Sy~1.5 galaxy POX~52. The soft X-ray excess emission is well described by the relativistic reflection from an ionized, higher density disc. The modeling of the reflection continuum prefers a density considerably higher than $10^{15}$~cm$^{-3}$ for eight out of thirteen sources in our sample with significance $\ge 90$~per~cent. The average value of the disc density as inferred from the reflection modeling is $\sim 10^{18}$~cm$^{-3}$. 

\medskip

\item The accretion discs of the low-mass active galaxies are found to be accreting matter at sub-Eddington or near-Eddington rates, and the surface density of the disc atmosphere depends both on the accretion rate and BH mass. We noticed an anti-correlation between $\log n_{\rm e}$ and the BH mass times accretion rate squared, $\log(M_{\rm BH}\dot{m}^{2})$, which is expected for a radiation pressure supported disc. The fraction of the total power released from the disc into the corona is estimated to be $\sim 70$~per~cent in the inner regions of the radiation pressure-dominated disc. When the disc power transferred to the corona is below 70~per~cent, the gas pressure starts dominating the disc.

\medskip
\item The variation of the ionizing radiation pressure with thermal pressure suggests a magnetic pressure gradient across the surface of the disc, and the incident radiation pressure is balanced by both thermal and excess magnetic pressure. When the magnetic pressure gradient across the disc surface is weak, the radiation pressure compression (RPC) mechanism compresses the X-ray illuminated disc atmosphere in the vertical direction and consequently increases the disc density.

\medskip
\item The illuminated gas in the inner accretion disc is found to be moderate to highly ionized and agrees with the ionization parameter calculated from the disc density and irradiating source flux at the ISCO within 2$\sigma$ confidence limits. The dependence of the disc ionization state on the Eddington ratio is consistent with the predictions of the radiation pressure-dominated SS73 disc model for a geometrically thick corona. The $\log\xi-L_{\rm bol}$/$L_{\rm E}$ relation prescribes that if the relativistic reflection originated from within 6~gravitational radii of the disc, the black hole spin should be in the range of $a\in [0.25,0.998]$.

\medskip
\item The nuclei of the low-mass active galaxies harbor mild to fast-spinning black holes, where the measured spin parameter is $a\gtrsim 0.25$, and the most compact corona is likely to be located around the most rapidly spinning black holes. The relatively low spin provides a hint that the merger of IMBH pairs is perhaps responsible for the formation of the low-mass end of SMBHs, which can be confirmed with the future gravitational wave detector {\it LISA}.

\end{itemize}

\section{Future work}
\label{sec5}
Our current high-density relativistic reflection model {\tt{relxillDCp}} allows the density parameter to vary within the range $\log[n_{\rm e}$/cm$^{\rm -3}$]$=15-20$ and can provide the physical understanding of the broad-band (0.3--10\keV{}) AGN spectra. The observations of these low-mass AGNs with \nustar{} in the 3--78\keV{} band are required to test the validity of the high-density reflection model for the description of the Compton hump, which is usually observed at around $15-30$\keV{}. 

We find that the density is pegged at the upper bound of the model for a few sources in our sample, which argues for the existence of an even higher density disc. We, therefore, plan to extend the density parameter of the {\tt{relxillDCp}} model up to $\log[n_{\rm e}$/cm$^{\rm -3}$]$=22$ in the future version.

\section*{Acknowledgments}
LM is supported by NASA ADAP grant 80NSSC21K1567. JAG acknowledges support from NASA grant 80NSSC21K1567 and from the Alexander von Humboldt Foundation. JAT acknowledges partial support from NASA ADAP grant 80NSSC19K0586. BDM acknowledges support via Ram\'on y Cajal Fellowship RYC2018-025950-I. AGM acknowledges partial support from Polish National Science Center (NCN) grant numbers 2016/23/B/ST9/03123 and 2018/31/G/ST9/03224. We thank the anonymous reviewer for a constructive report. We dedicate this paper to doctors and nurses fighting the COVID-19 global pandemic. This research has made use of the NASA/IPAC Extragalactic Database (NED), which is operated by the Jet Propulsion Laboratory, California Institute of Technology, under contract with the NASA. This research has made use of data, software and/or web tools obtained from the High Energy Astrophysics Science Archive Research Center (HEASARC), a service of the Astrophysics Science Division at NASA/GSFC and of the Smithsonian Astrophysical Observatory's High Energy Astrophysics Division.

\section*{Data availability}
All the data used in this article are publicly available from ESA \xmm{} Science Archive (XSA; \url{http://nxsa.esac.esa.int/}) and NASA High Energy Astrophysics Science Archive Research Center (HEASARC; \url{https://heasarc.gsfc.nasa.gov/}).

\bibliographystyle{mnras}

\appendix
\section{Additional tables and plots}
\label{sec:appendix}

\begin{table*}
\centering
\caption{The best-fit model parameters of the sample as obtained from the broad-band X-ray spectral fitting of the EPIC-pn+MOS data with the model, {\tt{Tbabs$\times$(relxillDCp$+$nthComp)}}. The fixed and tied parameters are denoted by symbols `(f)' and `$\ast$', respectively. The 90~per~cent confidence intervals of parameters are estimated through Monte Carlo simulations. The columns are (1) source name, (2) observation ID, (3) photon index of the primary continuum, (4) Primary continuum normalization in units of photons~cm$^{-2}$~s$^{-1}$~keV$^{-1}$, (5) inner emissivity index, (6) BH spin, (7) disc inclination angle, (8) disc ionization parameter, (9) disc density in logarithmic units, (10) normalization of {\tt{relxillDCp}}, (11) reflection strength $R_{\rm s}$, (12) fit statistic.}
\begin{center}
\scalebox{0.87}{%
\begin{tabular}{ccccccccccccc}
\hline 
Source  & Obs.~ID  & & & {\tt{nthComp}} & &  &  & {\tt{relxillDCp}} & & & &    \\
\hline 
        &   & $\Gamma$ &  $K_{\rm nth}$ & $q_{\rm in}$ & $a$ & $\theta^{\circ}$ & $\log[\frac{\xi}{\rm erg~cm~s^{-1}}]$ & $\log[\frac{n_{\rm e}}{{\rm cm^{-3}}}]$ & $K_{\rm ref}$ & $R_{\rm s}$ & $\frac{C-\rm {stat}}{\rm {d.o.f}}$ \\
   &     & & [$10^{-5}$] &  &   &   &   &   & [$10^{-6}$] &   &     \\ [0.1cm]
(1)  & (2) & (3) & (4) & (5) & (6)  & (7) & (8) & (9) & (10) & (11) & (12)   \\ [0.1cm]
\hline 
J0107 & 0305920101 & $2.03_{-0.03}^{+0.17}$ & $5.3_{-4.6}^{+2.0}$ & $\ge 7.1$ & $0.87_{-0.24}^{+0.08}$ & $41_{-25}^{+9}$ & $2.8_{-0.1}^{+0.6}$ & $\ge 17.8$ & $0.7_{-0.3}^{+0.5}$ & $1.2_{-0.2}^{+0.2}$ & $\frac{193.8}{240}$  \\ [0.15cm]

J0228 & 0674810101 & $1.91_{-0.12}^{+0.45}$ & $\le 4.3$ & $\ge 5.6$ & $0.82_{-0.09}^{+0.16}$ & $\le 45$ & $2.8_{-0.4}^{+0.2}$ & $\ge 16.3$  & $1.1_{-0.6}^{+0.5}$ & $7.7_{-5.8}^{+5.9}$ & $\frac{82.2}{116}$ \\ [0.15cm]

J0940 & 0306050201 & $2.06_{-0.07}^{+0.11}$ & $\le 14.4$ & $\ge 5.9$ & $0.996_{-0.015}^{+0.001}$ & $65_{-14}^{+4}$ & $3.0_{-0.3}^{+0.1}$ & $16.4_{-1.3}^{+0.9}$ & $9.5_{-5.2}^{+2.6}$ & $\le 33.2$ & $\frac{243.7}{232}$ \\ [0.15cm]

J1023 & 0108670101 & $2.28_{-0.06}^{+0.08}$ & $1.7_{-0.9}^{+0.7}$ & $\ge 4.0$ & $0.53_{-0.15}^{+0.39}$ & $27_{-8}^{+13}$ & $3.1_{-0.2}^{+0.3}$  & $\le 17.0$ & $0.3_{-0.1}^{+0.1}$ & $0.9_{-0.2}^{+0.2}$ & $\frac{586.4}{645}$ \\ [0.15cm]

     & 0605540201 & $2.28^{\ast}$ & $1.5_{-1.4}^{+1.2}$ & $-$ & $0.53^{\ast}$ & $27^{\ast}$ & $3.1^{\ast}$  & $-$ & $0.6_{-0.2}^{+0.2}$ & $2.1_{-0.4}^{+0.4}$ & -- \\ [0.15cm]

     & 0605540301 & $2.28^{\ast}$ & $1.4_{-1.0}^{+0.8}$ & $-$ & $0.53^{\ast}$ & $27^{\ast}$ & $3.1^{\ast}$  & $-$ & $0.3_{-0.1}^{+0.1}$ & $1.3_{-0.4}^{+0.4}$ & -- \\ [0.15cm]

J1140  & 0305920201 & $2.04_{-0.03}^{+0.02}$ & $1.3_{-1.2}^{+0.7}$ & $6.9_{-0.7}^{+1.3}$ & $0.975_{-0.016}^{+0.012}$ & $44_{-9}^{+6}$ & $2.7_{-0.01}^{+0.06}$  & $\ge 19.9$ & $2.6_{-0.3}^{+0.4}$ & $16.0_{-7.0}^{+7.1}$ & $\frac{1047.5}{1042}$   \\ [0.15cm]

        & 0724840101 & $2.04^{\ast}$ & $1.8_{-0.8}^{+0.4}$ & $6.9^{\ast}$ & $0.975^{\ast}$ & $44^{\ast}$ & $2.7^{\ast}$  & -- & $1.1_{-0.1}^{+0.2}$ & $4.7_{-1.0}^{+1.1}$ & -- \\ [0.15cm]

        & 0724840301 & $2.04^{\ast}$ & $2.4_{-1.4}^{+0.6}$ & $6.9^{\ast}$ & $0.975^{\ast}$ & $44^{\ast}$ & $2.7^{\ast}$  & -- & $2.2_{-0.2}^{+0.4}$ & $7.4_{-1.5}^{+1.6}$ & -- \\ [0.15cm]

J1347 & 0744220701 & $2.33_{-0.07}^{+0.07}$ & $15.5_{-2.1}^{+2.7}$ & $5.3_{-2.1}^{+4.0}$ & $0.77_{-0.43}^{+0.19}$ & $34_{-13}^{+19}$ & $1.7_{-0.6}^{+0.5}$ & $17.0_{-1.5}^{+0.8}$ & $5.0_{-2.0}^{+1.4}$ & $0.44_{-0.05}^{+0.05}$ & $\frac{296.6}{302}$   \\ [0.15cm]

J1357 & 0305920601 & $2.31_{-0.07}^{+0.05}$ & $14.8_{-1.5}^{+1.0}$ & $\ge 4.7$ & $0.35_{-0.09}^{+0.15}$ & $34_{-11}^{+15}$ & $1.3_{-0.1}^{+0.1}$ & $\le 17.6$ & $3.6_{-1.6}^{+2.4}$ & $0.22_{-0.04}^{+0.04}$ & $\frac{169.8}{178}$   \\ [0.15cm]

J1434 & 0305920401 & $2.10_{-0.15}^{+0.20}$ & $4.9_{-0.4}^{+0.2}$ & $\ge 4.5$ & $0.63_{-0.45}^{+0.27}$ & $48_{-26}^{+16}$ & $\le 0.7$ & $\le 17.0$ & $3.4_{-1.8}^{+0.3}$ & $0.25_{-0.09}^{+0.09}$ & $\frac{215.7}{224}$  \\ [0.15cm]

      & 0674810501 & $2.10^{\ast}$ & $5.1_{-0.5}^{+0.3}$ & $-$ & $0.63^{\ast}$ & $48^{\ast}$ & $-$  & $-$ & $4.0_{-2.3}^{+2.2}$ & $0.28_{-0.11}^{+0.12}$ & -- \\ [0.15cm]

J1541 & 0744220401 & $1.94_{-0.12}^{+0.13}$ & $9.1_{-1.6}^{+1.7}$ & $8.7_{-3.1}^{+1.2}$ & $0.91_{-0.21}^{+0.07}$  & $43_{-11}^{+11}$ & $2.0_{-1.6}^{+0.8}$ & $18.0_{-0.7}^{+0.7}$ & $2.2_{-1.4}^{+1.8}$ & $0.36_{-0.08}^{+0.09}$ & $\frac{160.1}{177}$  \\ [0.15cm]

J1626 & 0674811001 & $1.75_{-0.13}^{+0.31}$ & $2.9_{-1.2}^{+0.6}$ & $\ge 3.7$ & $0.68_{-0.21}^{+0.28}$ & $27_{-11}^{+14}$ & $2.0_{-1.6}^{+0.8}$ & $\le 19.1$  & $1.0_{-0.6}^{+1.8}$ & $0.4_{-0.1}^{+0.1}$ & $\frac{92.5}{73}$  \\ [0.15cm]

J1631 & 0674810601 & $1.96_{-0.15}^{+0.13}$ & $1.5_{-0.2}^{+0.6}$ & $\ge 3.9$ & $0.74_{-0.19}^{+0.16}$ & $26_{-8}^{+9}$ & $2.3_{-0.8}^{+0.5}$ & $\le 18.1$ & $1.0_{-0.8}^{+0.3}$ & $0.9_{-0.4}^{+0.4}$ & $\frac{38.9}{62}$  \\ [0.15cm]

\hline 
\end{tabular}}
\end{center}
\label{tab3}
\end{table*}

\begin{table*}
\centering
\caption{The best-fit spectral parameters as derived from the fitting of the EPIC-pn+MOS data with the model, {\tt{Tbabs$\times$gabs$\times$(zgauss$+$relxillDCp$+$nthComp)}}, and {\tt{Tbabs$\times$zxipcf$\times$(relxillDCp$+$nthComp)}}, respectively, for J1559 and POX52.}
\begin{center}
\scalebox{0.87}{%
\begin{tabular}{cccccccc}
\hline
 Model &  Source & J1559 & J1559 &  POX52 & Description  \\[0.1cm]
 Component  &  Obs.~ID     & 0744290101 & 0744290201 & 0302420101 &   \\[0.15cm]
\hline 
{\tt{zxipcf}} & $N_{\rm {H, wa}}$ [10$^{21}$~cm$^{-2}$] & &   & $7.9^{+0.3}_{-0.7}$ & Column density of the absorber \\ [0.15cm]

               & $\xi_{\rm {wa}}$ [erg~cm~s$^{-1}$] &  &   & $10^{+3}_{-7}$ & Ionization parameter  \\ [0.15cm]            
                  
               & $C_{\rm f}$[per~cent]  & &    & $97.9^{+0.6}_{-0.5}$ & Covering fraction \\ [0.15cm]   
{\tt{gabs}} & $E_{\rm abs}$ [keV] & $0.67^{+0.01}_{-0.02}$ & $0.67^{\ast}$  &  & Line energy [observed frame] \\ [0.15cm]

                & $\sigma_{\rm abs}$ [eV] & $83.1^{+13.9}_{-12.0}$  & $83.1^{\ast}$  &  & Line width \\ [0.1cm]

                & $\tau_{\rm abs}$ [10$^{-2}$] &  $3.6^{+1.0}_{-0.5}$ &  $3.6^{\ast}$  &  & Line depth  \\ [0.15cm]
{\tt{zgauss}} & $E_{\rm rest}$ [keV] & $6.43^{+0.04}_{-0.04}$ & $6.43^{\ast}$  &  & Rest-frame line centroid energy \\ [0.15cm]


                & $K_{\rm ga}$ [10$^{-6}$] &  $1.2^{+0.4}_{-0.4}$ &  $1.2^{\ast}$  &  &  Line normalization  \\ [0.15cm]
{\tt{nthComp}} & $\Gamma$   & $2.07_{-0.03}^{+0.02}$  & $2.07^{\ast}$   & $2.49_{-0.36}^{+0.19}$  & Photon index \\ [0.15cm] 
                  & $K_{\rm nth}$[$10^{-4}$]  & $3.5_{-0.6}^{+0.5}$ & $3.8_{-0.5}^{+0.4}$   & $2.6_{-0.3}^{+1.1}$ & Primary continuum normalization \\ [0.15cm]    
{\tt{relxillDCp}} & $q_{\rm in}$ & $6.1_{-1.3}^{+0.9}$ & $6.1^{\ast}$   & $6.8_{-3.5}^{+2.8}$ & Inner emissivity index \\ [0.15cm] 
                 
                  & $a$ & $0.998_{-0.023}^{+0.0}$ & $0.998^{\ast}$ &  $0.56_{-0.46}^{+0.36}$ &  BH spin \\ [0.15cm]
        
                  & $\theta^{\circ}$ & $59_{-5}^{+4}$ & $59^{\ast}$ & $28_{-14}^{+13}$  & Disc inclination angle \\ [0.15cm]
                  
                  & $\log[\frac{\xi}{\rm erg~cm~s^{-1}}]$  & $2.7_{-0.1}^{+0.1}$ & $2.7^{\ast}$ & $1.5_{-0.8}^{+0.5}$ & Disc ionization parameter \\ [0.15cm]
                  
                  & $\log[n_{\rm e}$/cm$^{\rm -3}$] & $17.1_{-0.2}^{+0.3}$ & $17.1^{\ast}$  & $17.6_{-0.5}^{+1.1}$ & Disc density  \\ [0.15cm]                                          
               & $K_{\rm ref}$[$10^{-5}$] & $2.6_{-0.8}^{+0.3}$ & $2.1_{-0.7}^{+0.3}$   & $1.1_{-0.6}^{+1.1}$ & Reflection normalization \\ [0.15cm]
                  
              & $R_{\rm s}$  & $2.0_{-0.07}^{+0.07}$ & $1.5_{-0.05}^{+0.05}$ & $1.0_{-0.1}^{+0.1}$ & Reflection strength \\[0.15cm]                                                        
                 & $\frac{C-\rm {stat}}{\rm {d.o.f}}$   & $\frac{961}{942}$ & -- &  $\frac{331}{295}$ & Fit statistic   \\ [0.15cm]      
\hline       
\end{tabular}}
\end{center}
\label{tab4}
\end{table*}

\begin{figure*}
\centering
\begin{center}
\includegraphics[scale=0.25,angle=-0]{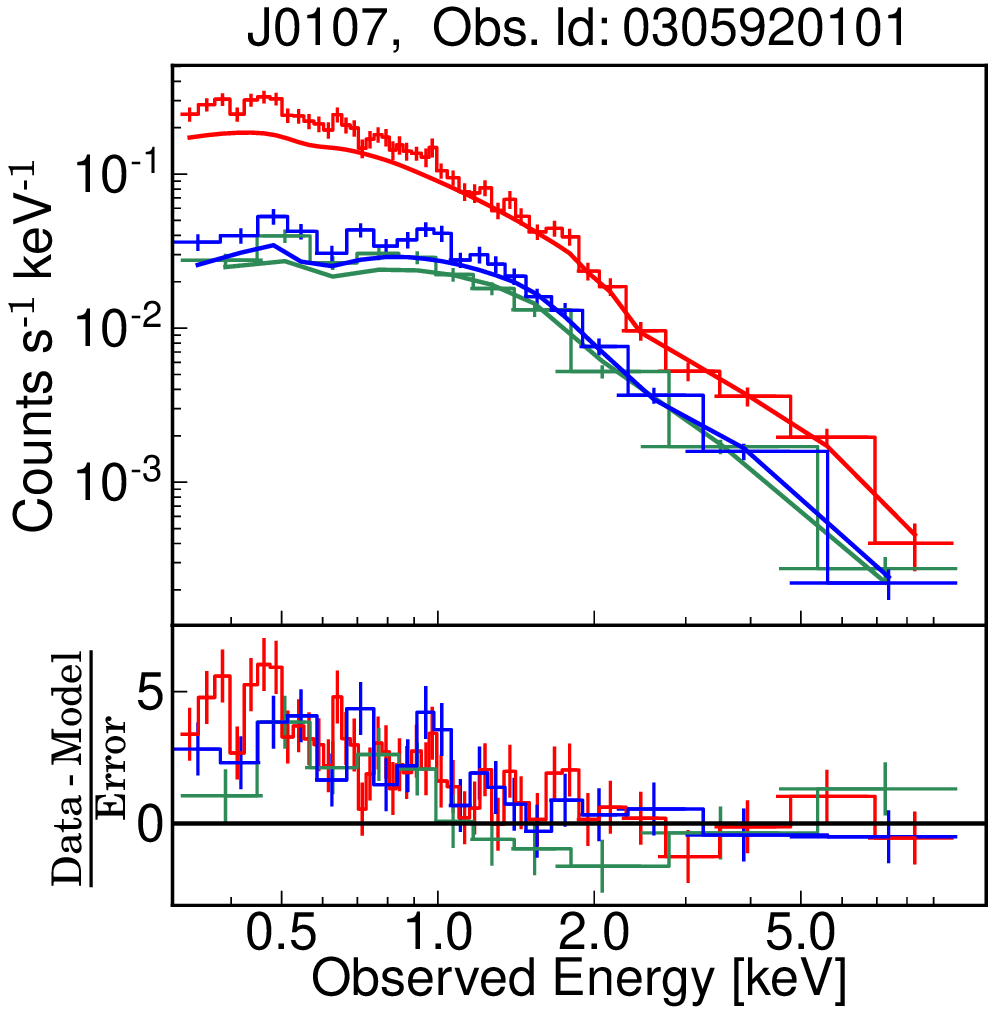}
\includegraphics[scale=0.25,angle=-0]{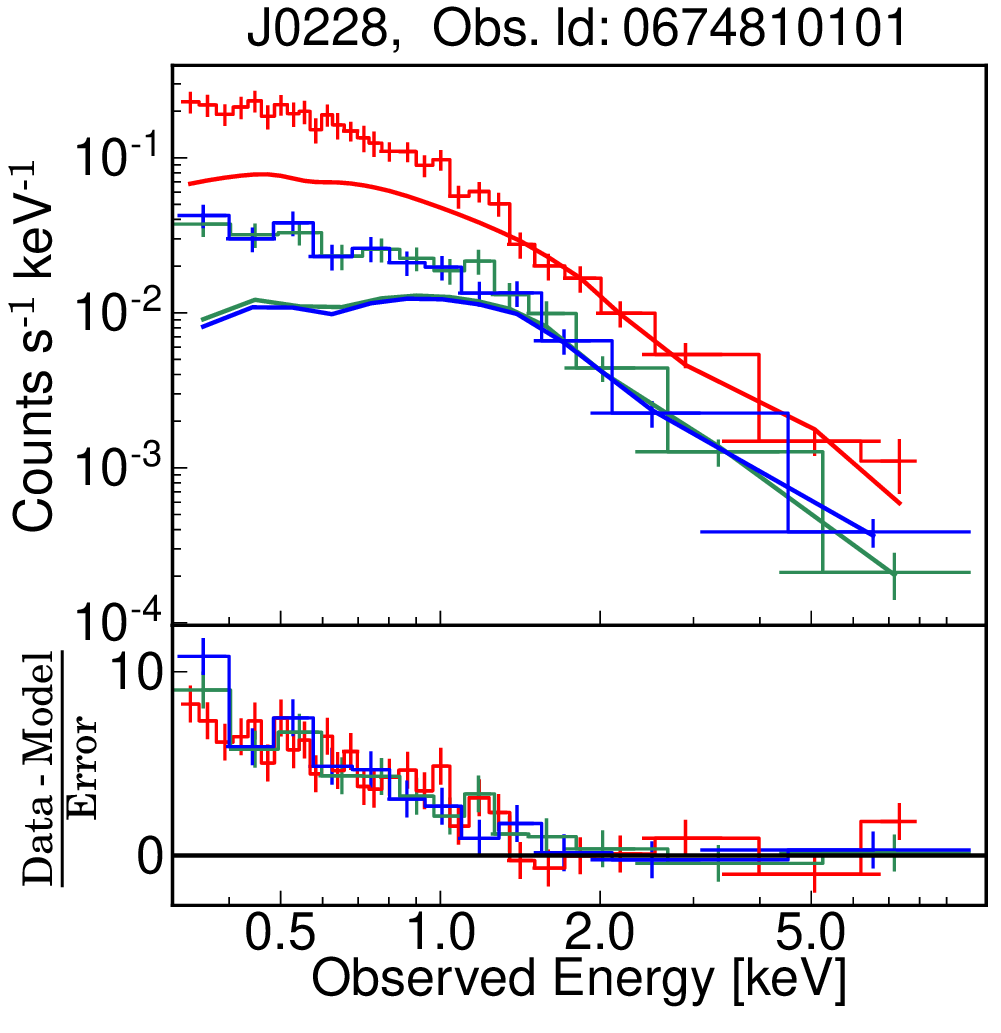}
\includegraphics[scale=0.25,angle=-0]{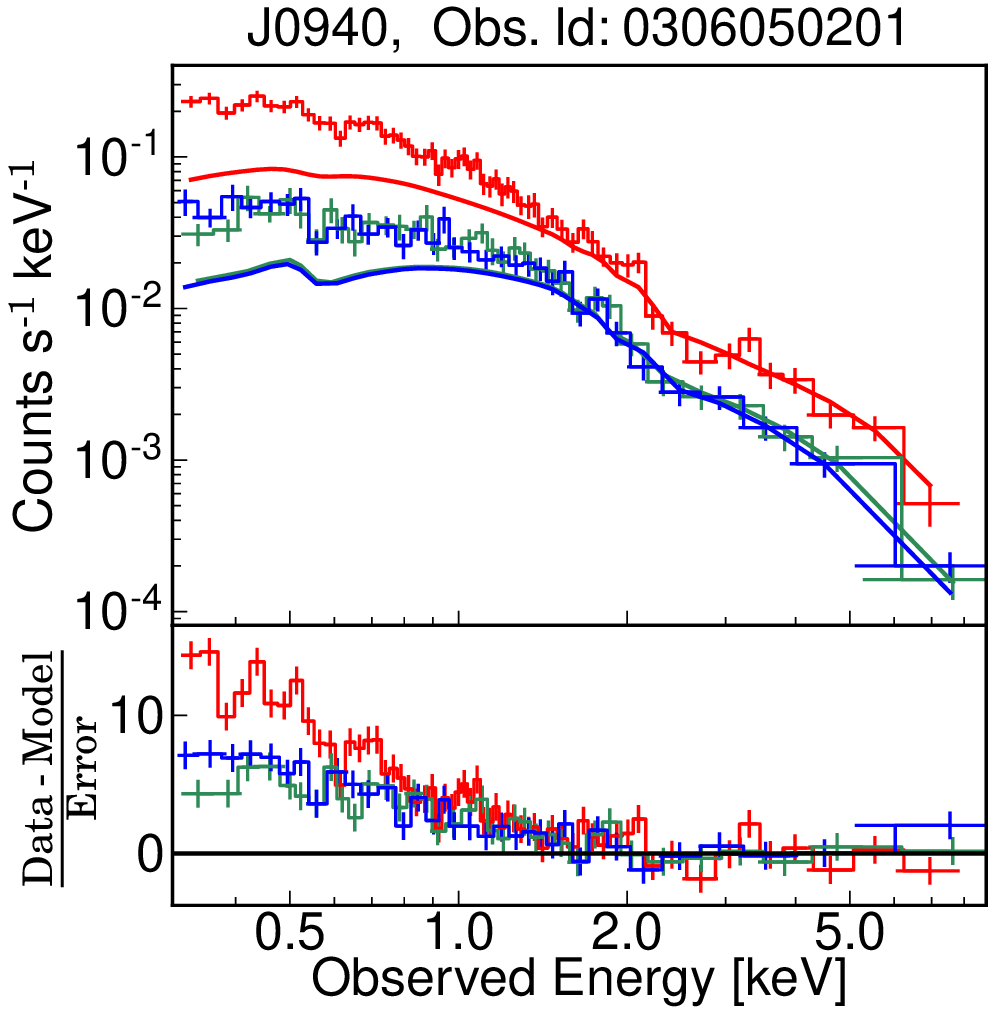}
\includegraphics[scale=0.25,angle=-0]{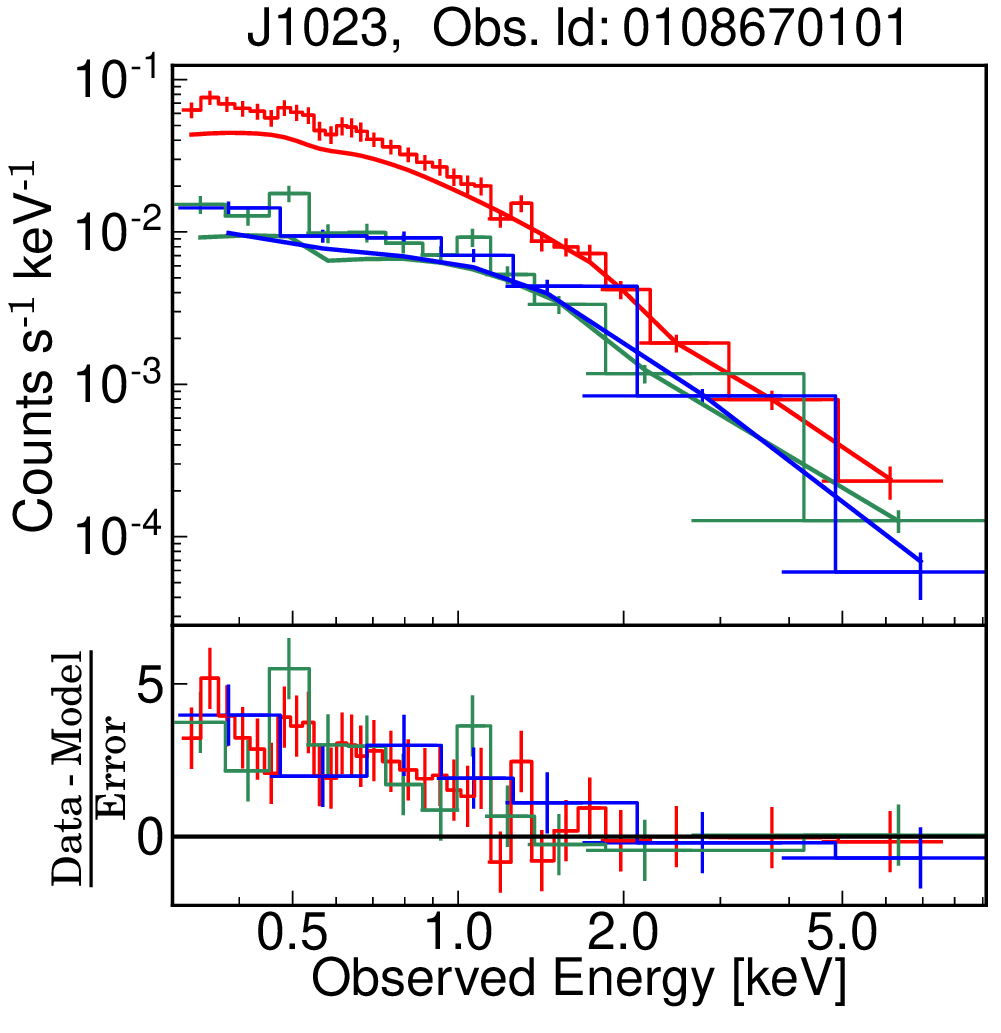}
\includegraphics[scale=0.25,angle=-0]{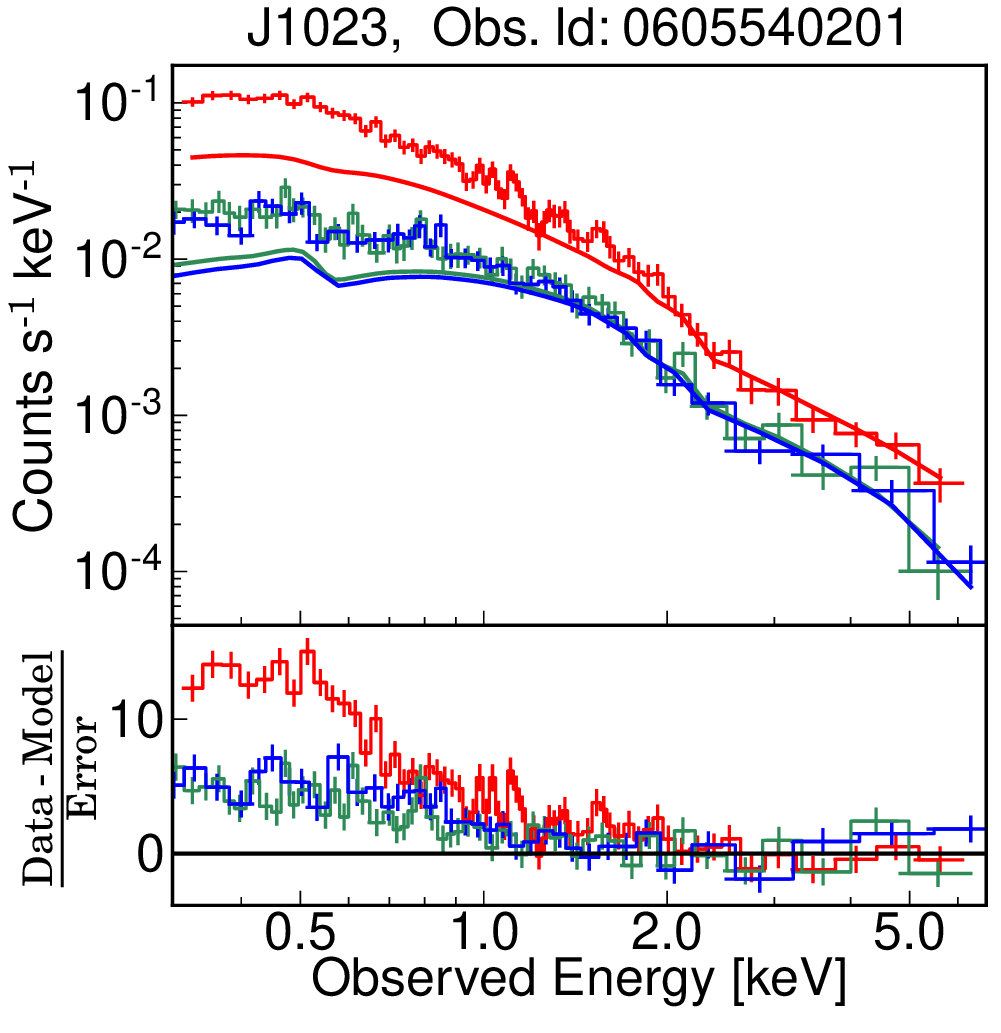}
\includegraphics[scale=0.25,angle=-0]{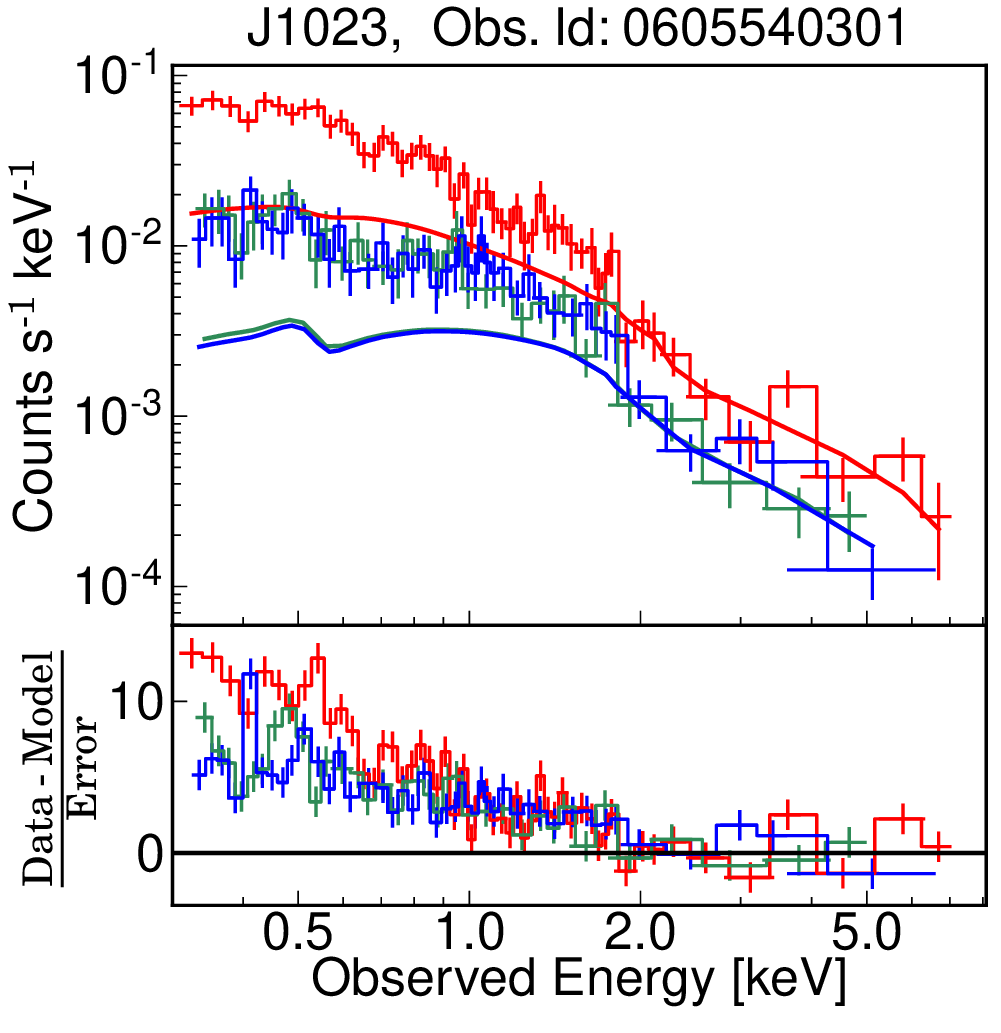}
\includegraphics[scale=0.25,angle=-0]{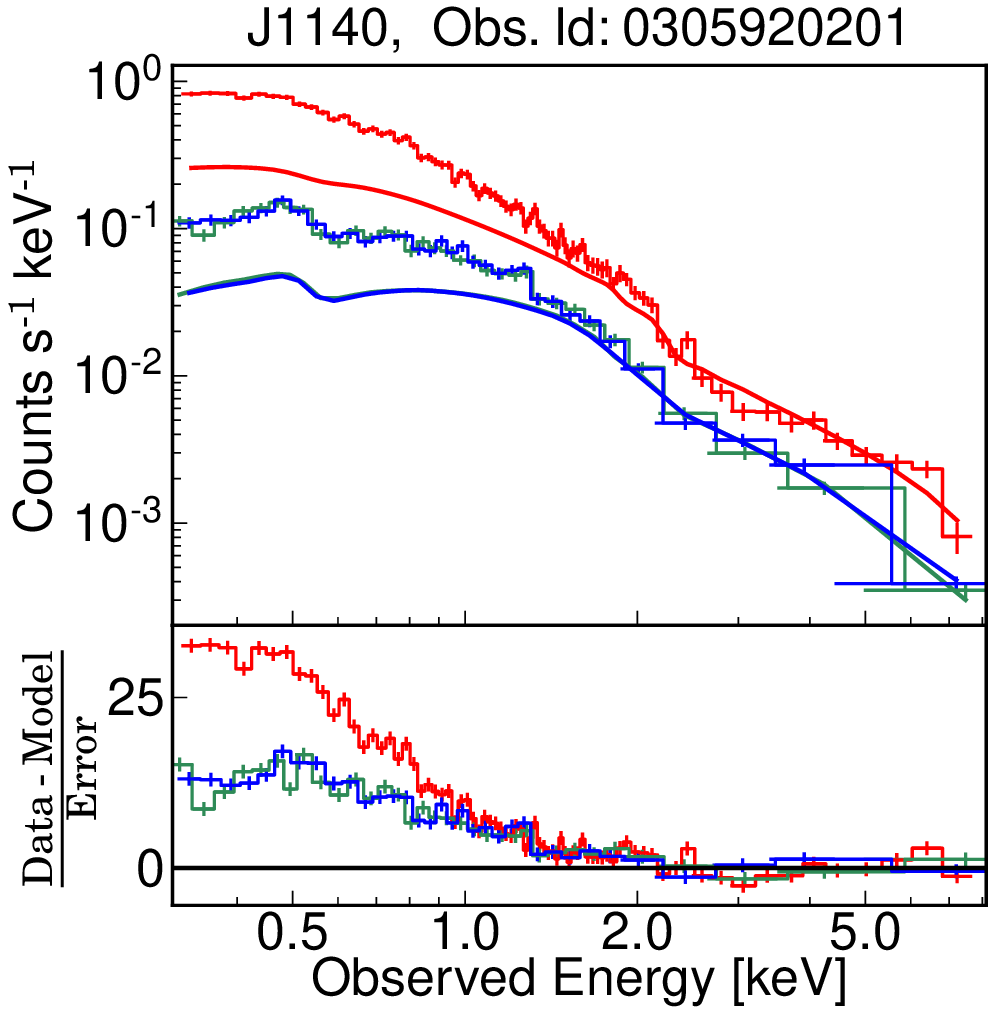}
\includegraphics[scale=0.25,angle=-0]{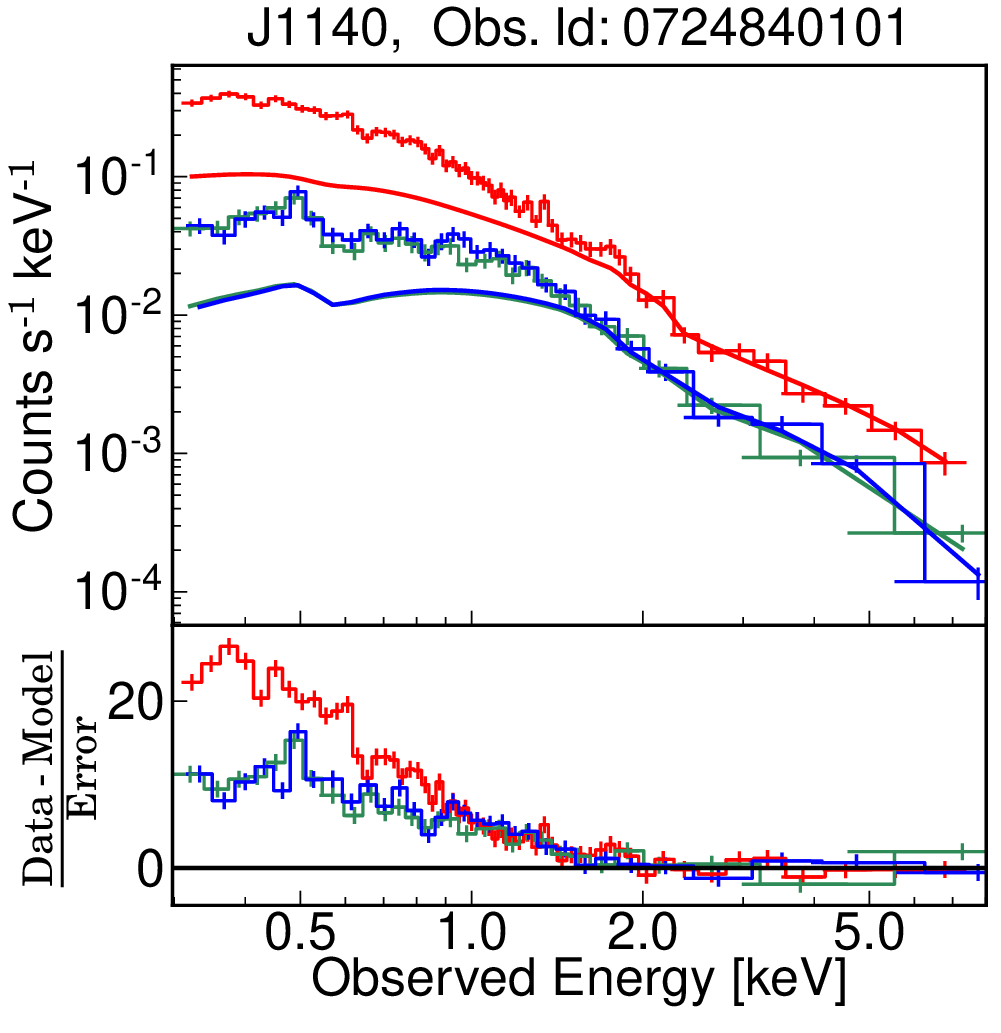}
\includegraphics[scale=0.25,angle=-0]{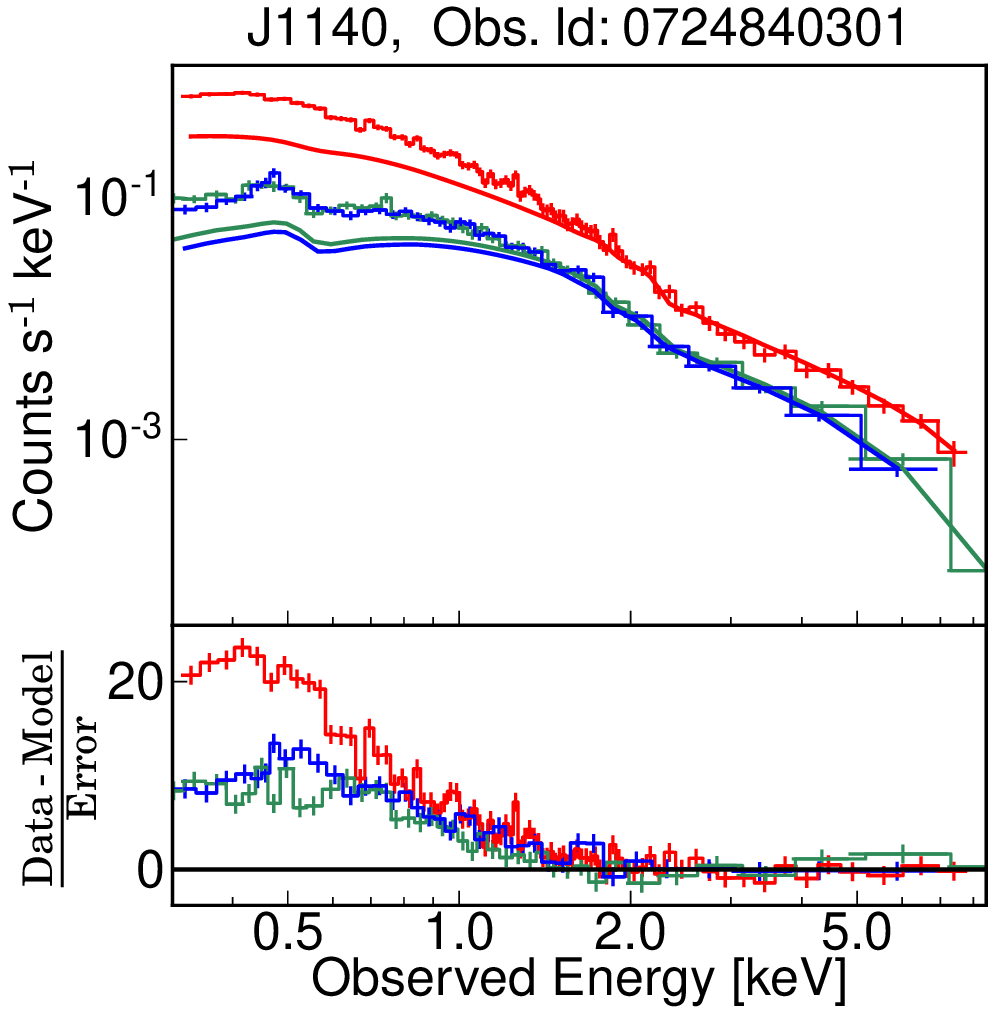}
\includegraphics[scale=0.25,angle=-0]{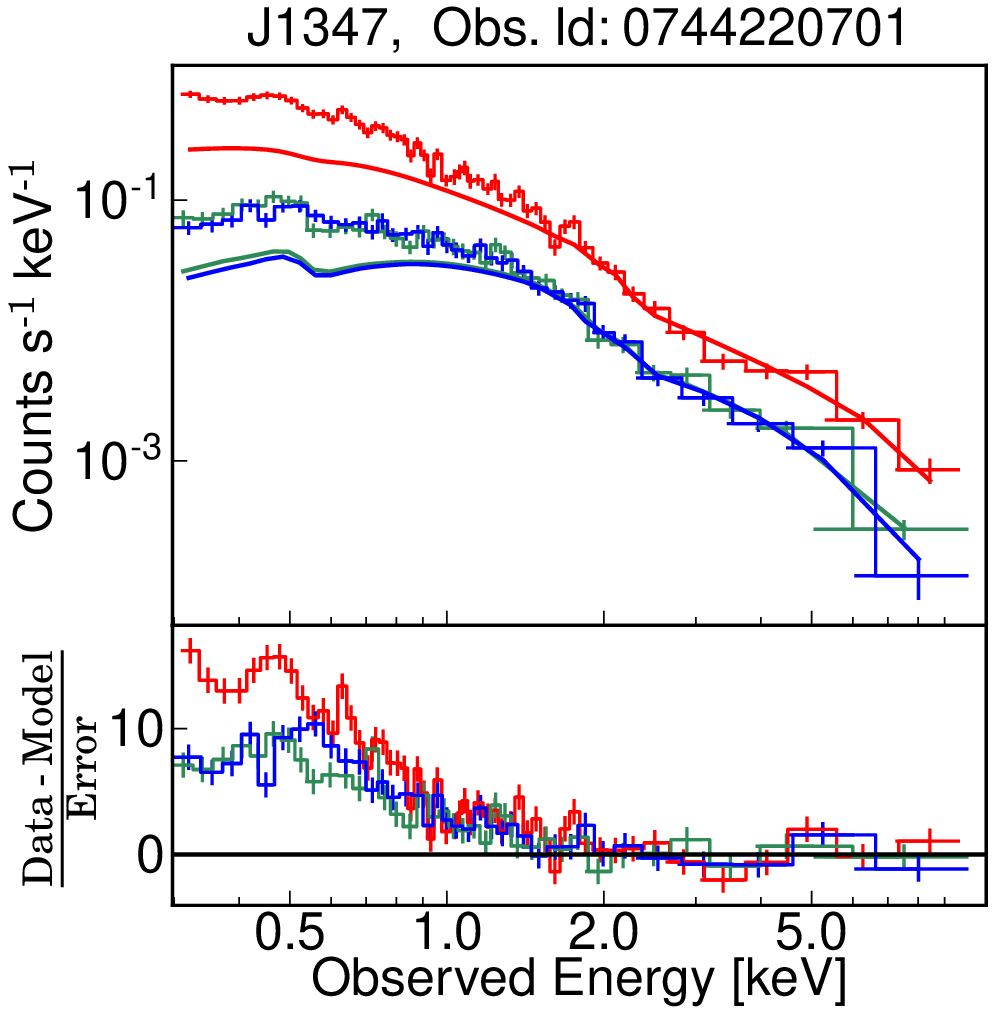}
\includegraphics[scale=0.25,angle=-0]{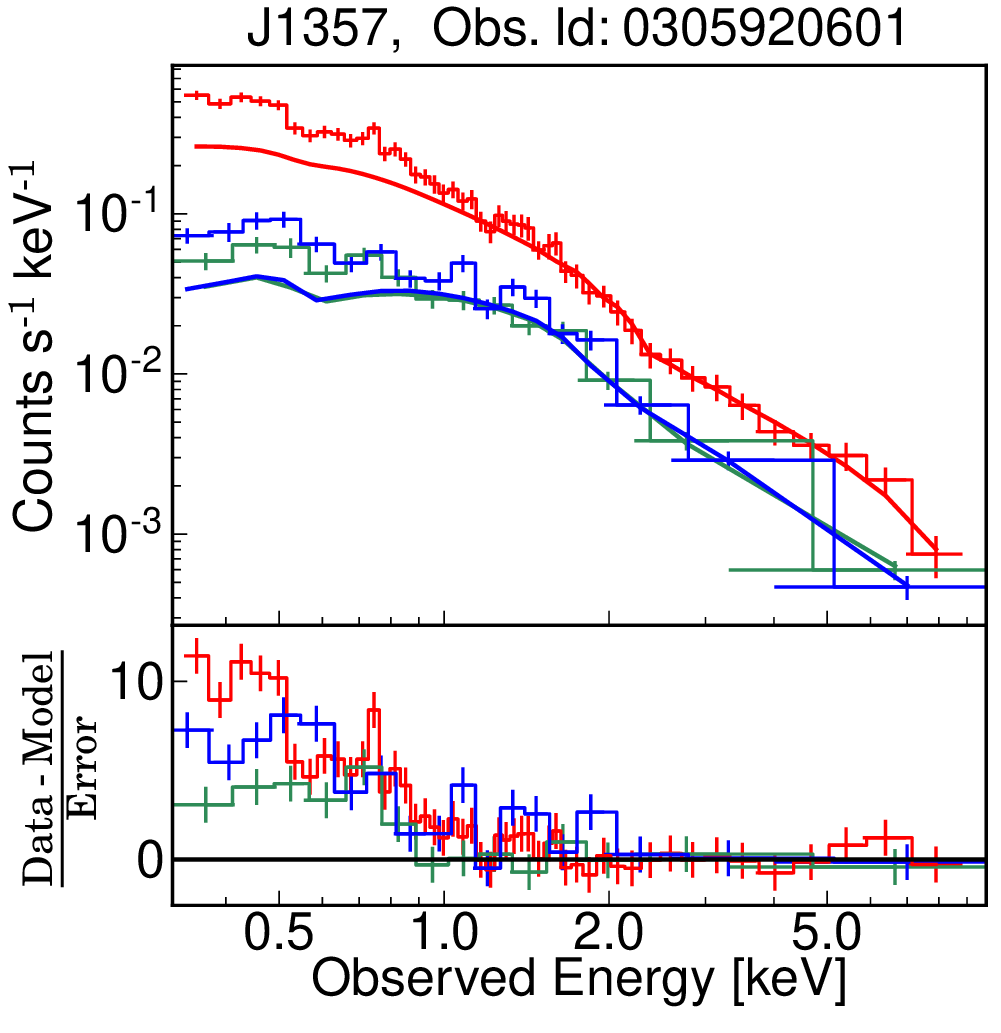}
\includegraphics[scale=0.25,angle=-0]{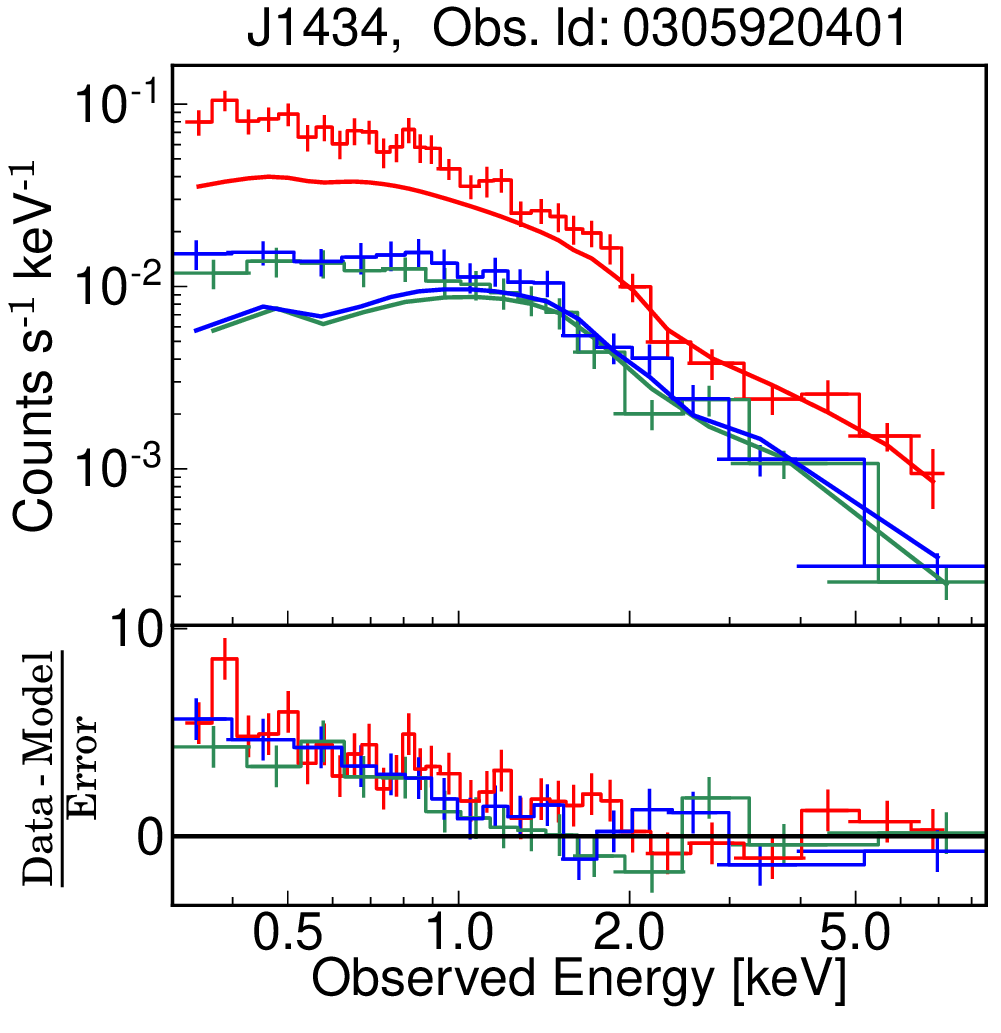}
\includegraphics[scale=0.25,angle=-0]{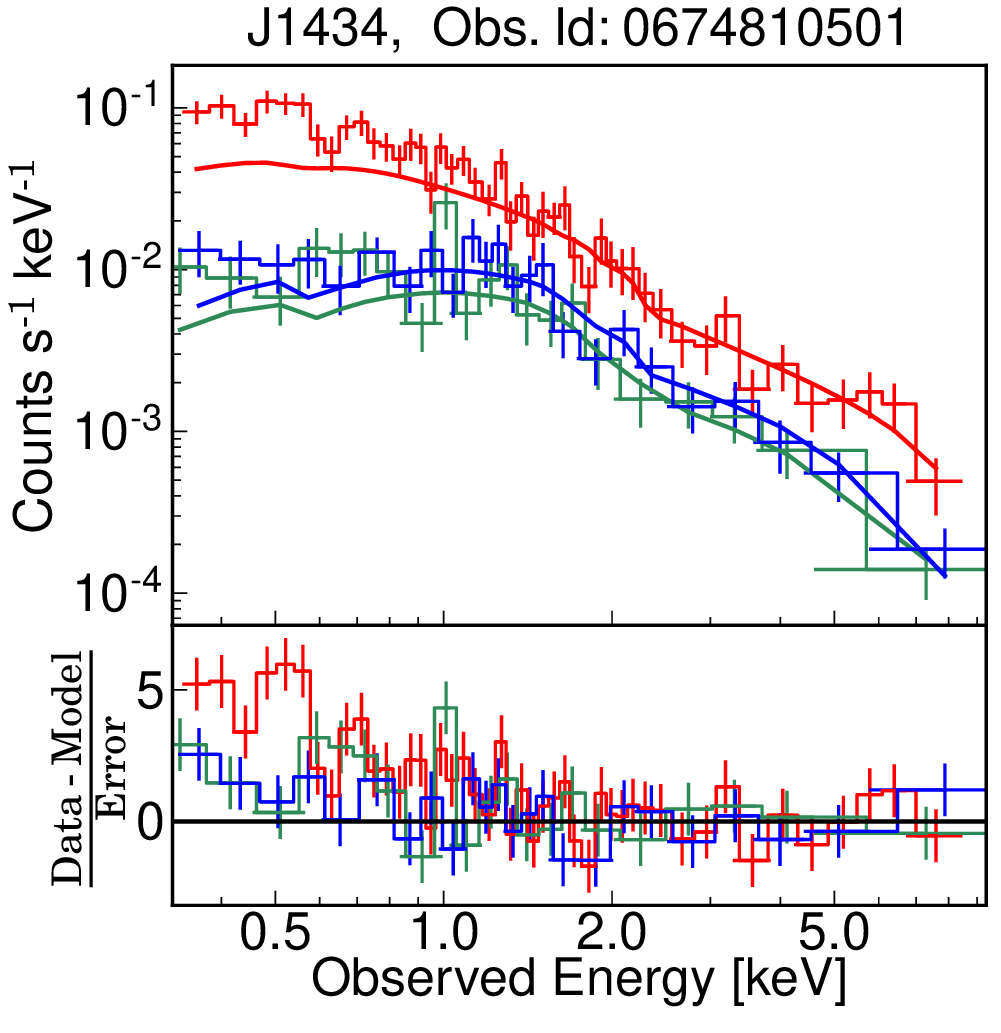}
\includegraphics[scale=0.25,angle=-0]{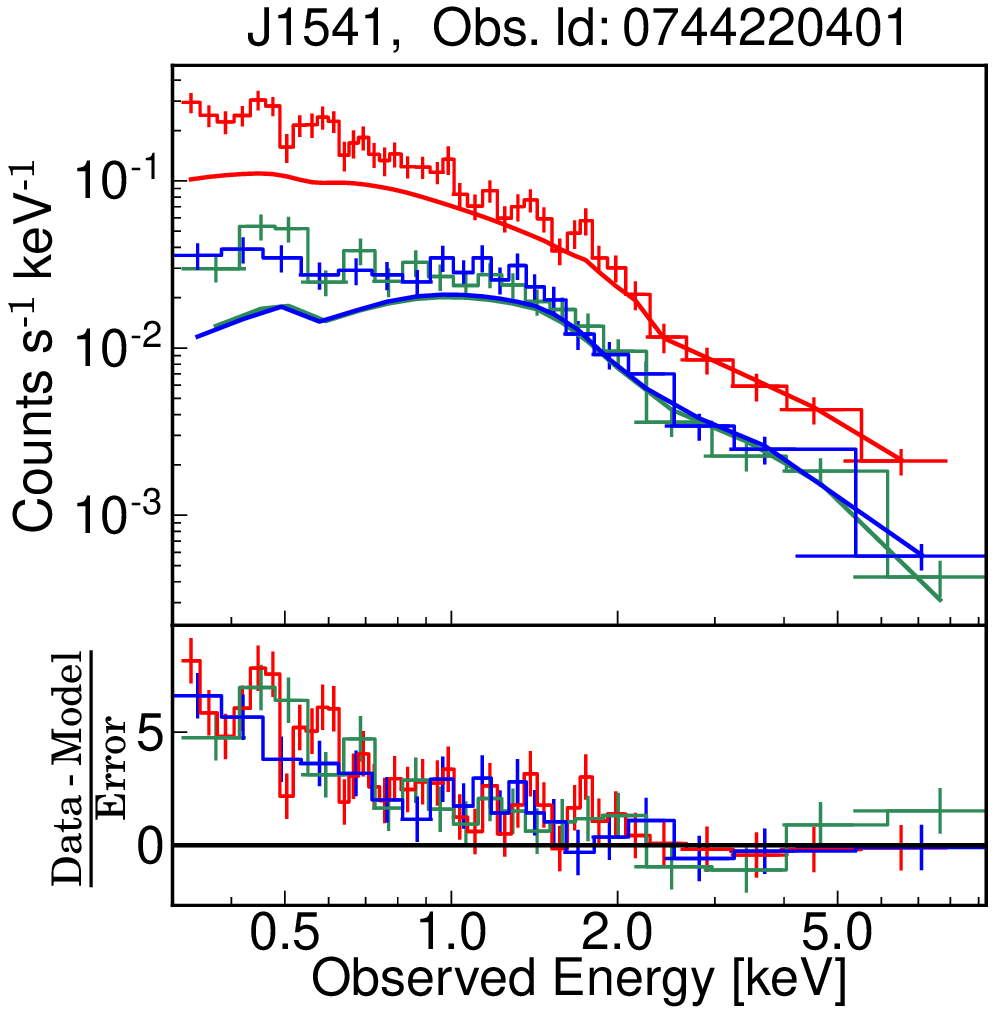}
\includegraphics[scale=0.25,angle=-0]{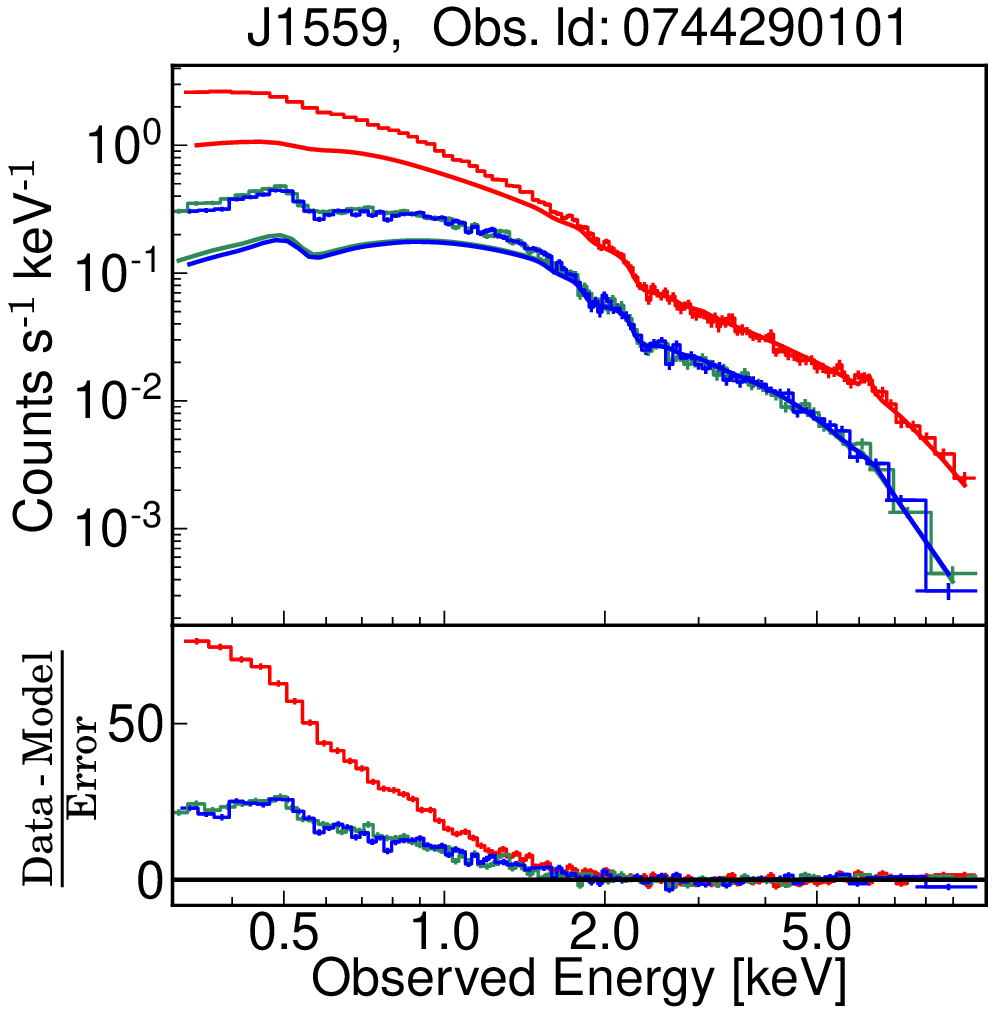}
\includegraphics[scale=0.25,angle=-0]{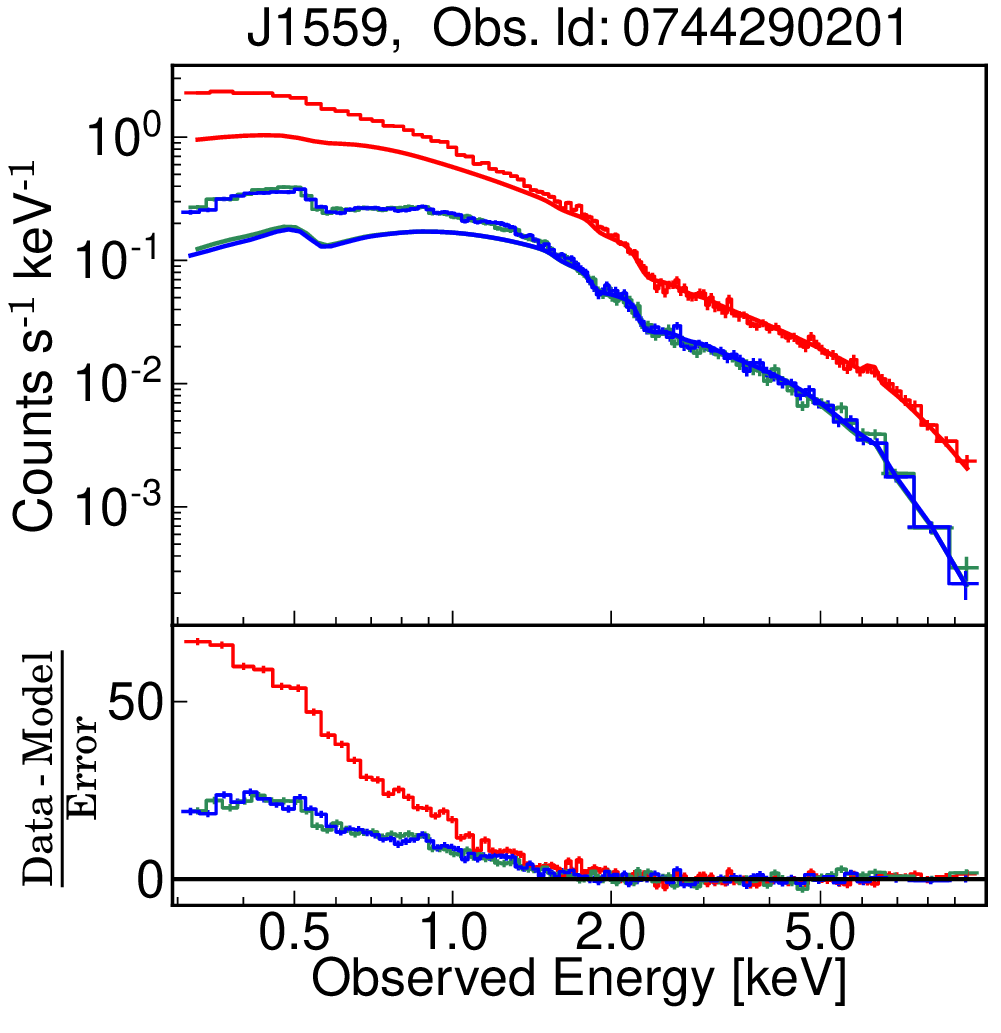}
\includegraphics[scale=0.25,angle=-0]{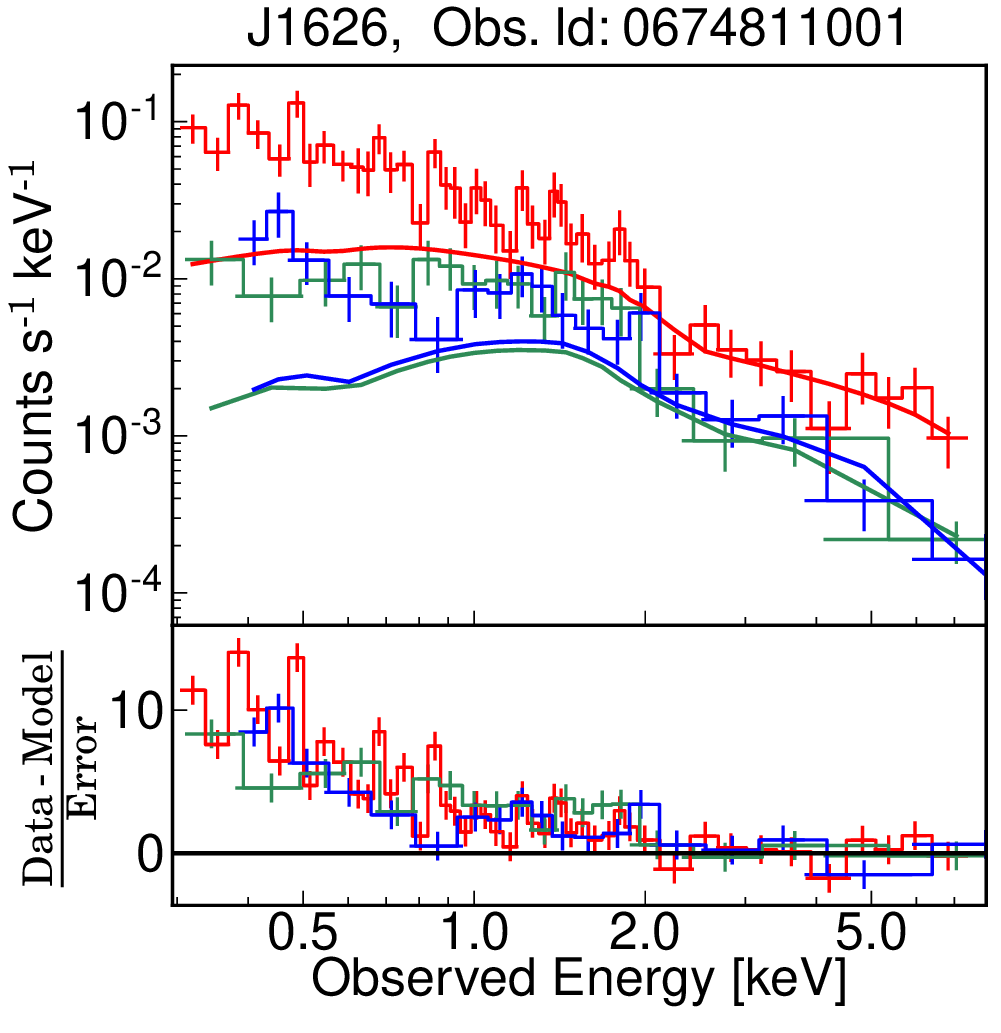}
\includegraphics[scale=0.25,angle=-0]{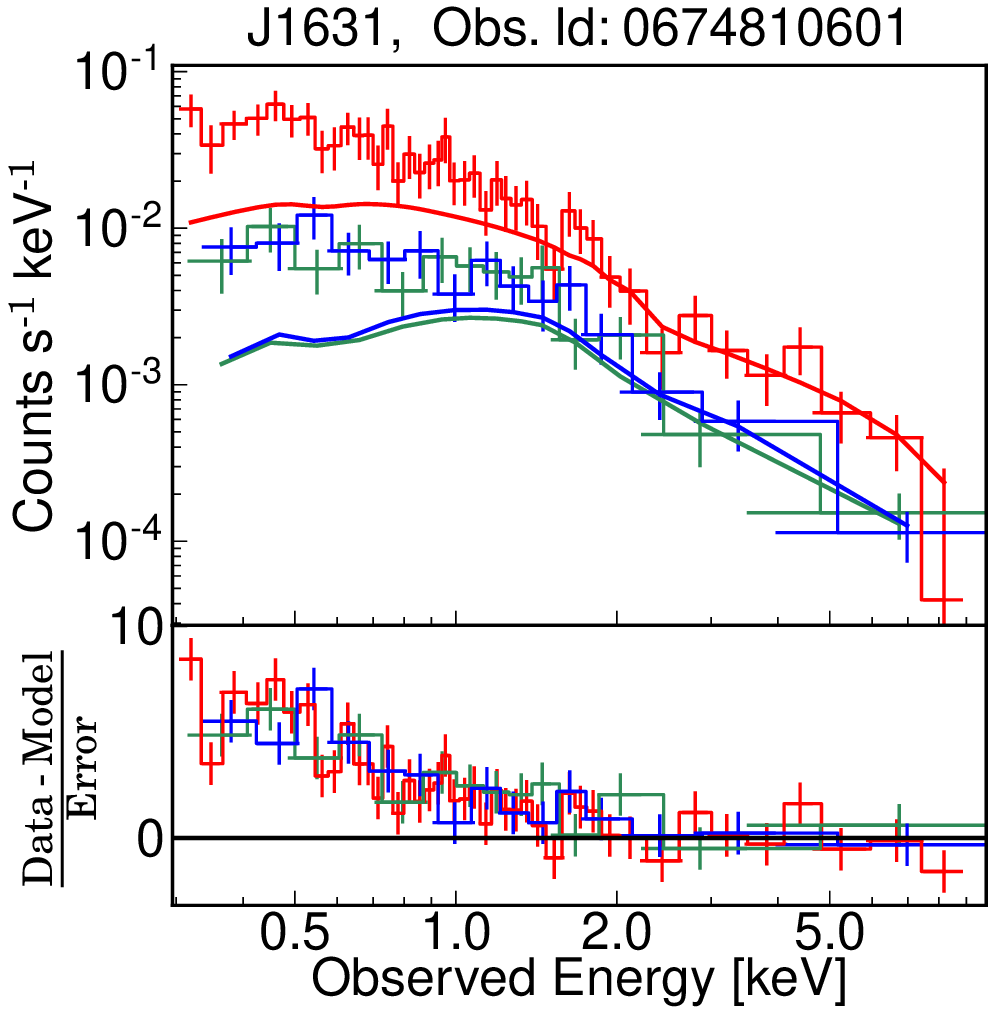}
\includegraphics[scale=0.25,angle=-0]{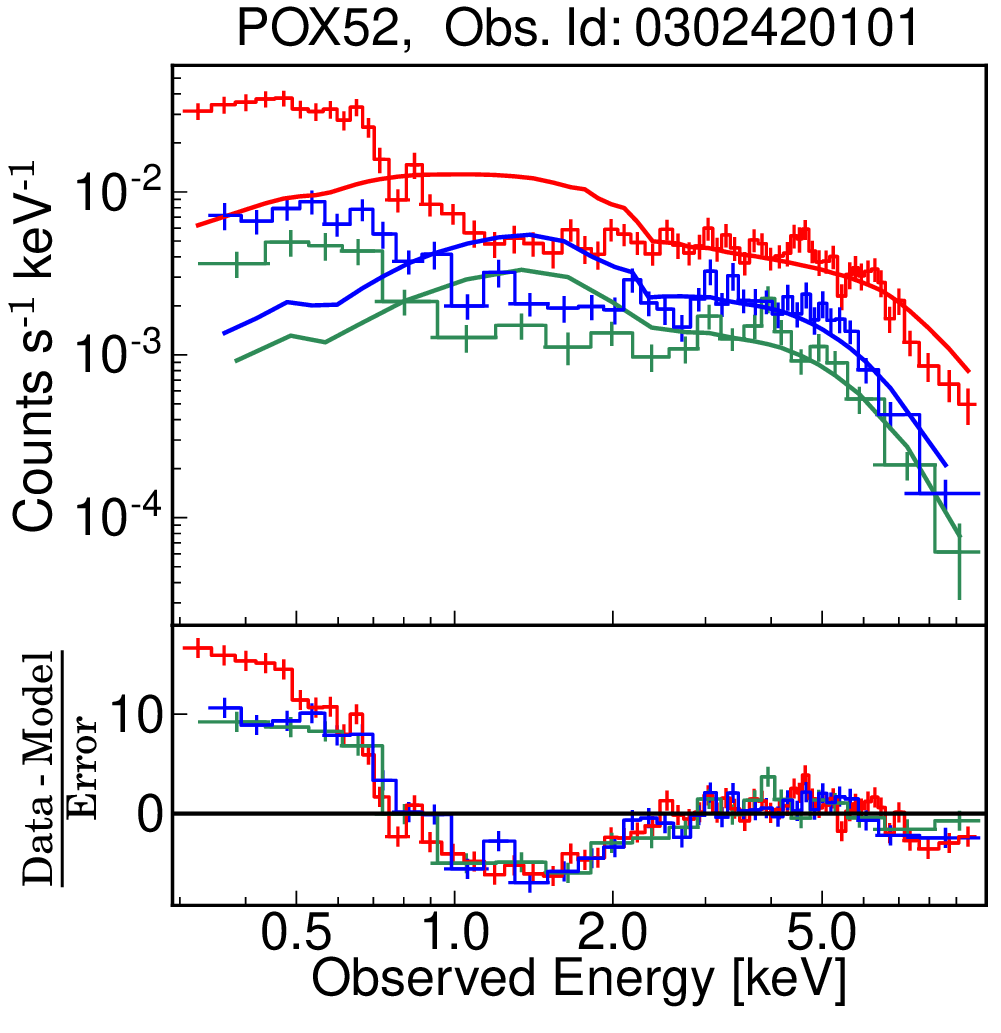}
\caption{The top panel shows the broad-band (0.3$-$10\keV{}) EPIC-pn (red), MOS1 (green), and MOS2 (blue) spectral data and hard band (2$-$10\keV{}) best-fit model ({\tt{Tbabs$\times$[zgauss$+$zpowerlw]}} for J1559 and {\tt{Tbabs$\times$zpowerlw}} for all other sources) extrapolated down to 0.3\keV{}. The bottom panels depict the residual plots, which reveal excess emission below around $\sim 1$\keV{}. For POX~52, we notice an absorption curvature in the range of $\sim 1-2$\keV{}, which is a sign of partial covering absorption. The spectra are binned up for plotting purposes only.}
\end{center}
\label{spec1}
\end{figure*}

\begin{figure*}
\centering
\begin{center}
\includegraphics[scale=0.25,angle=-0]{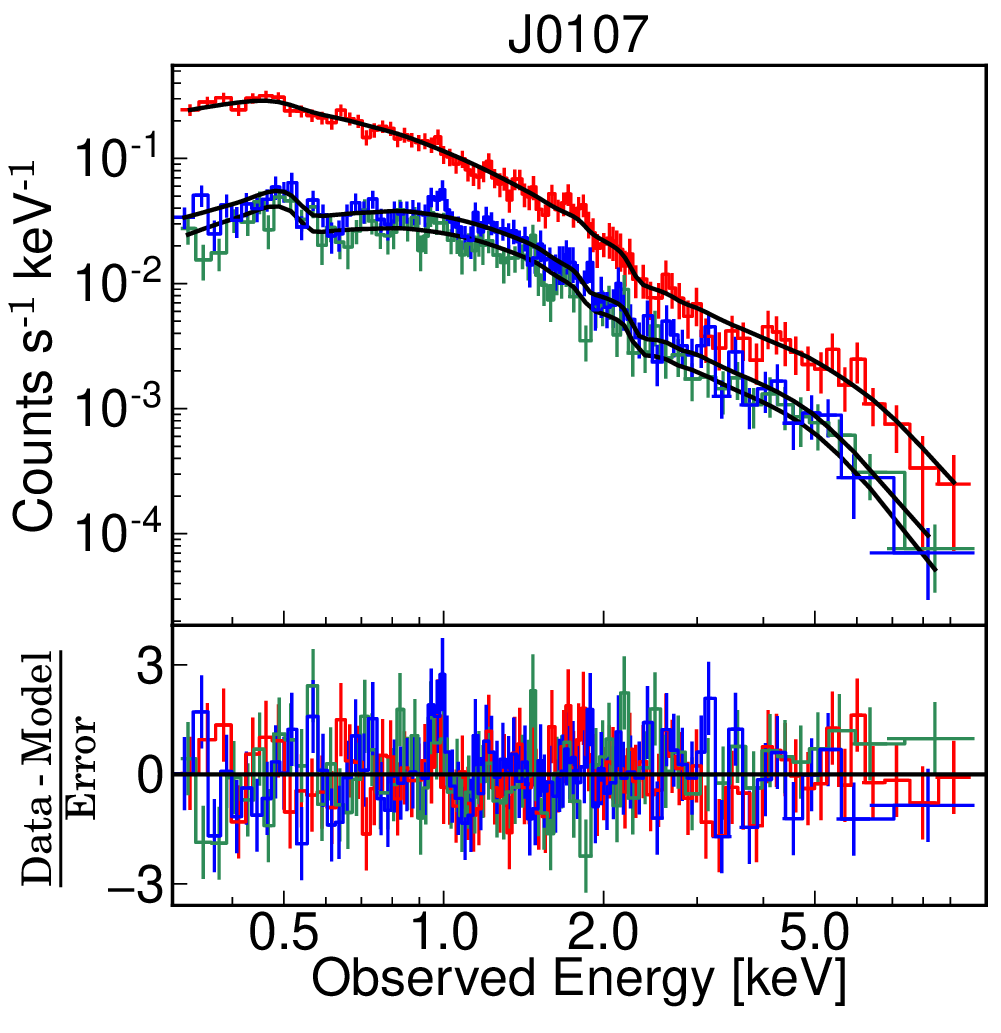}
\includegraphics[scale=0.25,angle=-0]{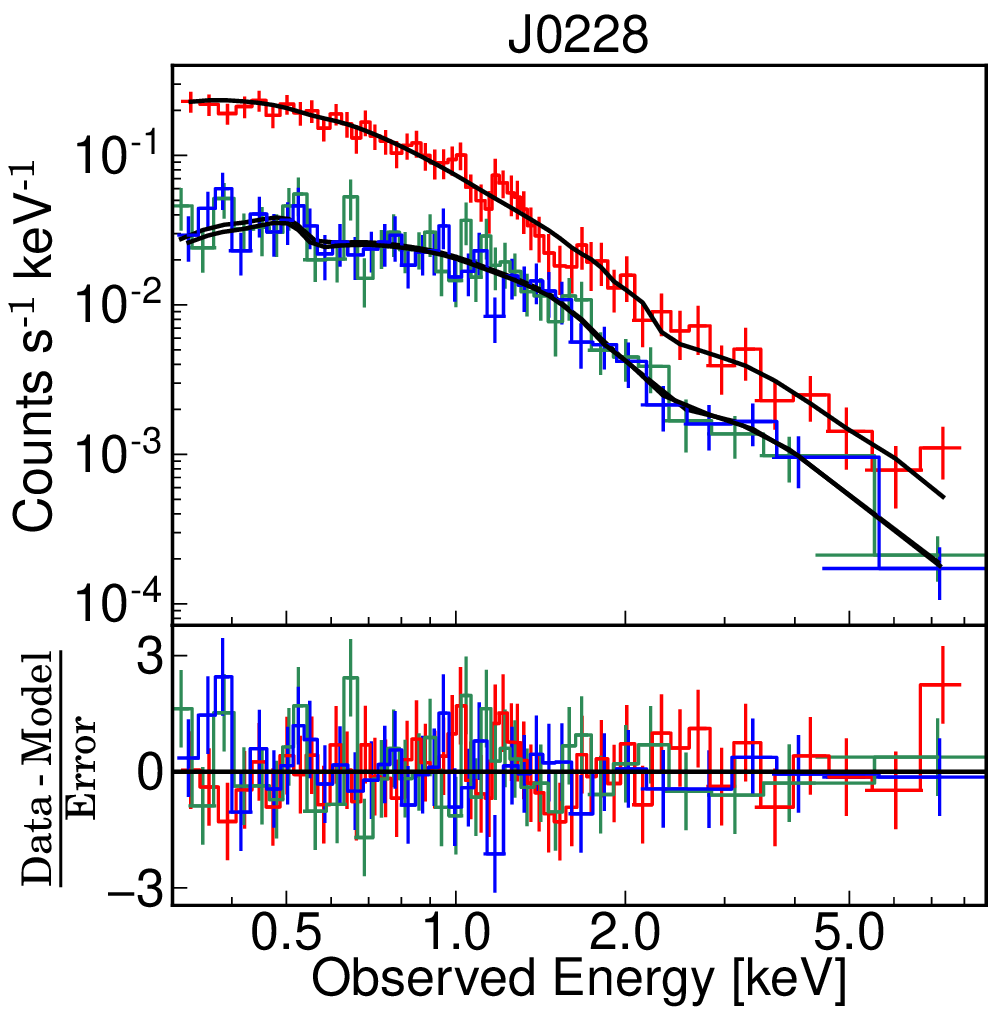}
\includegraphics[scale=0.25,angle=-0]{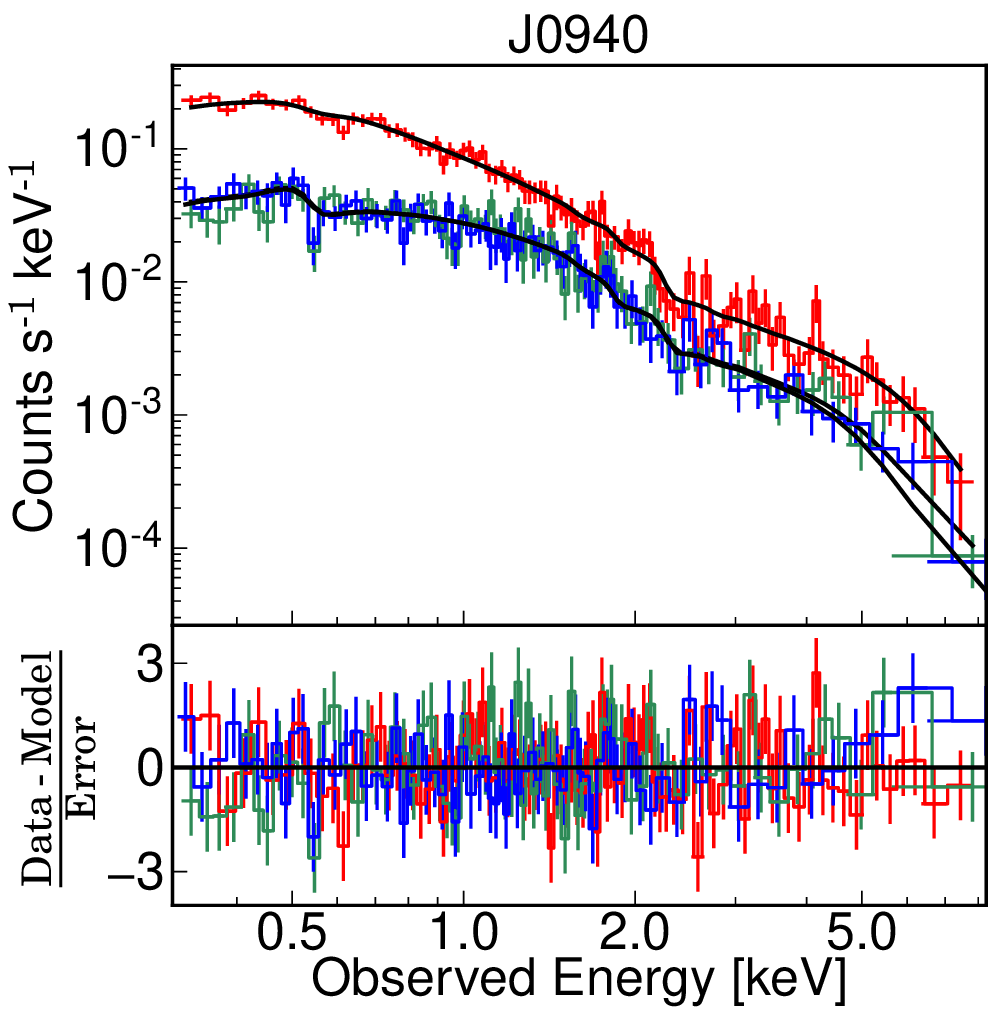}
\includegraphics[scale=0.25,angle=-0]{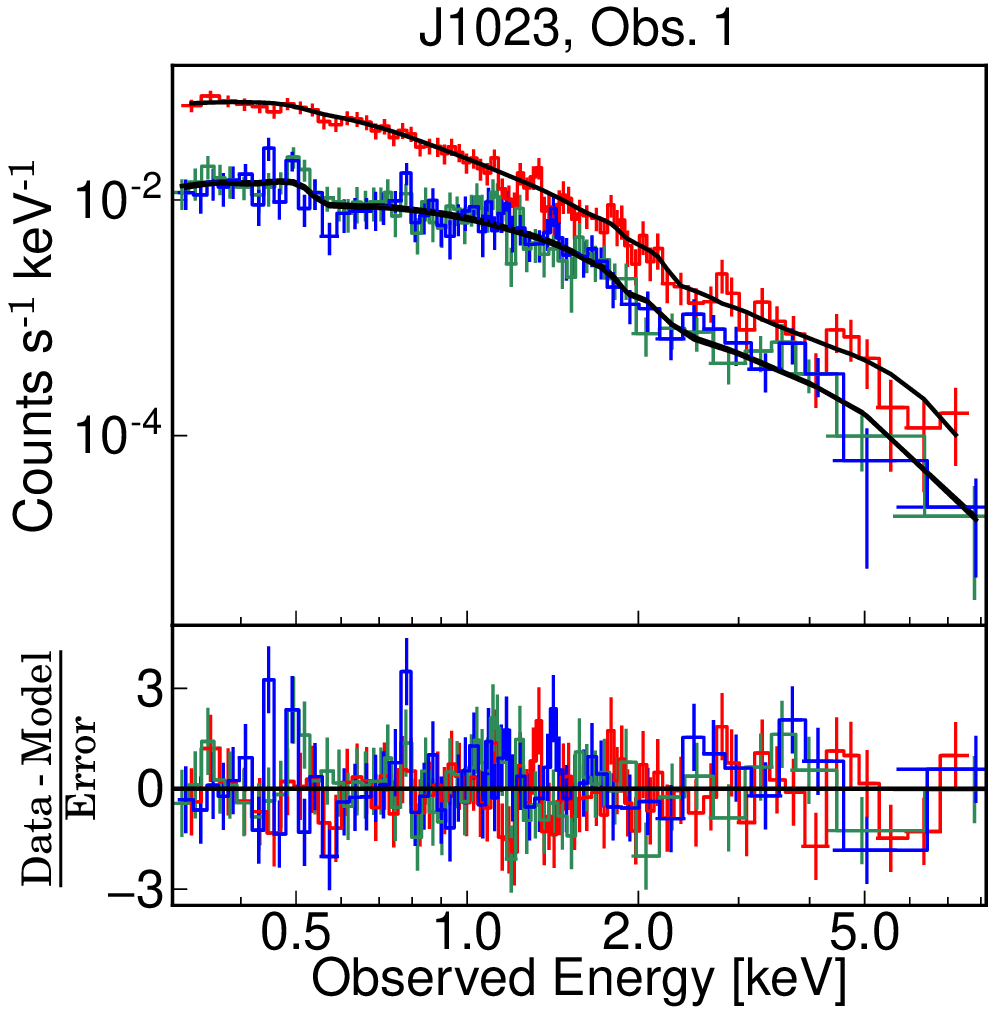}
\includegraphics[scale=0.25,angle=-0]{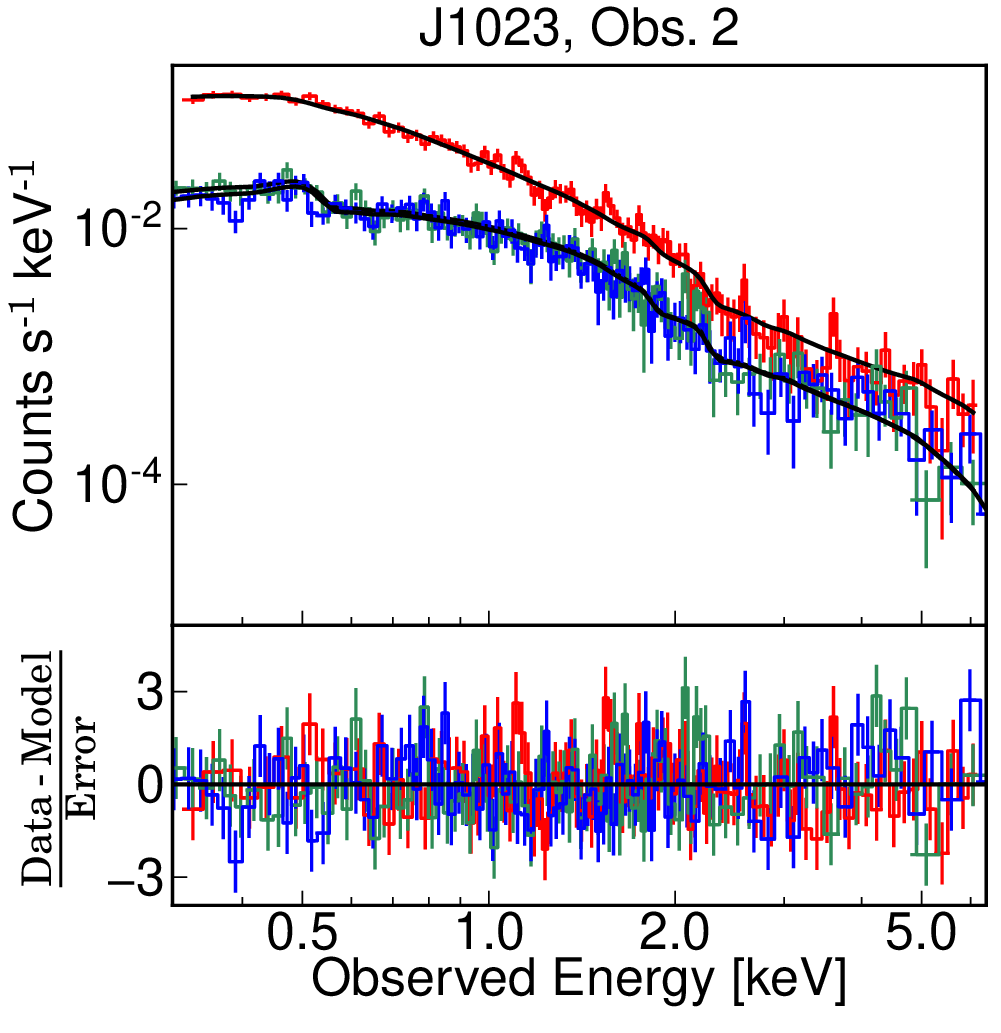}
\includegraphics[scale=0.25,angle=-0]{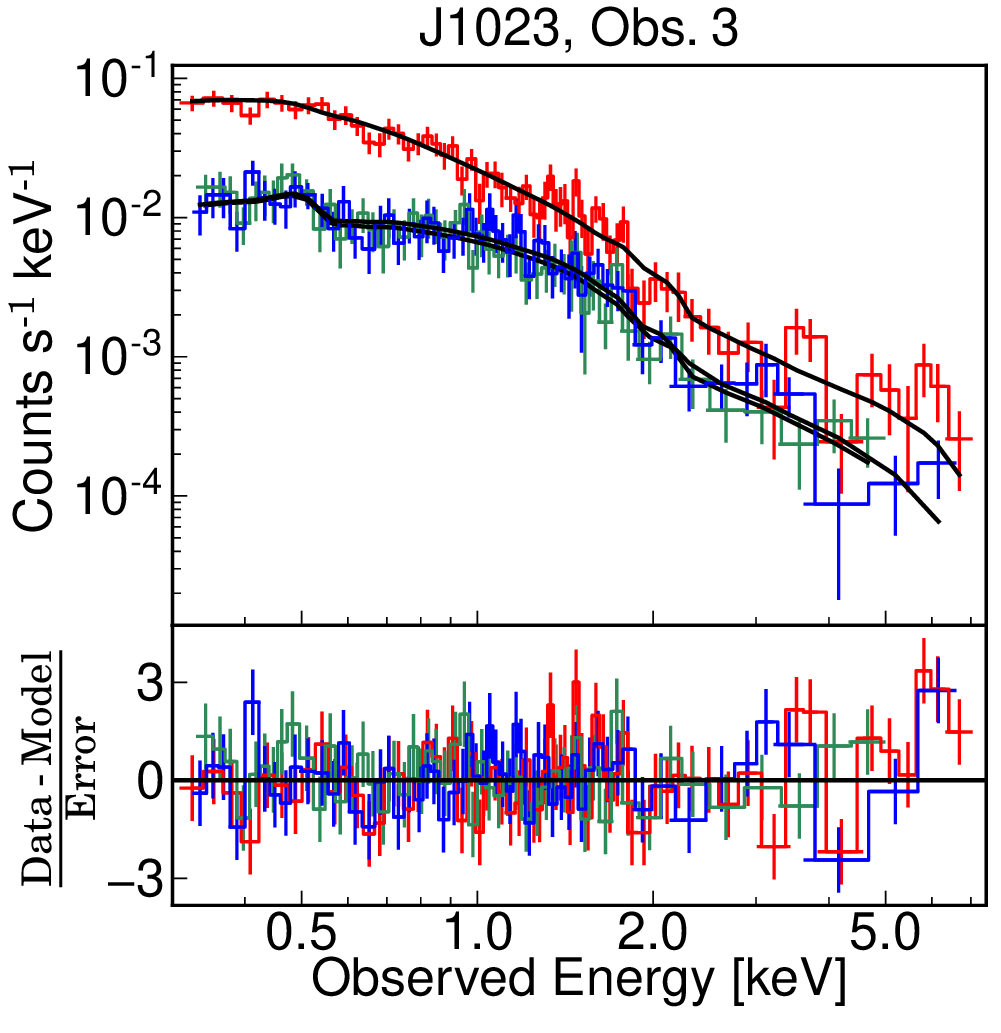}
\includegraphics[scale=0.25,angle=-0]{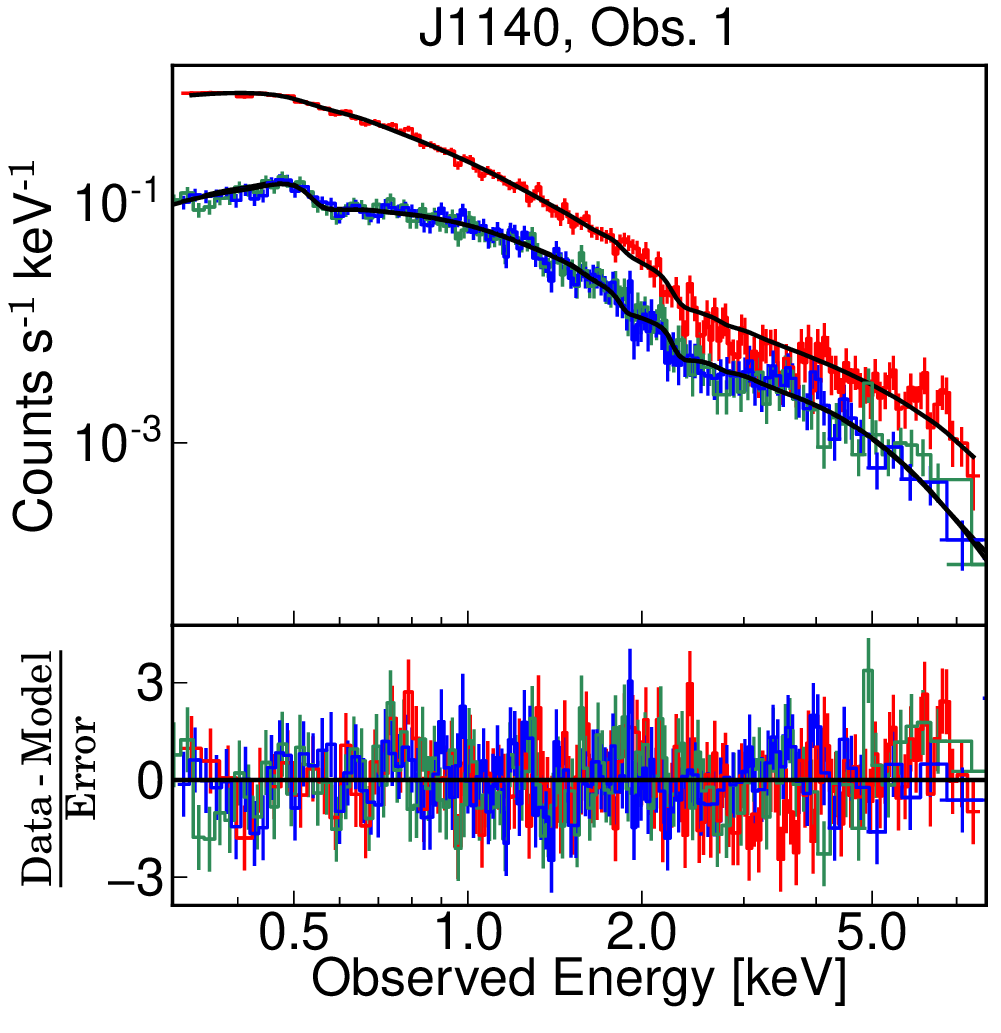}
\includegraphics[scale=0.25,angle=-0]{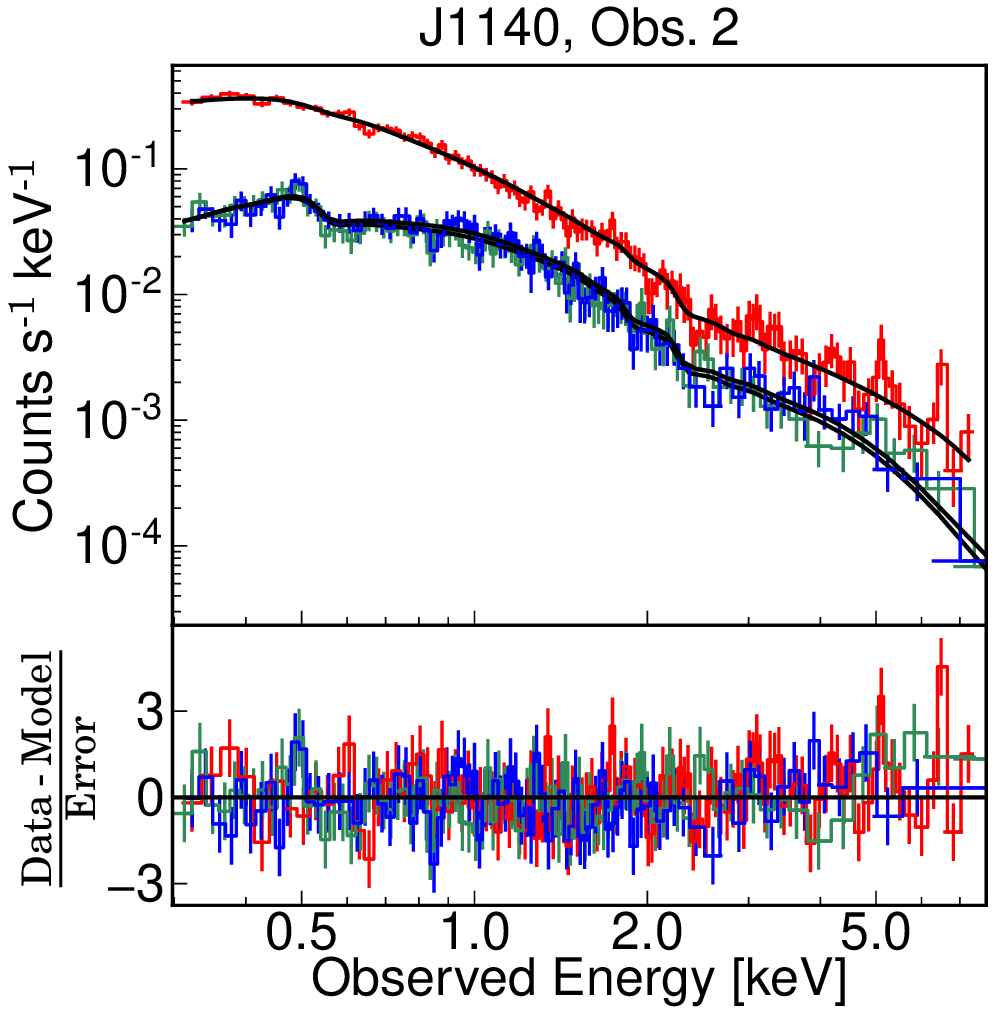}
\includegraphics[scale=0.25,angle=-0]{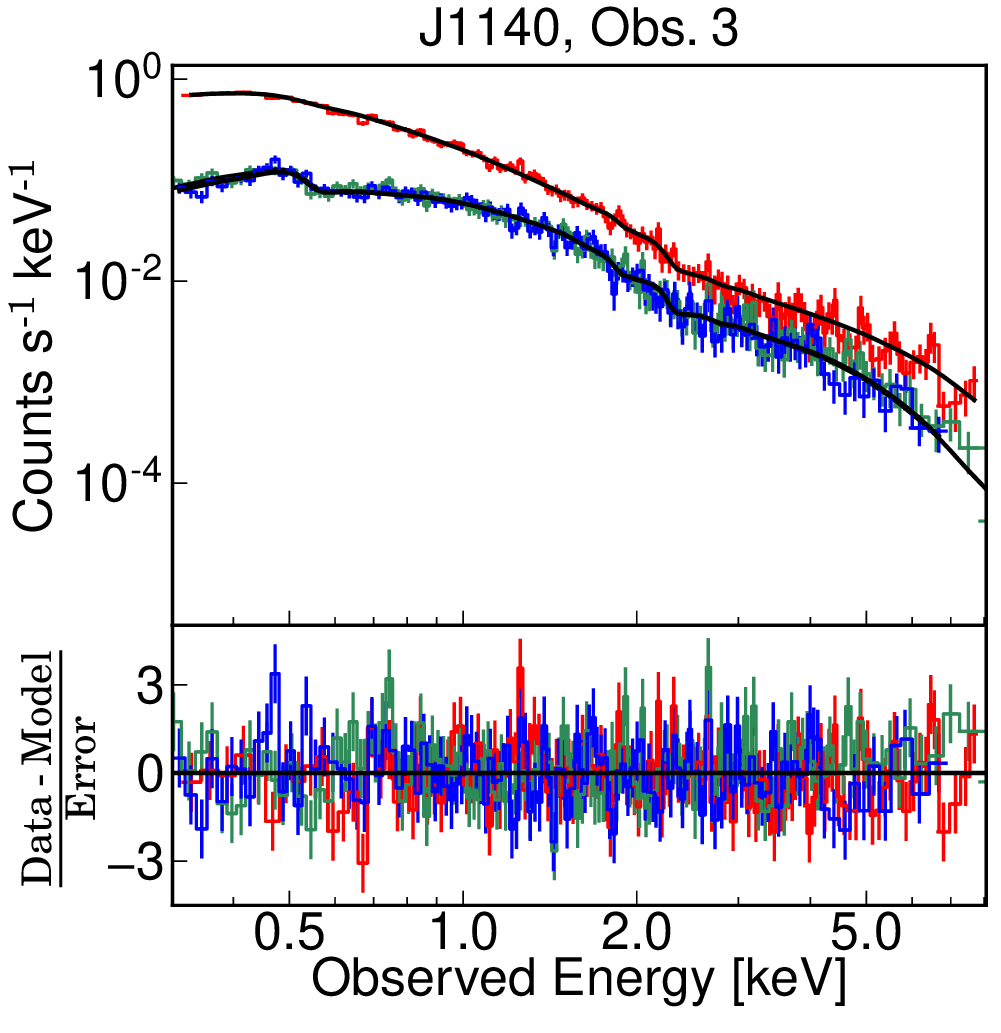}
\includegraphics[scale=0.25,angle=-0]{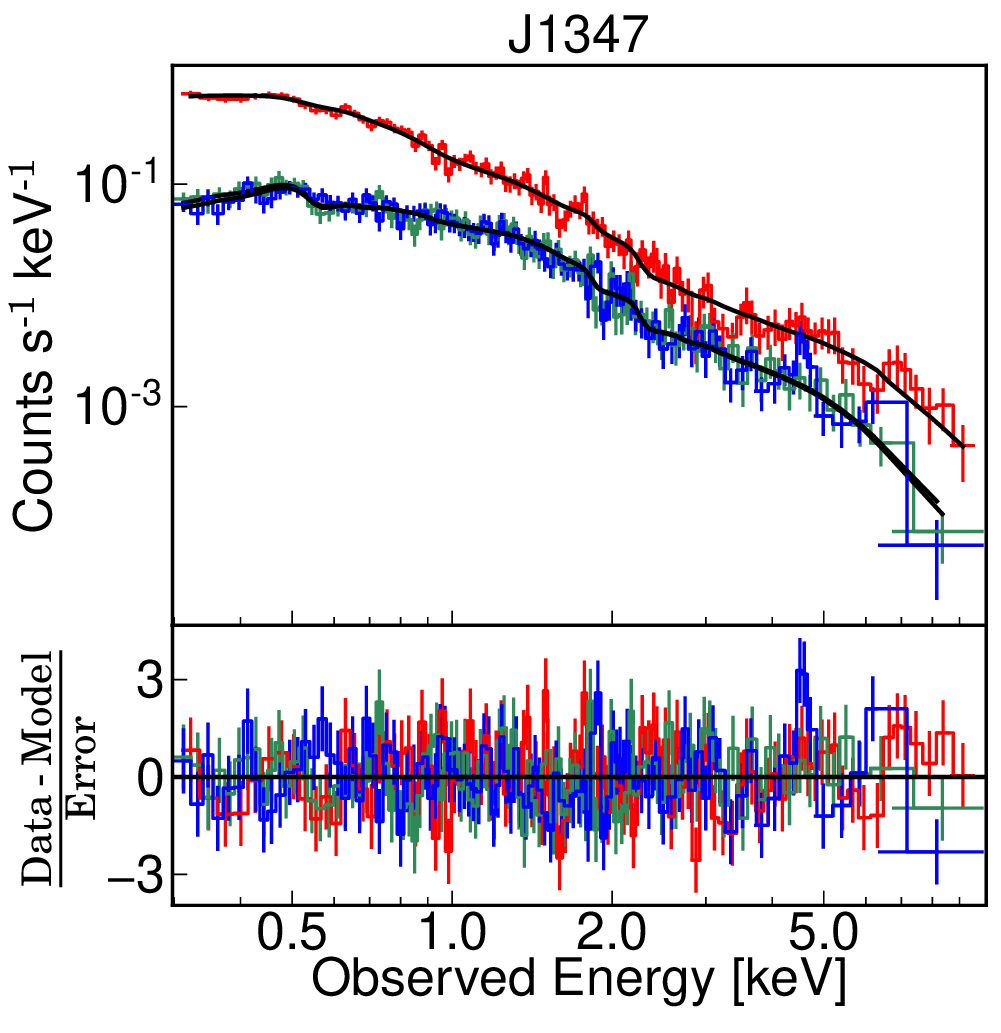}
\includegraphics[scale=0.25,angle=-0]{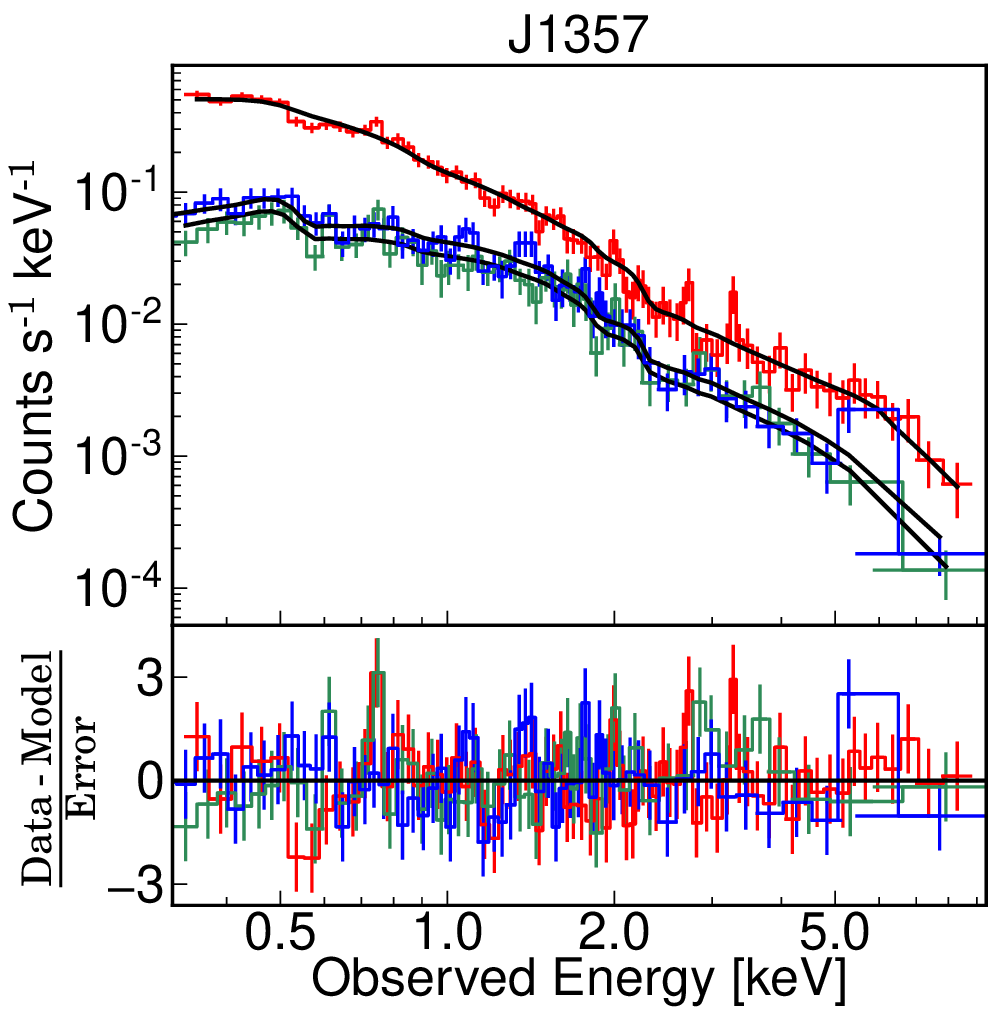}
\includegraphics[scale=0.25,angle=-0]{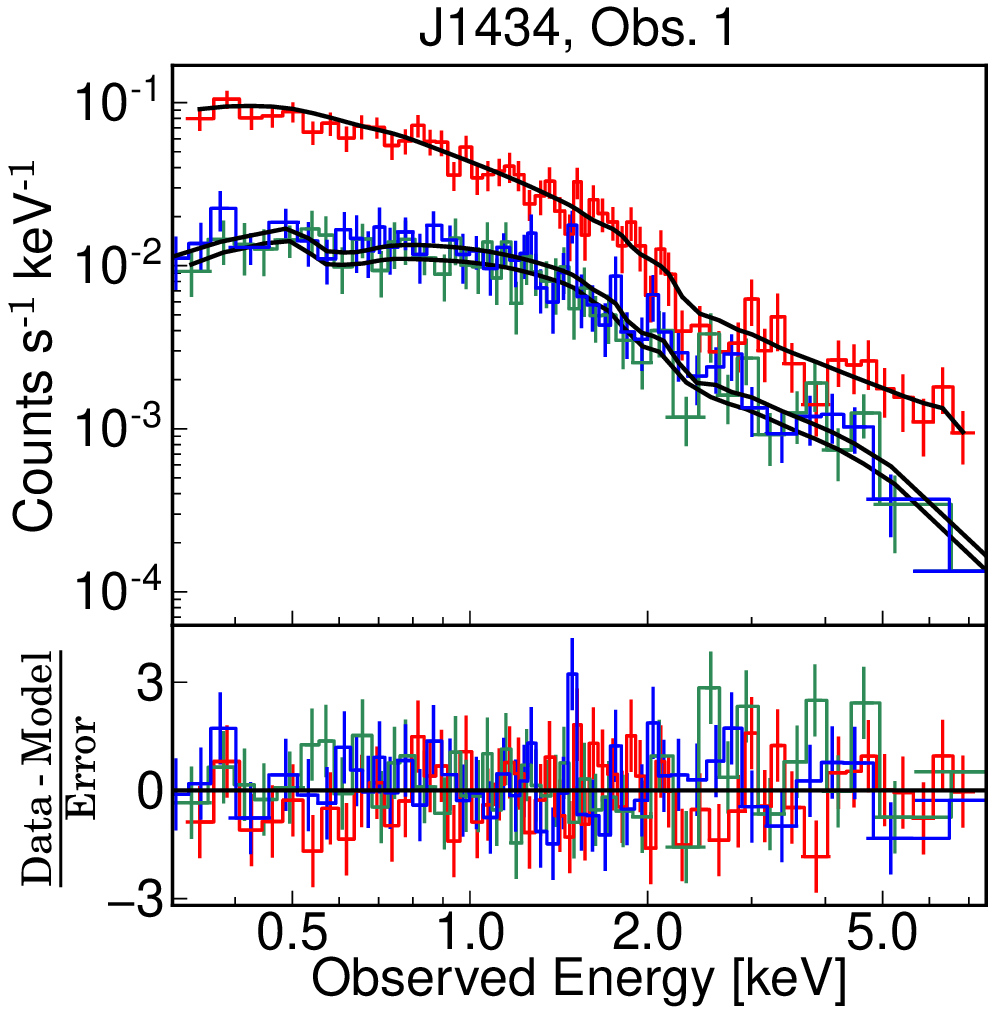}
\includegraphics[scale=0.25,angle=-0]{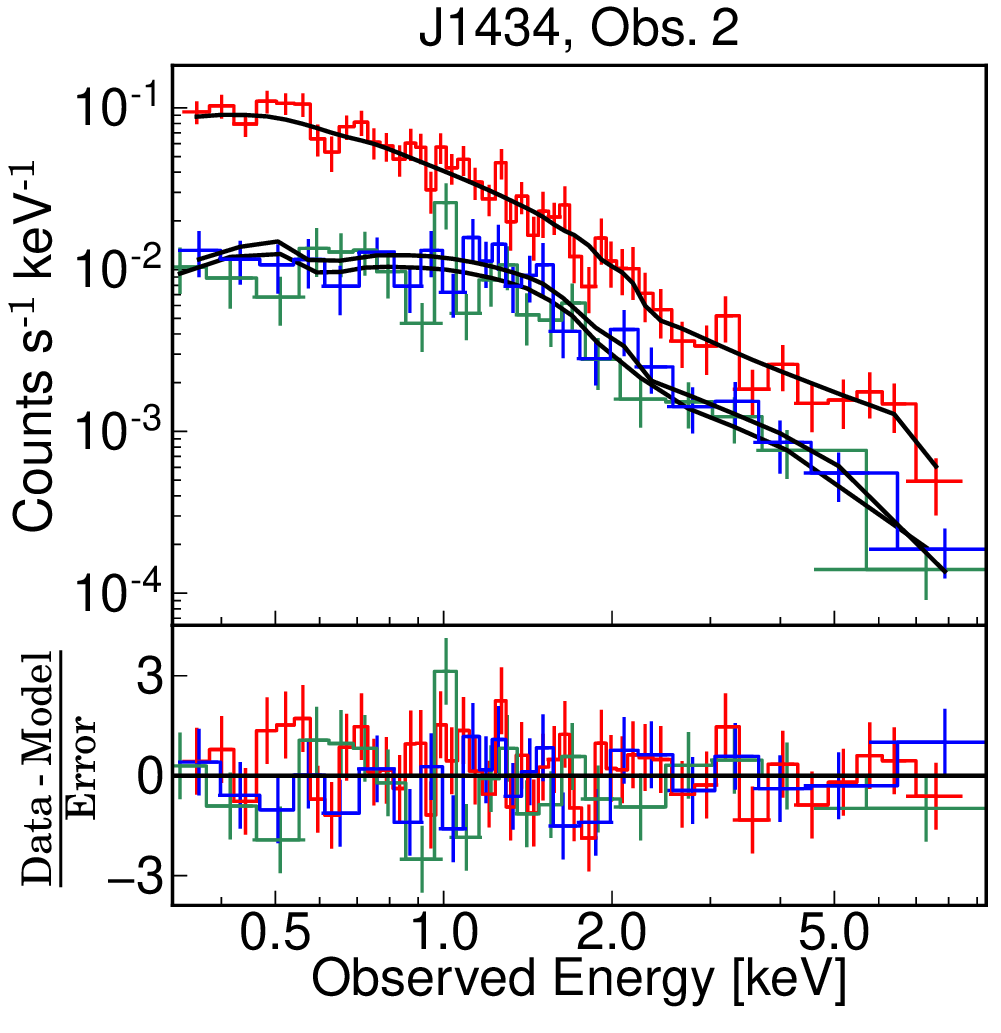}
\includegraphics[scale=0.25,angle=-0]{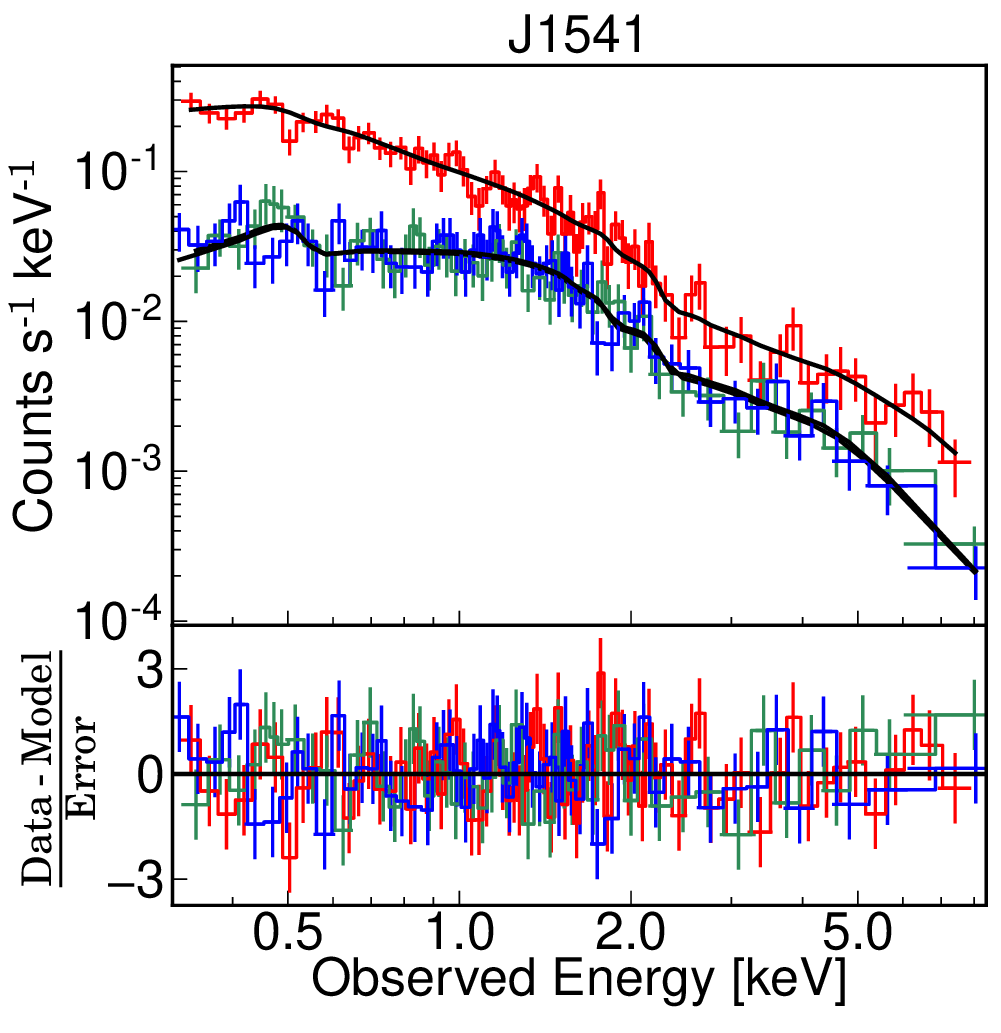}
\includegraphics[scale=0.25,angle=-0]{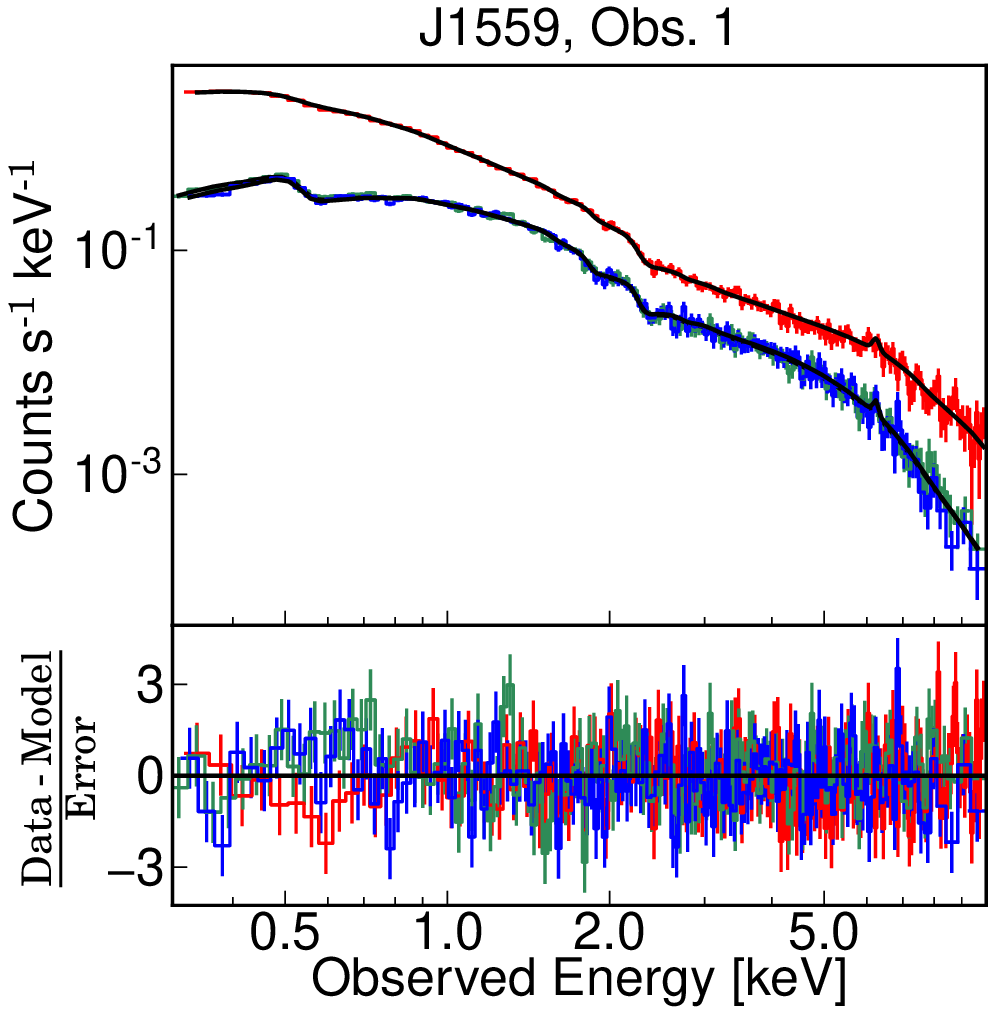}
\includegraphics[scale=0.25,angle=-0]{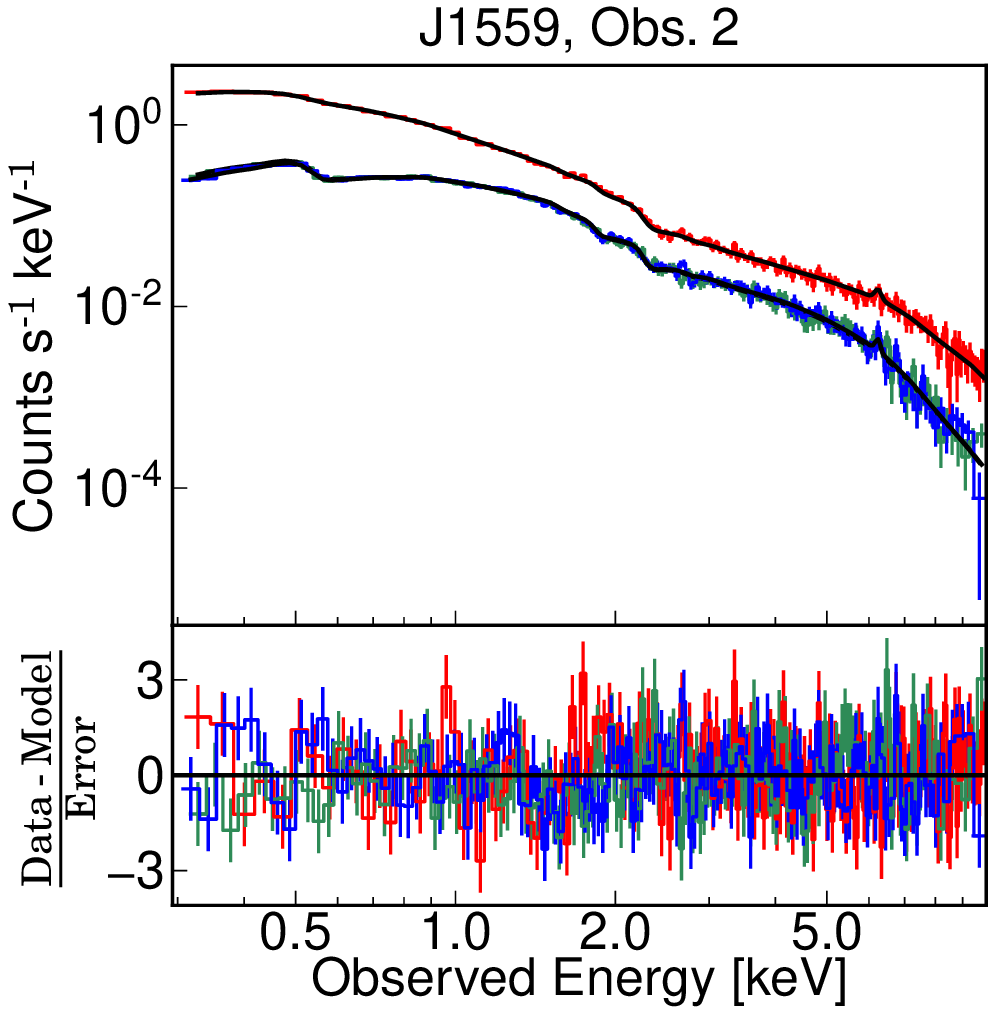}
\includegraphics[scale=0.25,angle=-0]{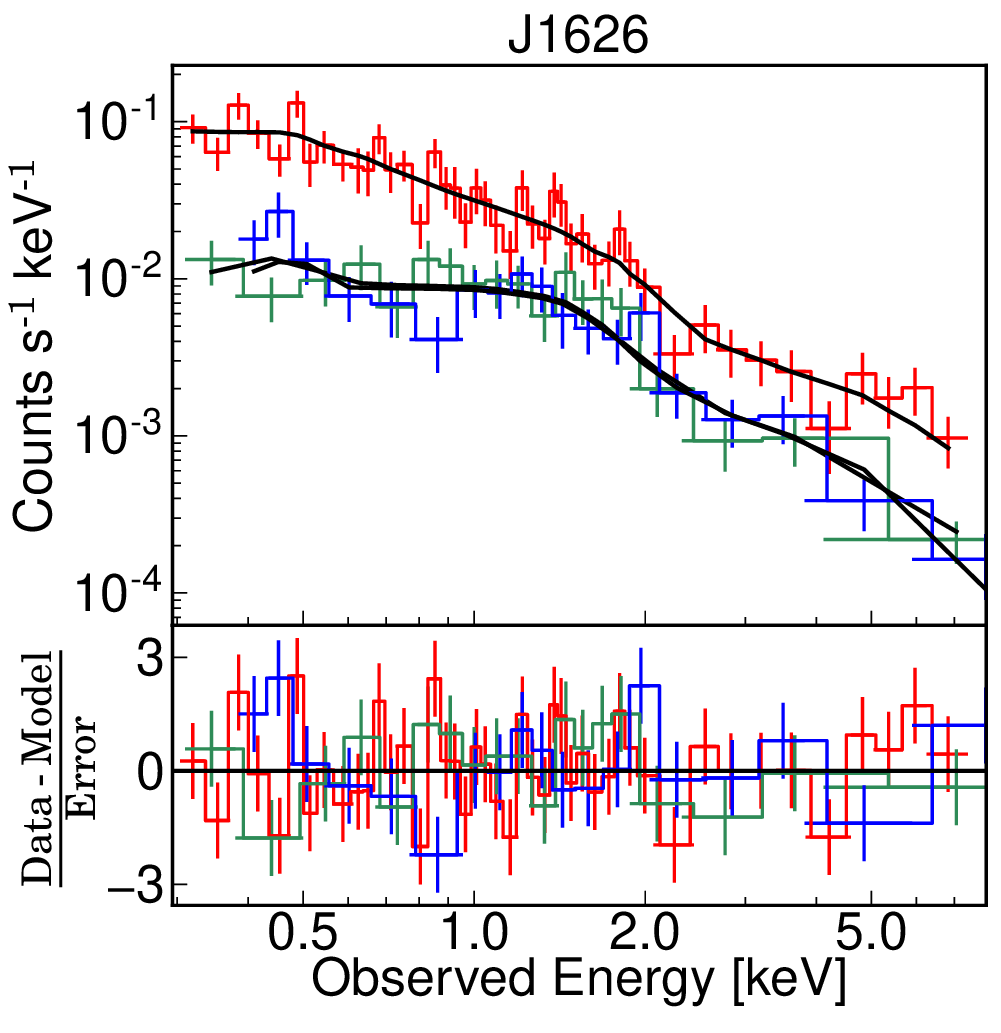}
\includegraphics[scale=0.25,angle=-0]{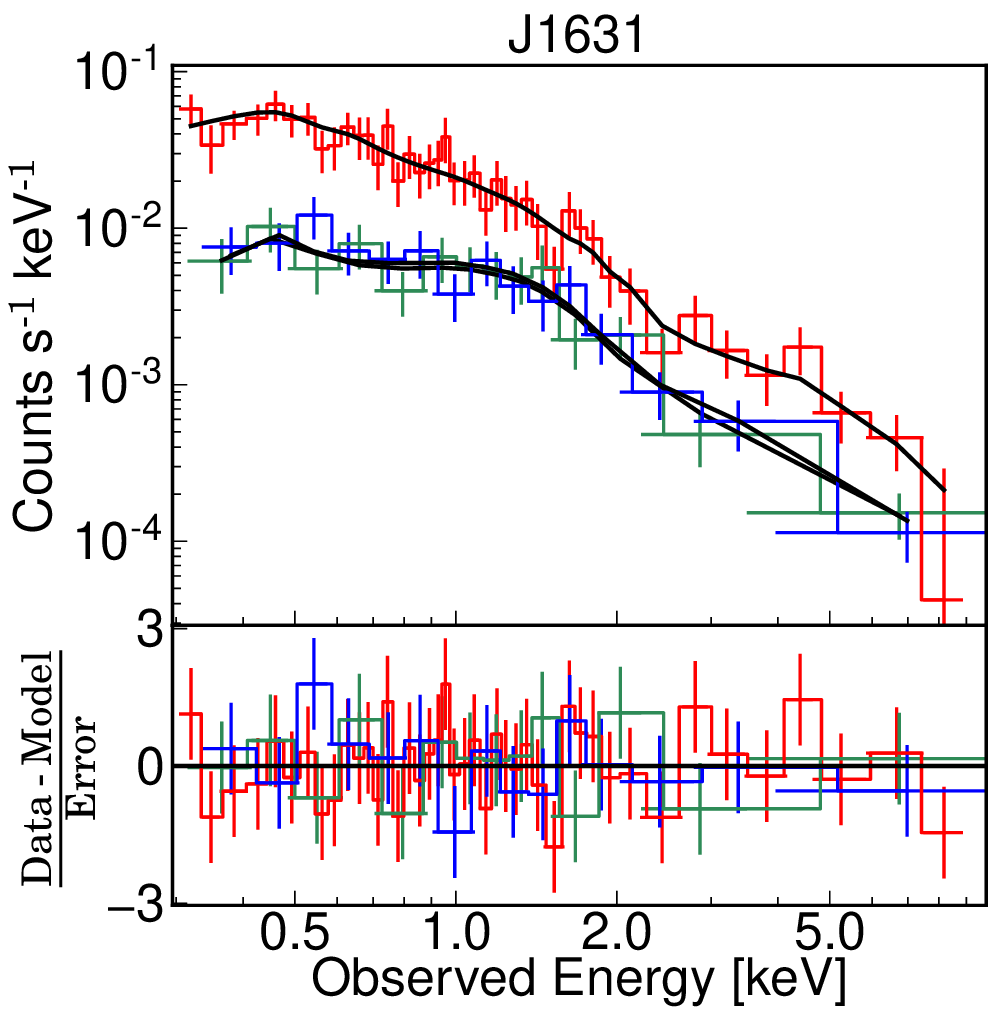}
\includegraphics[scale=0.25,angle=-0]{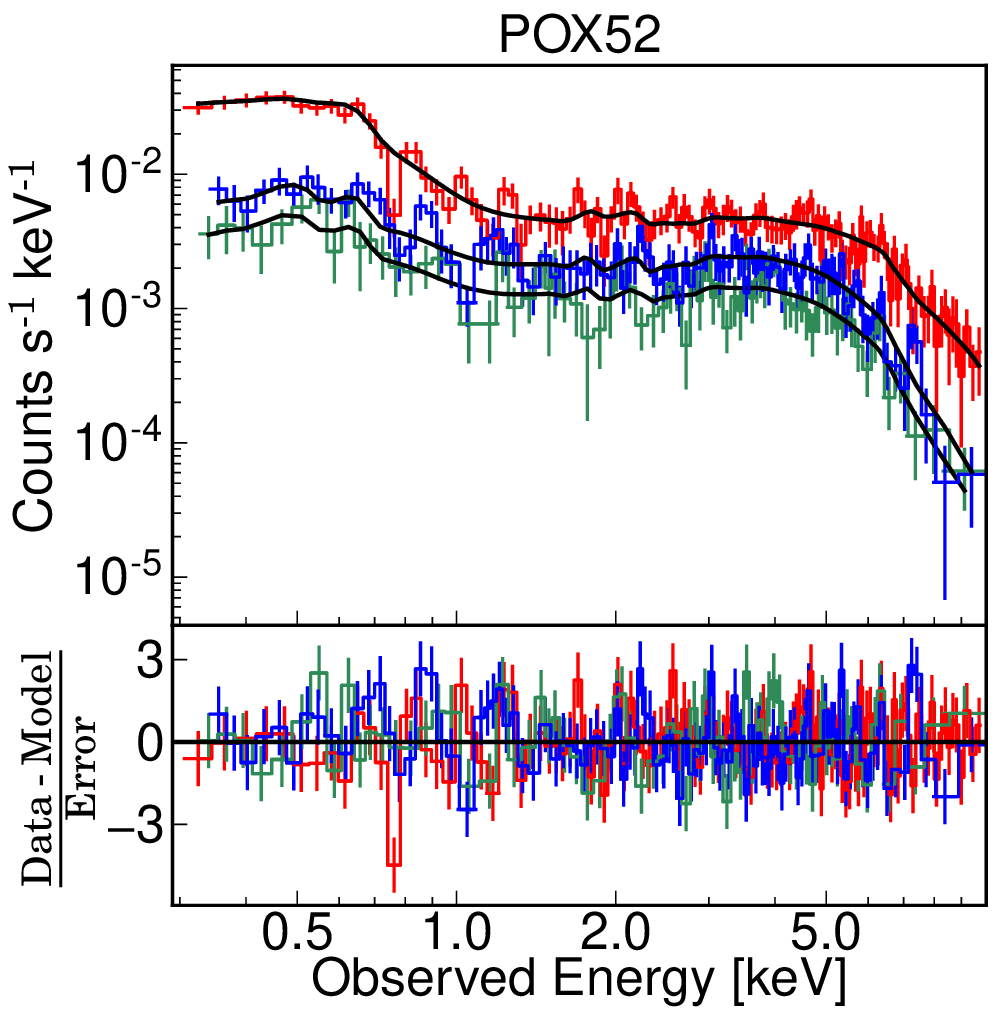}
\caption{The broad-band (0.3$-$10\keV{}) EPIC-pn (red), MOS1 (green), and MOS2 (blue) photon count spectra, the best-fit high-density disc reflection model, {\tt{Tbabs$\times$(relxillDCp$+$nthComp)}} and the residuals. For J1559 and POX52, the best-fit spectral models are {\tt{Tbabs$\times$gabs$\times$(zgauss$+$relxillDCp$+$nthComp)}} and {\tt{Tbabs$\times$zxipcf$\times$(relxillDCp$+$nthComp)}}, respectively.}
\end{center}
\label{spec2}
\end{figure*}

\begin{figure*}
\centering
\begin{center}
\includegraphics[scale=0.3,angle=-0]{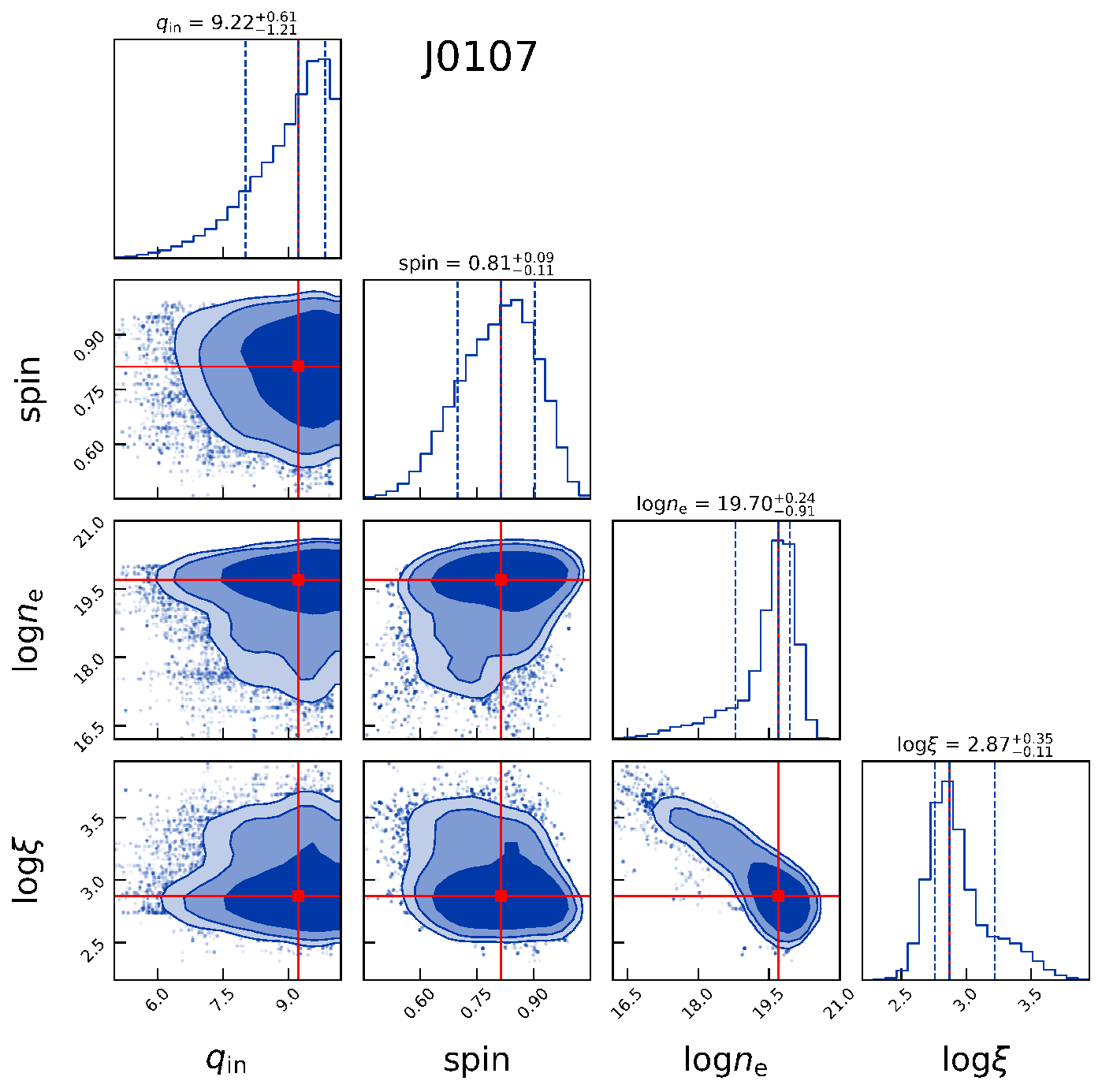}
\includegraphics[scale=0.3,angle=-0]{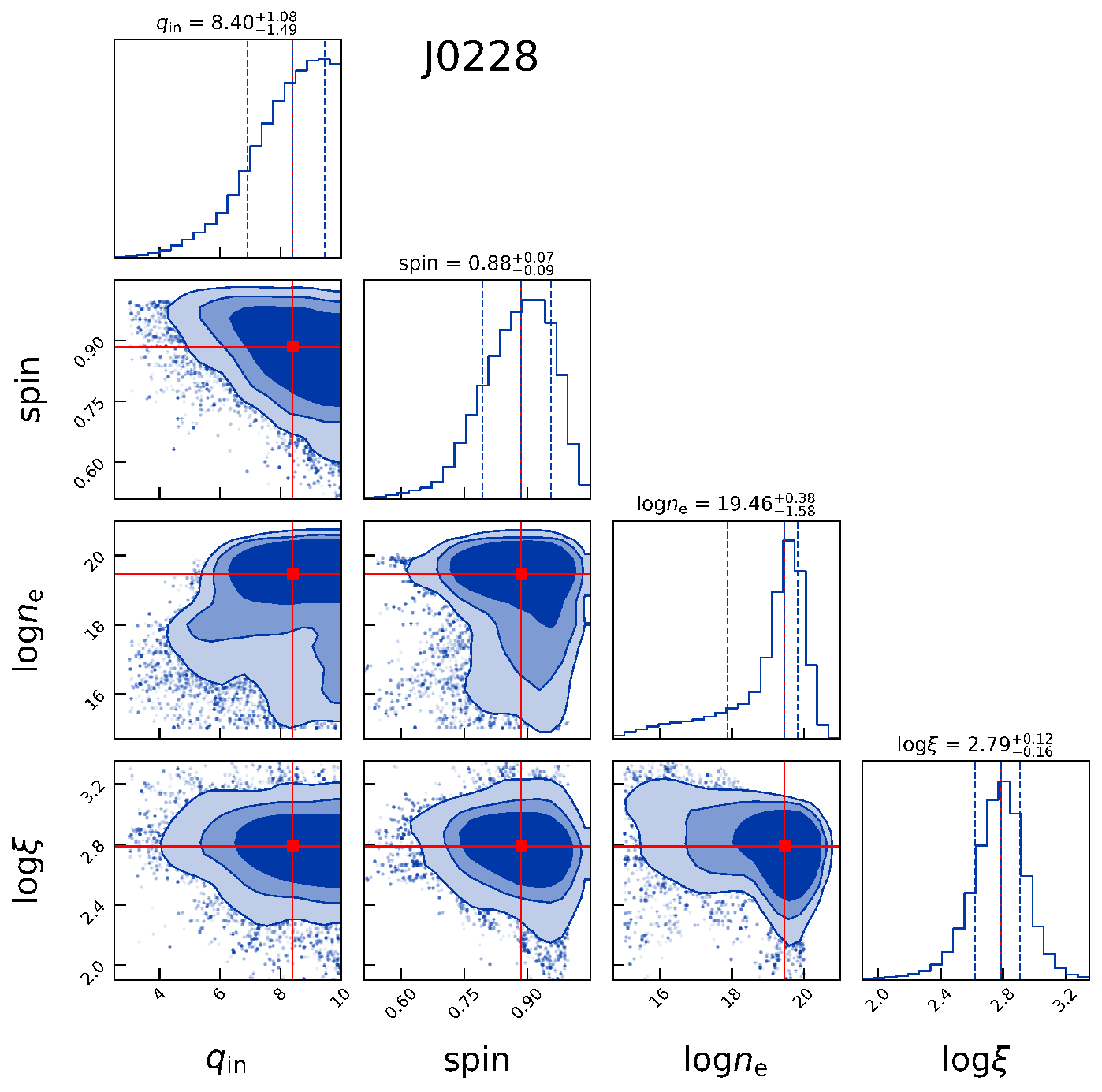}
\includegraphics[scale=0.3,angle=-0]{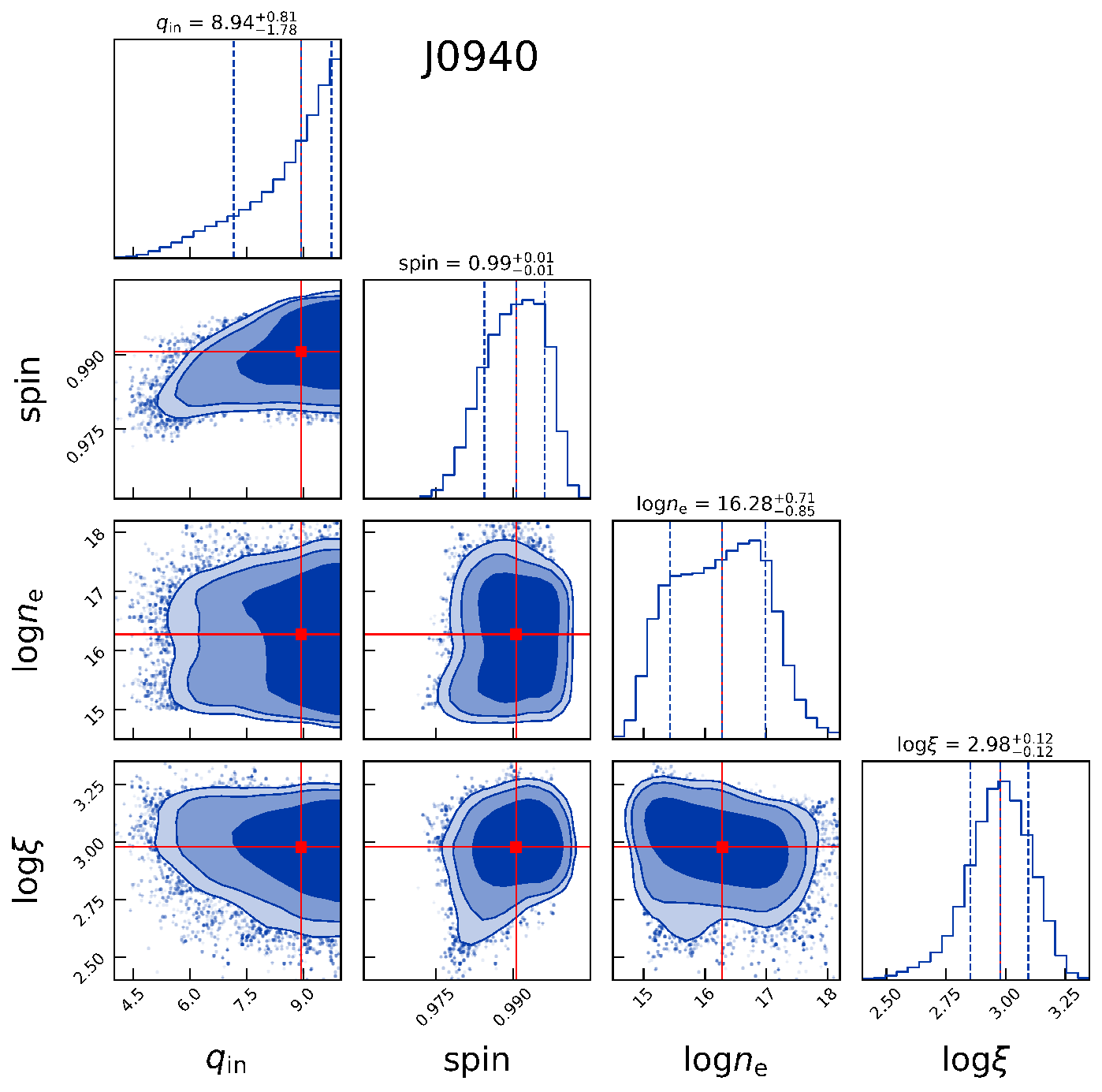}
\includegraphics[scale=0.3,angle=-0]{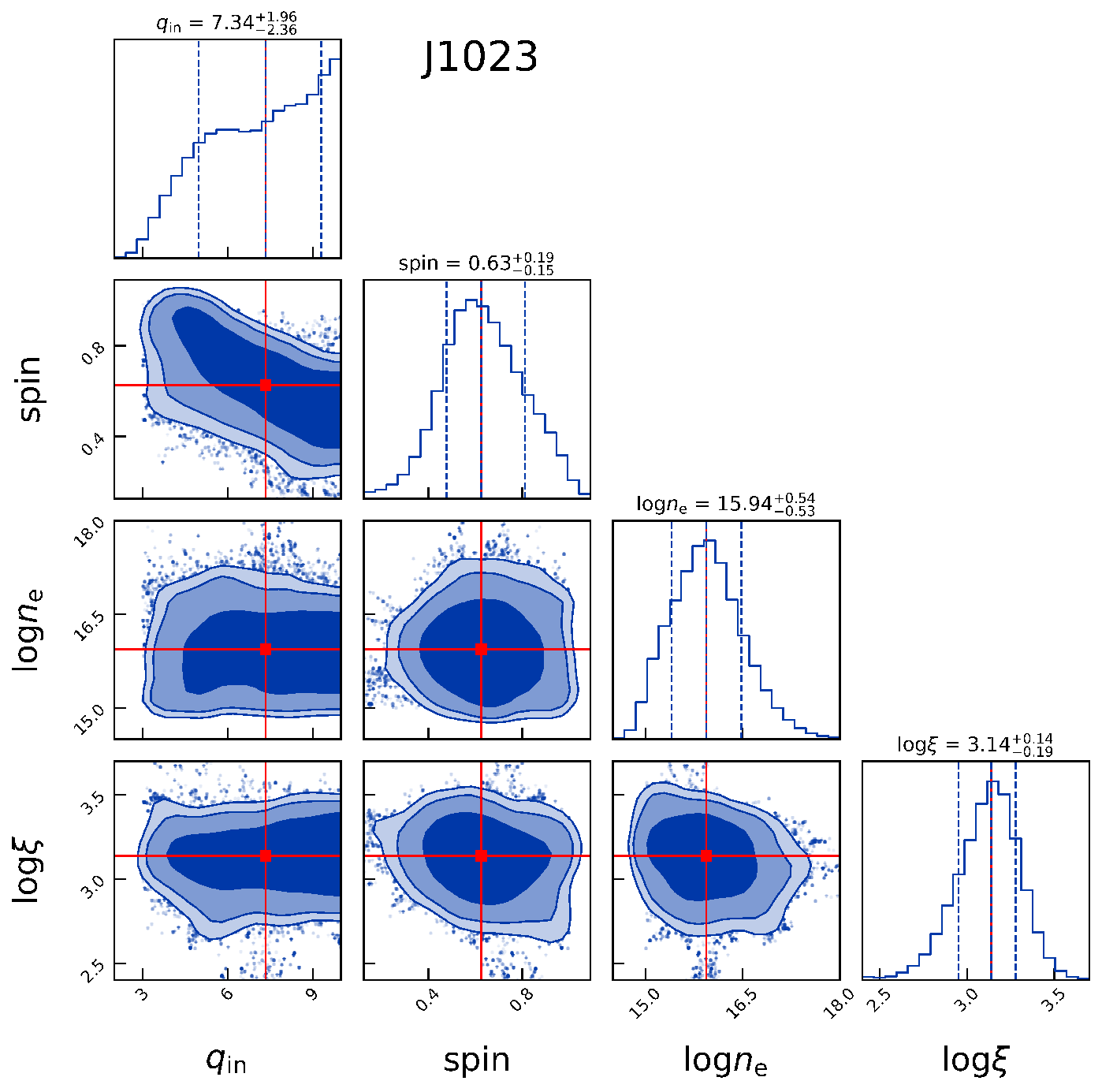}
\includegraphics[scale=0.3,angle=-0]{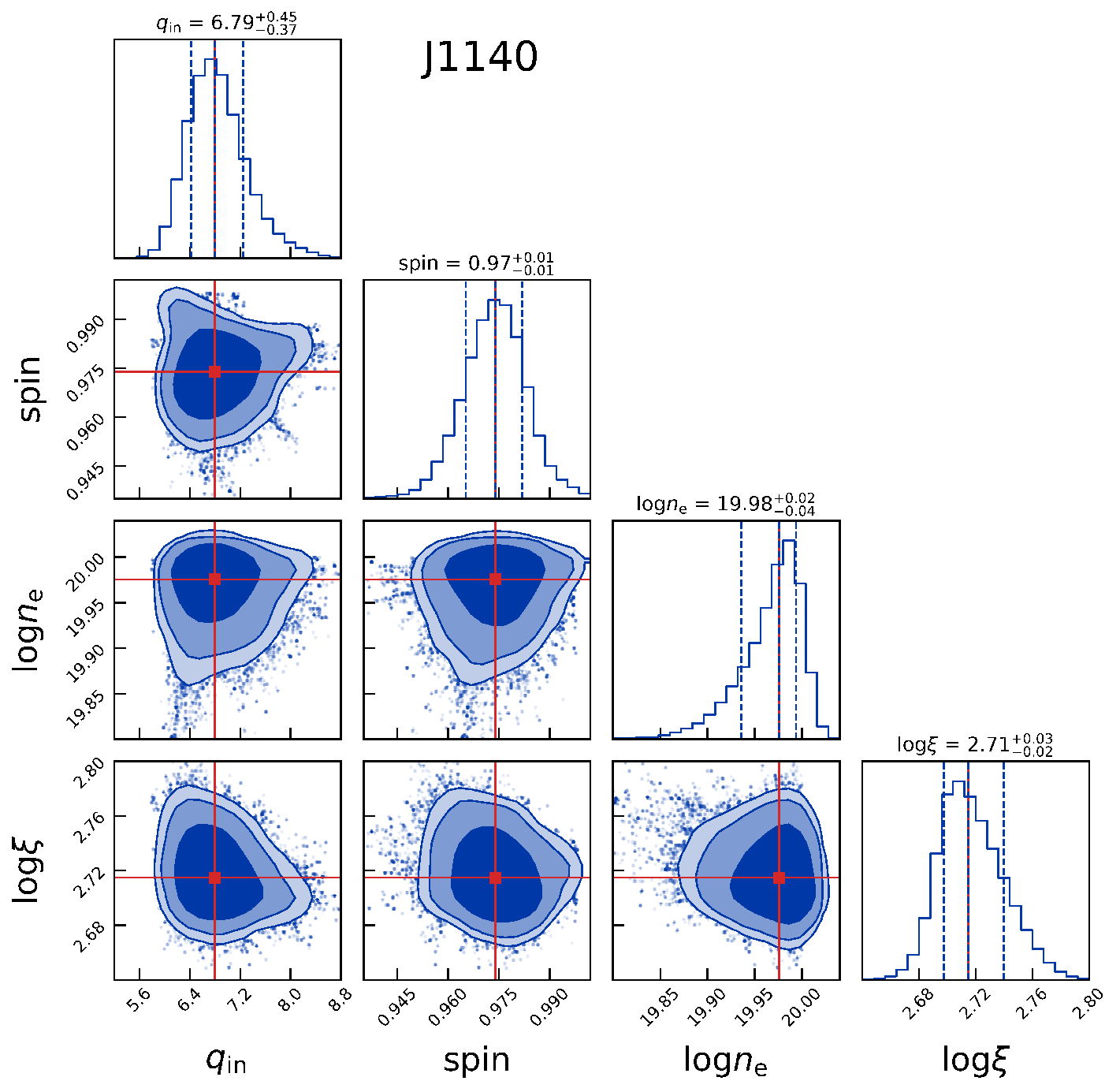}
\includegraphics[scale=0.3,angle=-0]{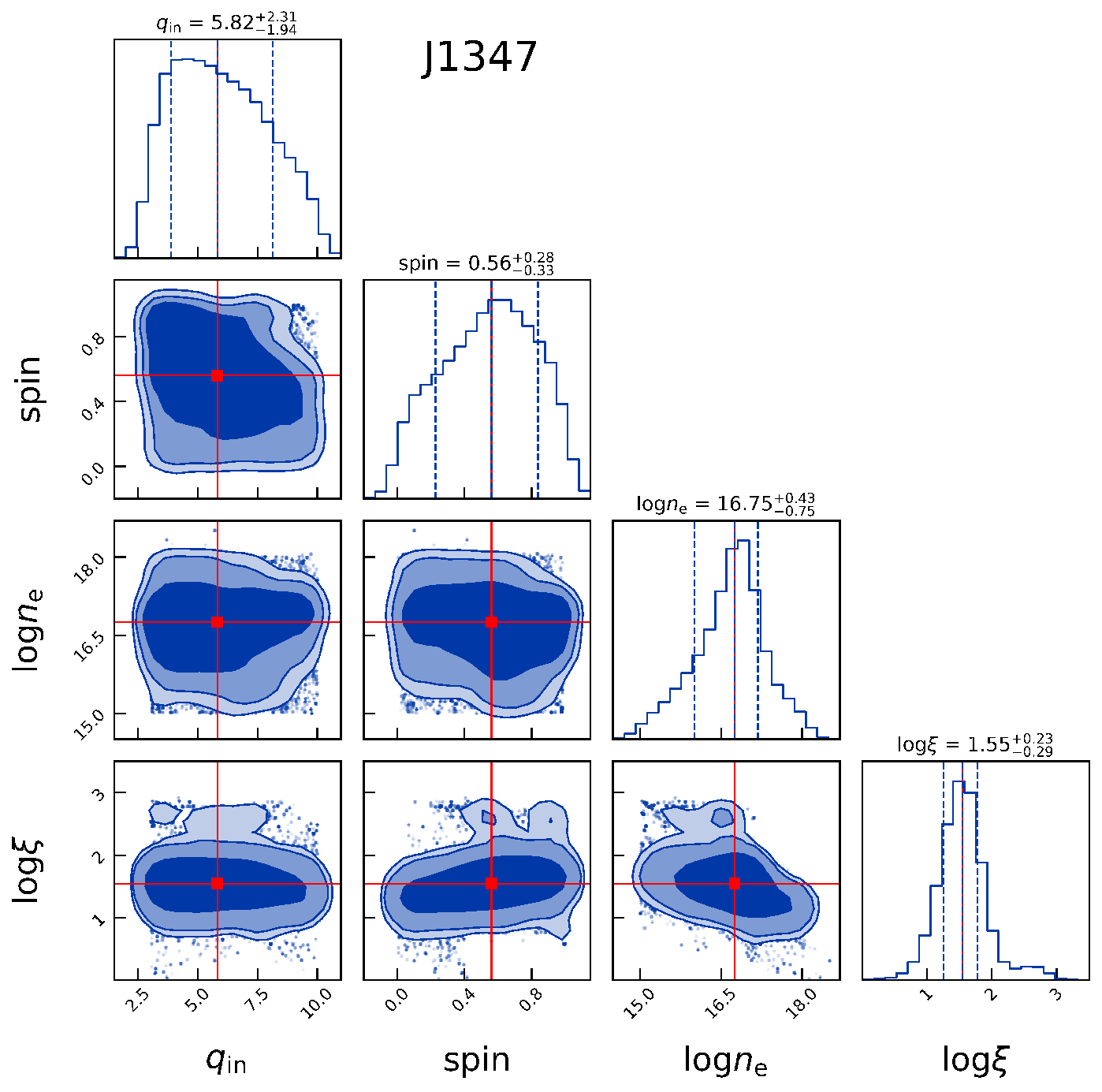}
\caption{Corner plots of various spectral parameters resulting from the MCMC analysis. The histograms on the diagonal show the probability distributions for each parameter with medians and their 1$\sigma$ uncertainties (vertical dashed lines). The off-diagonals show contour plots for each pair of parameters, with dark blue, medium blue, and light blue regions representing 68.3\%, 90\%, and 95\% confidence contours, respectively, and square symbols indicating medians of parameters.}
\end{center}
\label{fig1_mcmc}
\end{figure*}

\begin{figure*}
\centering
\begin{center}
\includegraphics[scale=0.3,angle=-0]{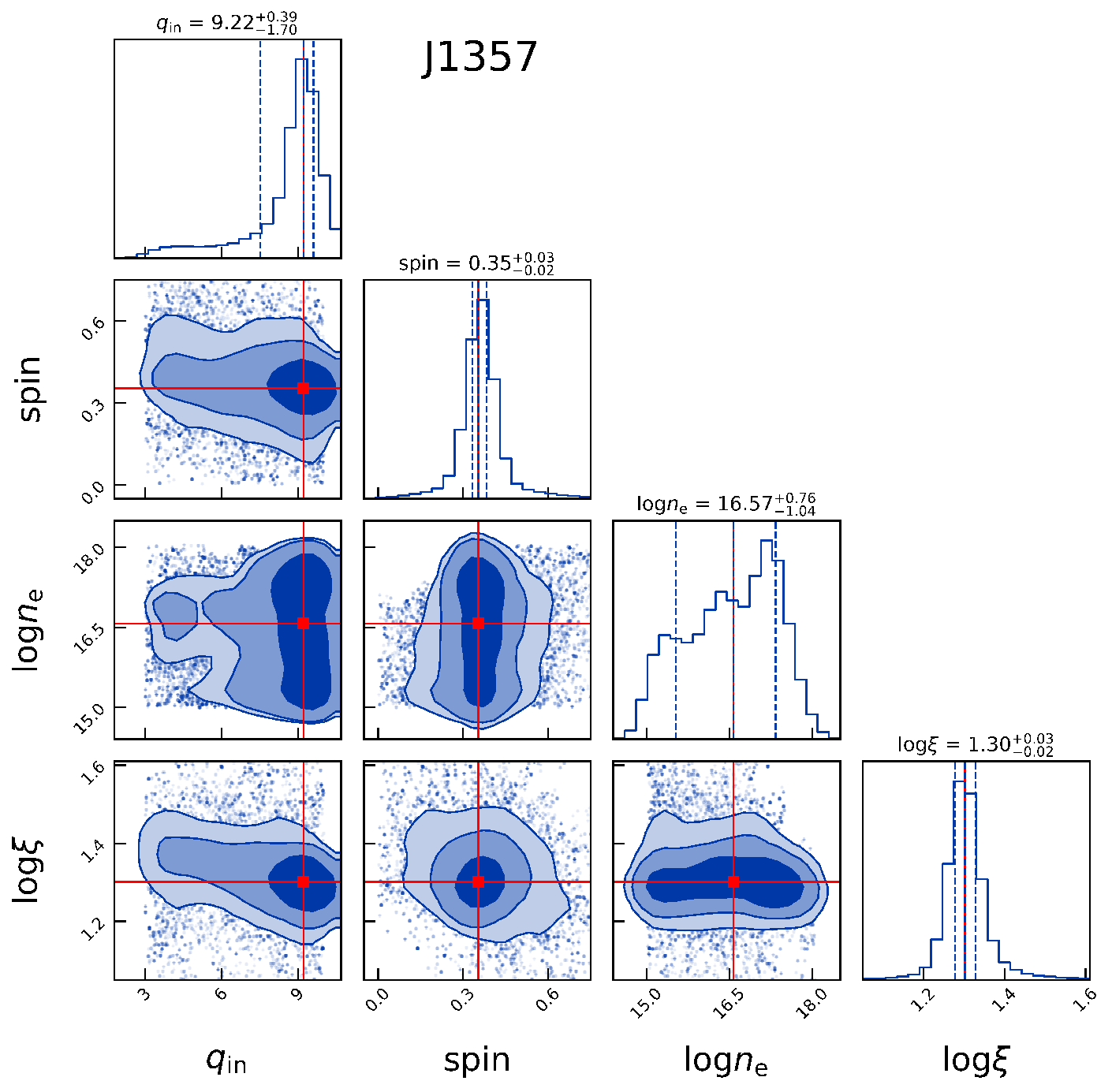}
\includegraphics[scale=0.3,angle=-0]{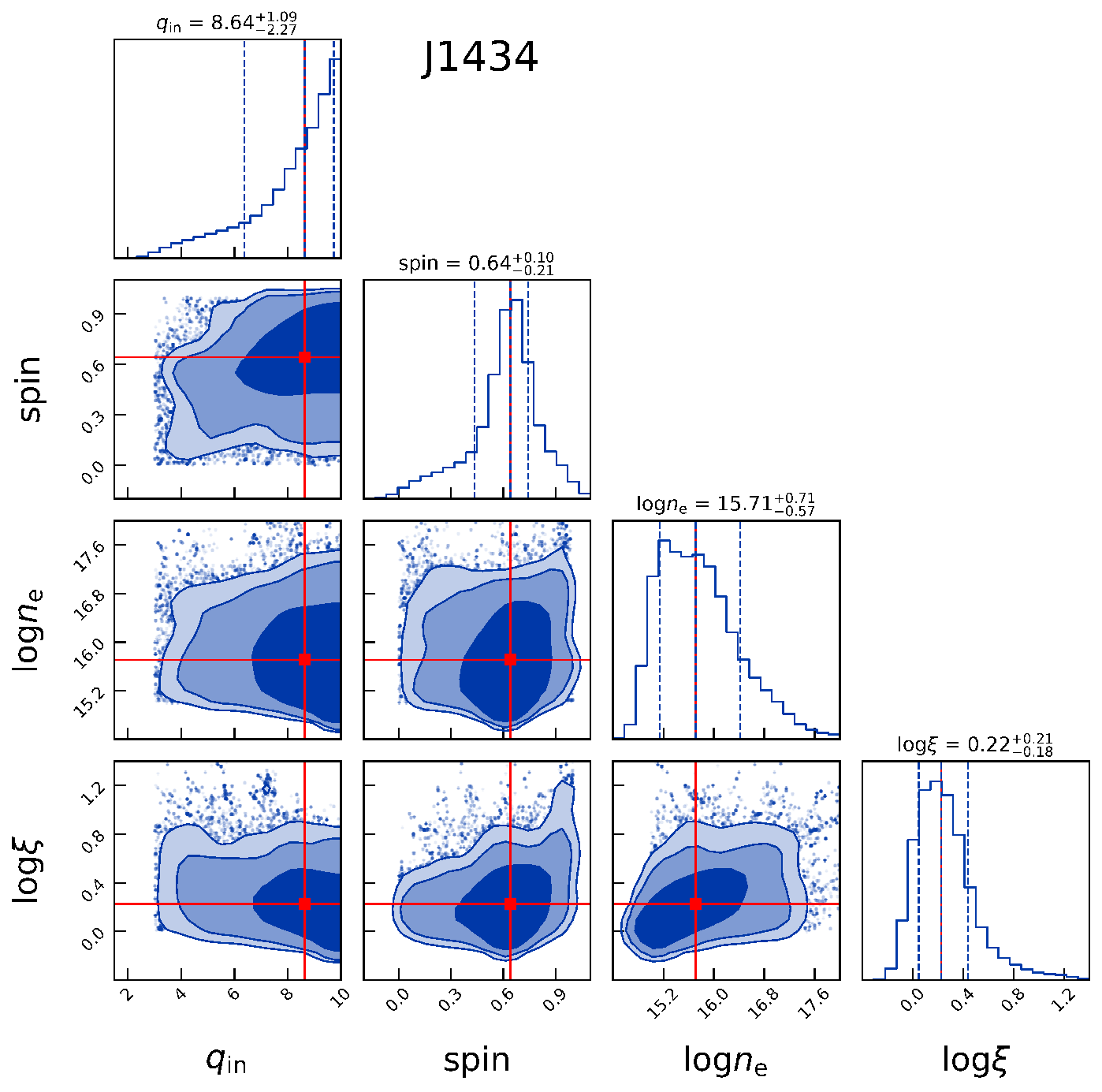}
\includegraphics[scale=0.3,angle=-0]{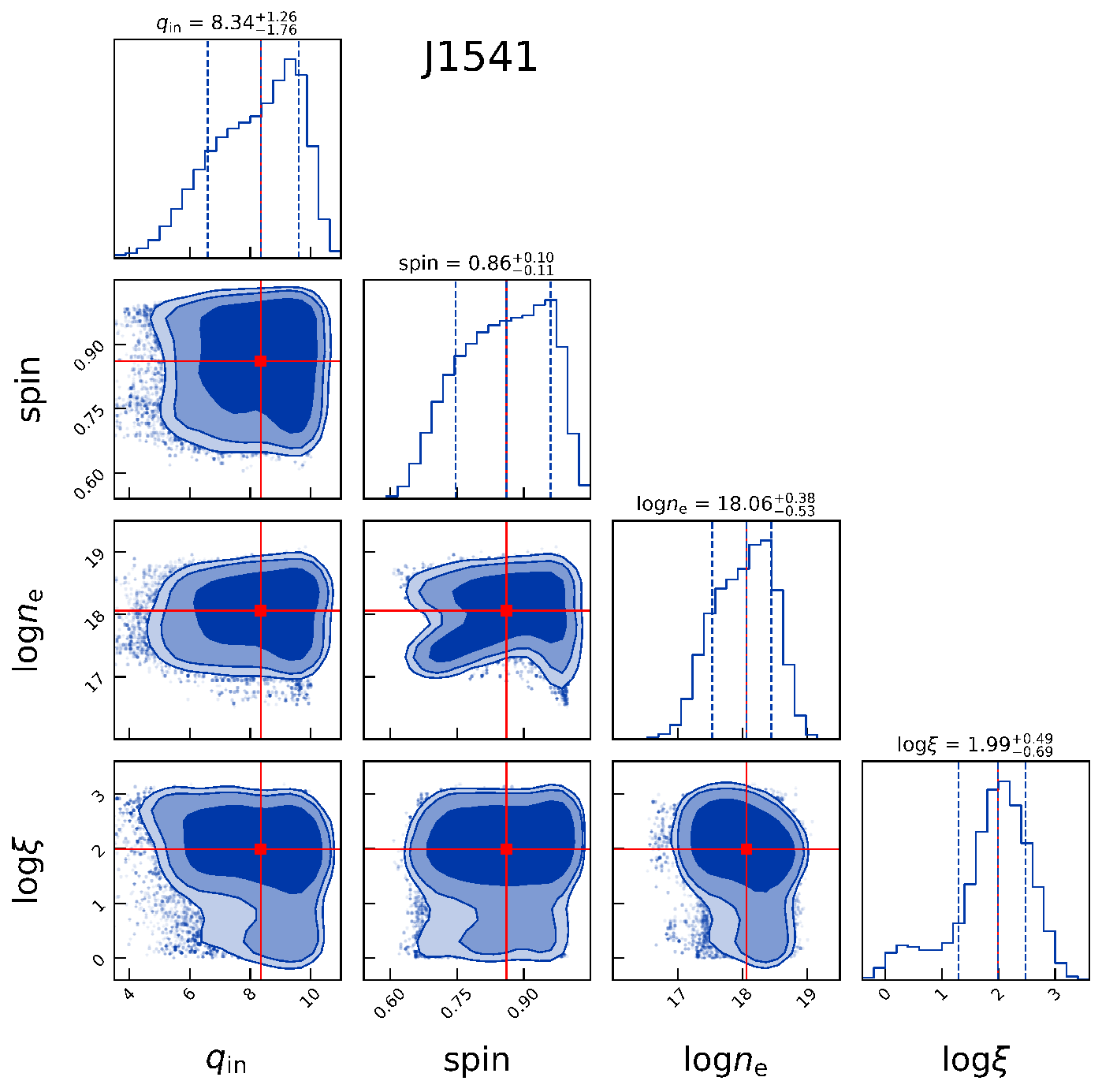}
\includegraphics[scale=0.3,angle=-0]{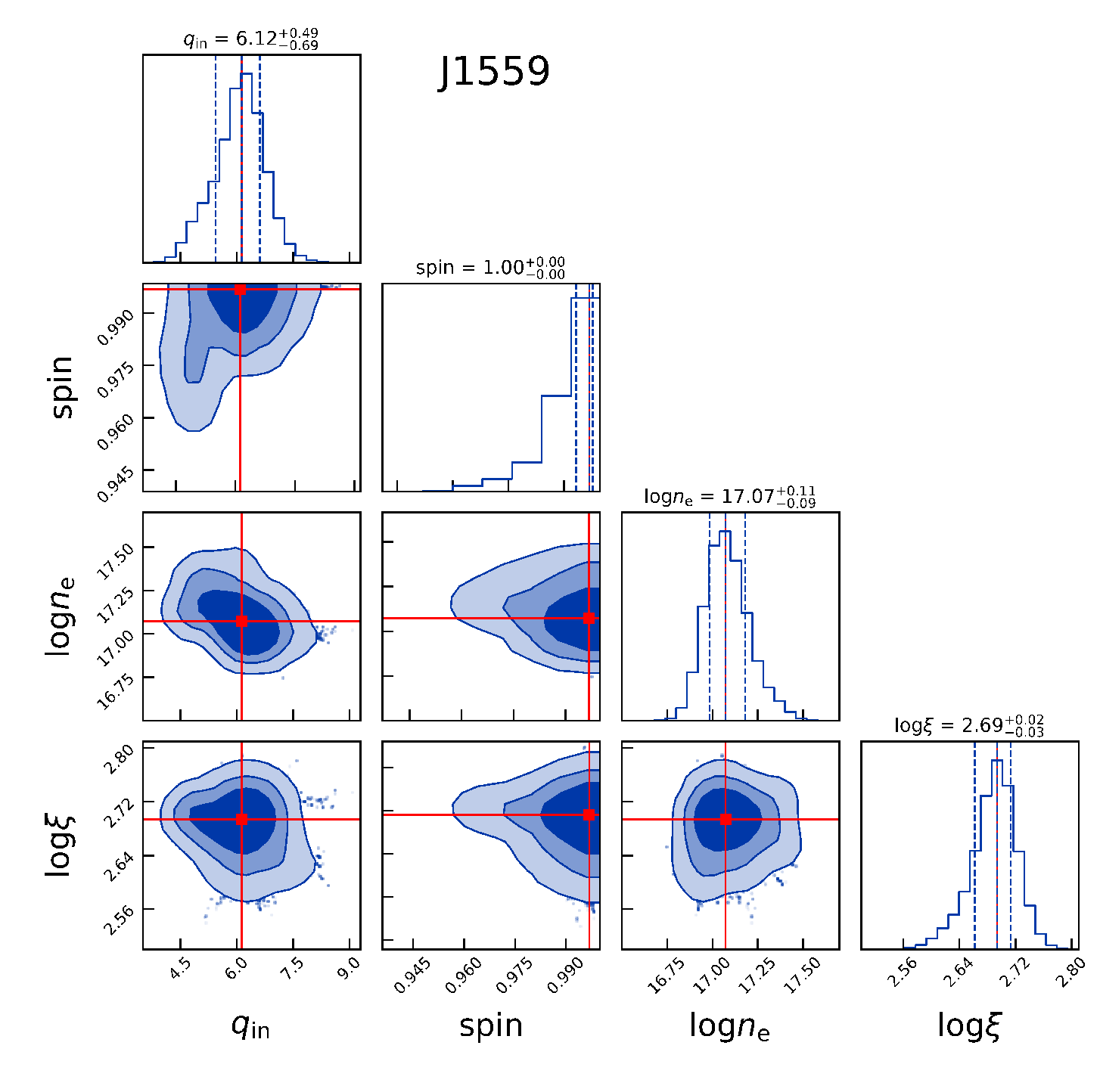}
\includegraphics[scale=0.3,angle=-0]{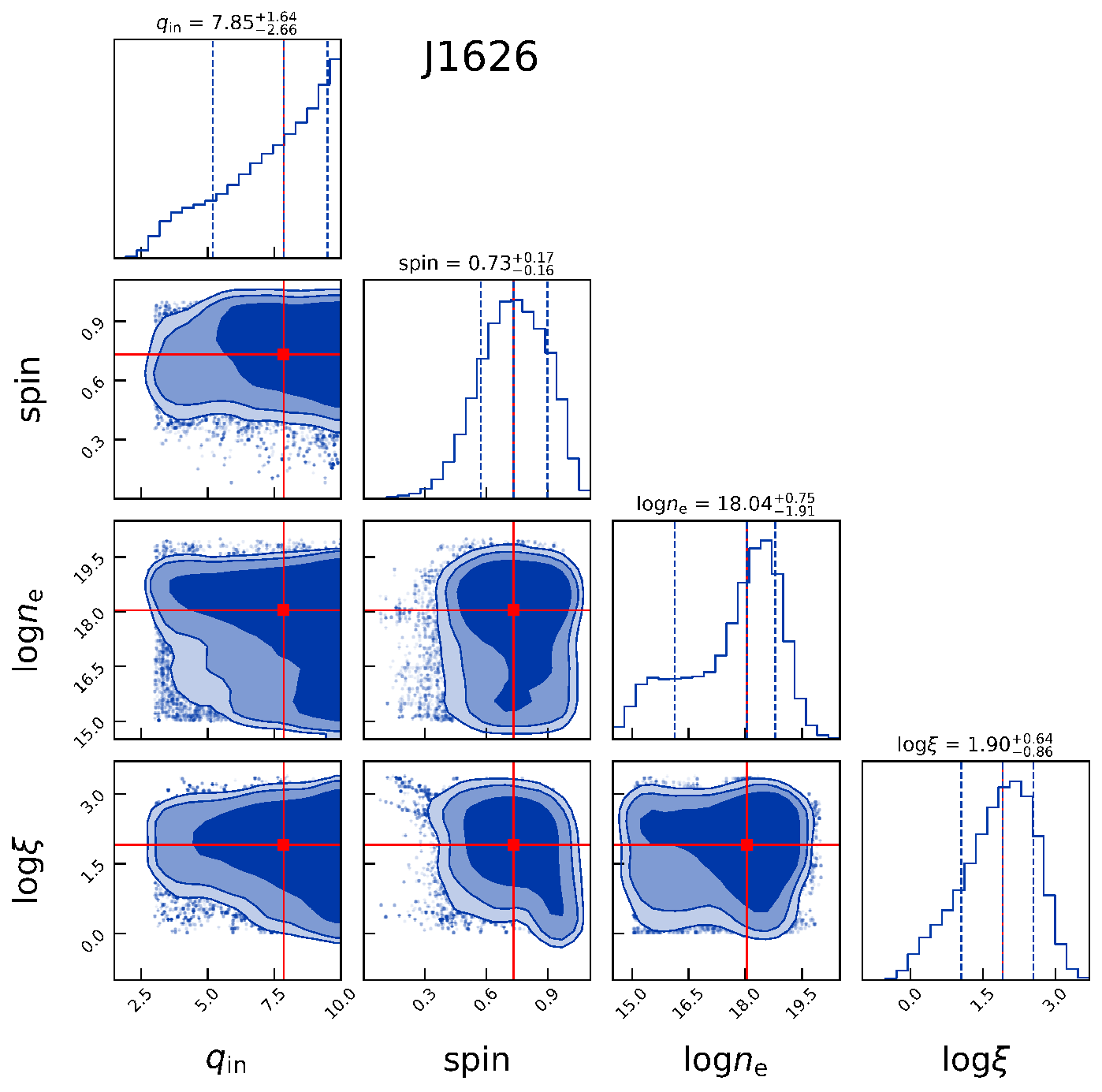}
\includegraphics[scale=0.3,angle=-0]{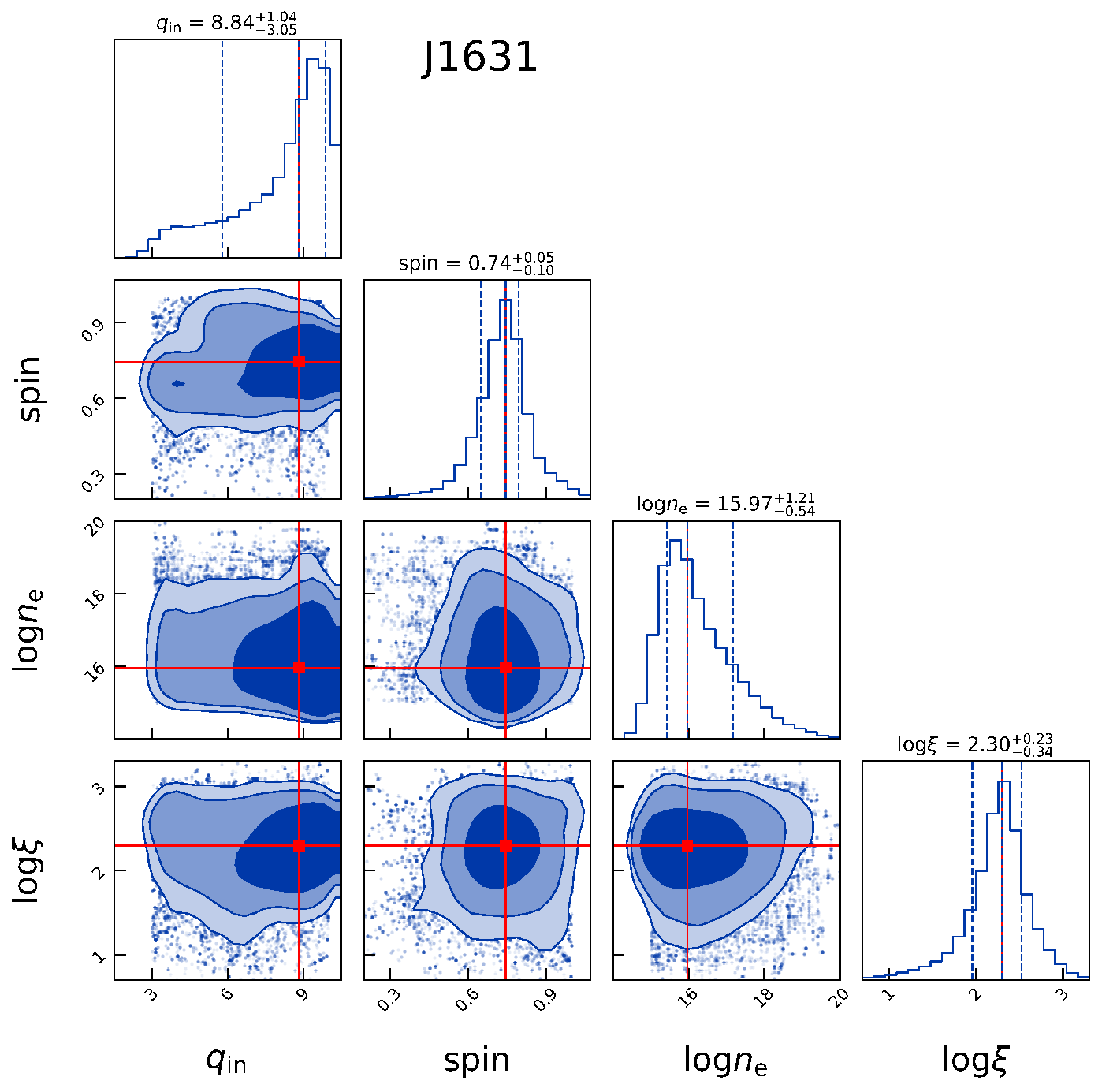}
\caption{Fig.~A3 is continued for J1357, J1434, J1541, J1559, J1626, and J1631.}
\end{center}
\label{fig2_mcmc}
\end{figure*}

\begin{figure*}
\centering
\begin{center}
\includegraphics[scale=0.4,angle=-0]{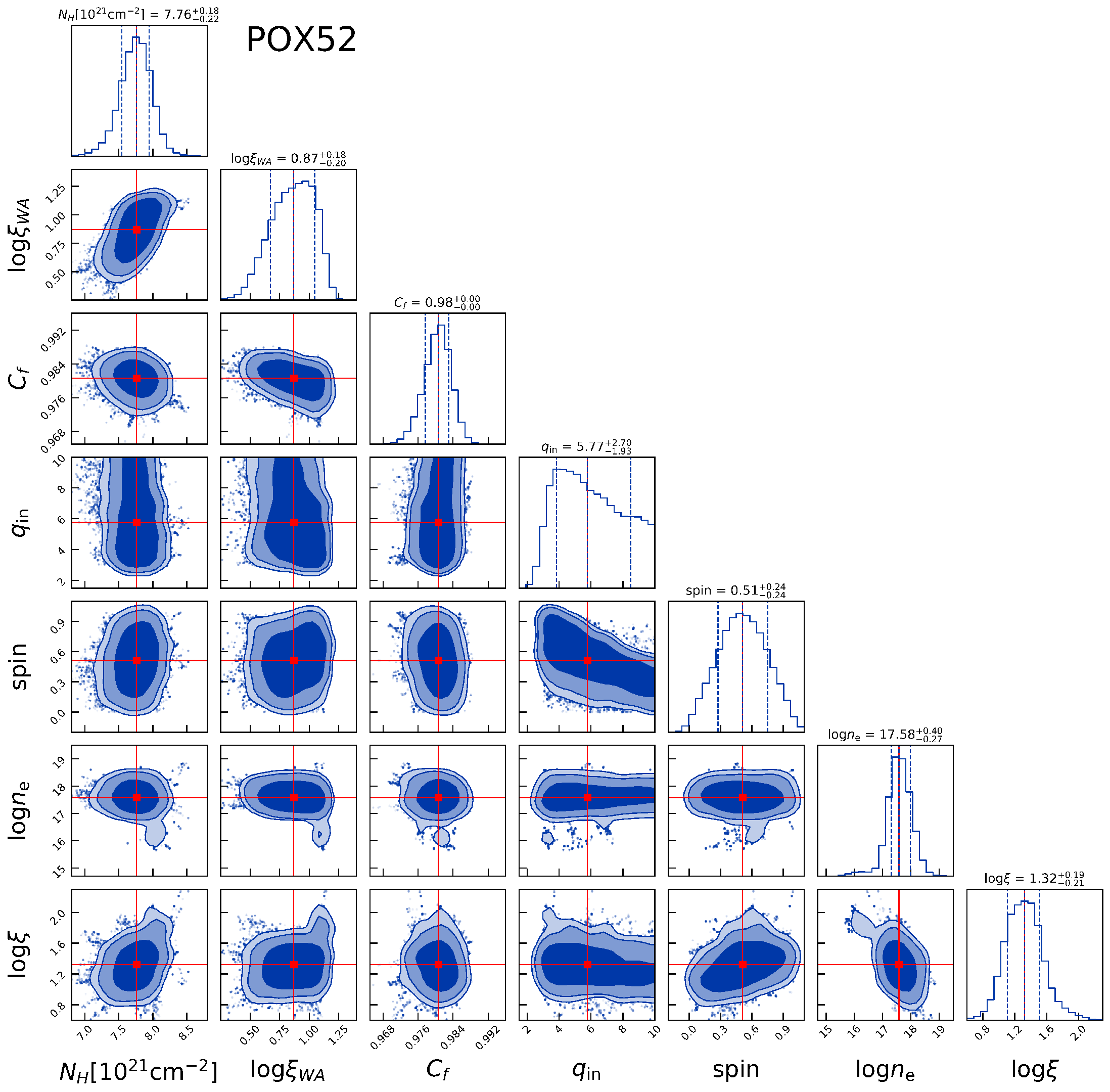}
\caption{Fig.~A3 is continued for various model parameters of POX~52.}
\end{center}
\label{fig3_mcmc}
\end{figure*}

\bsp	
\label{lastpage}
\end{document}